\documentclass[showpacs,showkeys,prc,twocolumn,floatfix,aps]{revtex4-1}
\usepackage{amssymb}
\usepackage{enumerate}
\usepackage[intlimits]{amsmath}
\usepackage{graphicx}
\usepackage{dcolumn}
\usepackage{amsfonts}
\usepackage{bm}

\begin{document}

\title{Investigations of three, four, and five-particle exit channels 
of levels in light nuclei created using a $^{9}$C beam}
\author{R. J. Charity}
\author{J. M. Elson}
\author{J. Manfredi}
\author{R. Shane}
\author{ L. G. Sobotka}
\affiliation{Departments of Chemistry and Physics, Washington University, 
St.~Louis, Missouri 63130, USA.}
\author{B. A. Brown}
\author{Z. Chajecki}
\author{D. Coupland}
\author{H. Iwasaki}
\author{M. Kilburn}
\author{ Jenny Lee}
\author{W. G. Lynch}
\author{A. Sanetullaev}
\author{M. B. Tsang}
\author{J. Winkelbauer}
\author{M. Youngs}
\affiliation{National Superconducting Cyclotron Laboratory and 
Department of Physics and Astronomy, Michigan State University, 
East Lansing, MI 48824, USA.}
\author{S. T. Marley}
\author{D. V. Shetty}
\author{A. H. Wuosmaa}
\affiliation{Department of Physics, Western Michigan University, Kalamazoo, 
Michigan 49008, USA.}
\author{T. K. Ghosh}
\affiliation{Variable Energy Cyclotron Centre, 1/AF Bidhannagar, 
Kolkata 700064, India}
\author{M. E. Howard}
\affiliation{Department of Physics and Astronomy, Rutgers University, 
New Brunswick, New Jersey 08903, USA}

\begin{abstract}
The interactions of a $E/A$=70-MeV $^9$C beam  with a Be target 
was used to populate
levels in Be, B, and C isotopes which undergo decay into many-particle 
exit channels. The decay products were detected in the HiRA array and the 
level energies were identified from their invariant mass. Correlations 
between the decay products were examined to deduce the nature 
of the decays, 
specifically  to what extent all the fragments were created in one prompt 
step or whether the disintegration proceeded in a sequential fashion through long-lived 
intermediate states.  In the latter case, information on the spin of the level 
was also 
obtained.  Of particular interest is the 5-body decay of the 
$^8$C ground state which was found to disintegrate in two steps of two-proton 
decay passing through the $^6$Be$_{g.s.}$ intermediate state. 
The isobaric analog of $^8$C$_{g.s.}$ in $^8$B was also found to undergo 
two-proton decay to the isobaric analog of $^6$Be$_{g.s.}$ in $^6$Li. A 9.69-MeV state
 in $^{10}$C was found to undergo prompt 4-body decay to the 
2\textit{p}+2$\alpha$ exit channel. The two protons were found to have a strong 
enhancement
 in the diproton region and the relative energies of 
all four \textit{p}-$\alpha$ pairs were consistent 
with the $^5$Li$_{g.s.}$ resonance.  
\end{abstract}

\maketitle

\section{INTRODUCTION}

\label{sec:intro}

The ground and excited states of many light proton-rich nuclei decay by
 the emission of protons and other charged particles. A number of these 
states disintegrate into just protons and alpha particles 
and other very light fragments.
Well known examples are the ground states of $^{6}$Be and $^{9}$B 
which decay to the 2\textit{p}+$\alpha$ and the \textit{p}+2$\alpha$ 
exit channels. Such states can be studied with resonance decay spectroscopy 
where all of the decay fragments are detected. The excitation energy of the 
parent state can be determined from the invariant mass of the detected 
decay products. For exit channels with more than two fragments, the angular 
and energy correlations between the fragments can provide information on the 
spin of the state and the nature of its decay, i.e., whether the decay proceeds 
through a series of sequential 2-body decay steps or via a more prompt process.

In this work we have utilized a $^9$C beam to study a number of such states which were
 formed via nucleon knockout and more complicated reactions.
Of particular interest is the ground state of $^8$C which is known to be
unstable to the 5-body exit channel 4\textit{p}+$\alpha$. Information about $^8$C 
is rather sparse and the 5-body exit channel has never been observed experimentally.
A particular question for $^8$C disintegration and other many-particle exit channels
detected in this work is whether the final particles were created in one 
fast process or in a series of sequential steps with the creation of long-lived 
intermediate states? Of course there is not always a clear demarcation between 
such processes. If the lifetime of the intermediate states becomes too short, 
does it make sense to still describe the decay as two or more sequential steps?

One situation that is particularly clear is the two-proton decay 
described by Goldansky \cite{Goldansky60} where no intermediate state can be 
energetically accessed. Examples of this are the two-proton decays of 
$^{45}$Fe \cite{Miernik07} and $^{54}$Zn\cite{Blank05}. 
In cases where this condition is not reached, Bochkarev \textit{et al.} 
considered a distinction in the decay type based on the width of the potential 
intermediate state.
For a 3-body exit channel, if the potential intermediate state has a width 
$\Gamma_{I}$  which is of the same magnitude as the kinetic energy $E_{k}^1$ released 
in the 
decay to this state, then its decay was deemed democratic, i.e., 
all energy scales in the subsystems and the three-body systems are 
comparable. We have constructed the ratio 
\begin{equation}
R_{E} = \frac{\Gamma_{I}}{E_{k}^1}
\end{equation}
 to quantify this. Thus if $R_{E} \ll$ 1, the decay is sequential, 
where as if $R_{E} \geq$ 1,  then the decay is democratic. Large values of 
$\Gamma_{I}$ and thus of $R_{E}$ imply a short-lived intermediate state 
and thus a difficulty in distinguishing the two decay steps. 
Another way to quantify this is to estimate the  separation $d_{E}$ 
of the fragments from the first step at average time of the second decay step, 
i.e., the decay of the the intermediate state. This separation is 
\begin{equation}
d_{E} = \sqrt{ \frac{2 E_k^1} {\mu}} \frac{\hbar} {\Gamma_{I}} ,       
\end{equation}
where $\mu$ is the reduced mass of products produced in the first decay step. 
If $d_{E}$ is much larger than the typical nuclear diameter ($\sim$5~fm), 
then the two decay steps are well separated and cannot influence each 
other through nuclear processes. 
 
Again we emphasize that the quantities $R_{E}$ and $d_{E}$ do not provide 
sharp distinctions between prompt and sequential processes.  
For intermediate values of these quantities, one may consider a decay 
process that is basically sequential, but where 
final-state interactions between 
the products from the different decay steps are still important. If these final-state interactions do not washout sequential signals such as 
the  invariant mass of the intermediate state deduced from its decay 
products, or the angular correlations between the sequential decay steps, 
then one might 
still claim that the decay has a strong sequential character. However 
detailed predictions for the correlations may 
require a many-body calculation. 

The angular correlations in sequential decay are a consequence that the 
system passes through an intermediate state of well defined spin. Not all 
potential sequential decay scenarios have such correlations, for example if 
the intermediate state is $J$=0, or the orbital angular momentum removed 
in either step is zero. Thus the search for these correlations cannot be 
performed for all cases.      

An example of a truly sequential process is the decay
of the ground state of $^9$B to the \textit{p}+2$\alpha$ exit channel.
The decay begins by a proton emission to the ground state of $^8$Be
 which has a very small width ($\Gamma_{I}$=6.8~eV). Here, 
 $R_{E}$=3.6$\times$10$^{-5}$ and $d_{E}$=6.1$\times$10$^5$~fm which 
indicates that there are no significant interactions between the proton 
from the first step and the $\alpha$ particles from the second step.

The nucleus $^6$Be$_{g.s.}$ which decays to the 2\textit{p}+$\alpha$ channel 
provides an example of democratic decay. The possible intermediate state 
$^5$Li$_{g.s.}$ is
very wide ($\Gamma_{E}$=1.23~MeV), but most of the strength associated with 
this state is energetically inaccessible in $^6$Be$_{g.s.}$ decay 
(see Fig.~\ref{fig:c8Level}). Sequential decay is only possible through the 
low-energy tail of this resonance. Thus $^6$Be$_{g.s.}$ is thus almost a 
Goldansky-type decay. In a sequential scenario,  the mean energy released 
in the first step can be determined in an R-matrix approximation 
(Sec.~\ref{sec:simul})
as $E_k^1$=0.64 MeV  and this gives $R_{E}$=1.92 and $d_{E}$=6.5~fm, which 
is not in the realm of a truly sequential decay. 
Experimental studies show no 
evidence of the angular correlations expected in a sequential scenario 
\cite{Geesaman77} (see also Sec.~\ref{sec:8C}) and a good description of the decay 
correlations, the decay energy and its width can be obtained 
with a 3-body cluster model \cite{Grigorenko09}.   

In this work we report on an experimental investigation of 
particle-unstable states in Be, B, and C isotopes 
all of which undergo disintegration into 3 or more final fragments. 
In each case, we have measured correlations between these fragments to deduce 
the nature of the decay  and to determine to what extent 
sequential and prompt many-body processes are involved.   
The states reported on were all created as projectile-like fragments 
after the interaction of  a secondary $E/A$=70 MeV $^{9}$C beam with a 
Be target. One case of particular interest is the 4\textit{p}+$\alpha$ 
decay of $^8$C$_{g.s}$. As $^6$Be$_{g.s.}$ is a possible 
intermediate state in the decay of this level, 
we have obtained improved data for the $^6$Be 
system using a $E/A$=70~MeV $^7$Be 
secondary beam to help in the interpretation of the $^8$C data. 
Some of the $^8$C results have already been 
published in Ref.~\cite{Charity10}. 
The details of the experiment are discussed in 
Sec.~\ref{sec:experiment}. Results for each of the examined states are given
in Sec.~\ref{sec:results} and the conclusions of this work are presented in 
Sec.~\ref{conclusion}.

\section{EXPERIMENTAL METHOD}
\label{sec:experiment}

A primary beam of $E/A$=150-MeV $^{16}$O was extracted from the Coupled
Cyclotron Facility at the National Superconducting Cyclotron Laboratory at
Michigan State University with an intensity of 125 pnA. 
This beam bombarded a $^{9}$Be target producing 
$E/A$=70.0-MeV $^{9}$C and $^7$Be projectile-fragmentation products which 
were selected 
by the A1900 separator with a momentum acceptance of $\pm $0.5\%. 
The $^9$C secondary beam had an intensity 
of 1.6$\times $10$^{5}$~s$^{-1}$ with a
purity of $\sim$65\% with the main contaminant being $^6$Li.  
The $^7$Be beam had an intensity 
of 4$\times $10$^{7}$~s$^{-1}$ with a purity of $\sim$90\%.

\begin{figure}[tbp]
\includegraphics*[ scale=.4]{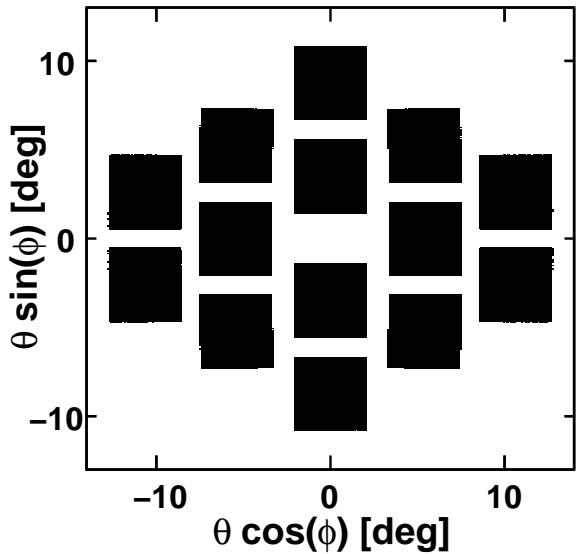}
\caption{The angular coverage of the HiRA array used in this experiment.} 
\label{fig:array}
\end{figure}

The two secondary beams impinged on a 1-mm-thick target of $^9$Be.
Charged particles produced in the reactions with this target 
were detected in the HiRA array 
\cite{Wallace07}. For this experiment, the array consisted of 
14 $E$-$\Delta E$ [Si-CsI(Tl)] 
telescopes located at a distance 90~cm downstream from the target. 
The angular coverage of the array is illustrated in Fig.~\ref{fig:array}
and it subtended polar angles from 1.4$^\circ$ to 13$^\circ$.
Each telescope consisted of a 1.5-mm thick, double-sided Si strip $\Delta E$
 detector followed by a 4-cm thick, CsI(Tl) $E$ detector. The $\Delta E$
detectors are 6.4~cm$\times $6.4~cm in area with each of 
the faces divided into 32
strips. Each $E$ detector consisted of four separate CsI(Tl) elements each
spanning a quadrant of the preceding Si detector. Signals produced in the
896 Si strips were processed with the HINP16C chip electronics \cite%
{Engel07}. For the $^9$C beam, the time of flight measured between 
a thin scintillator foil in the A1900
extended focal plane and the HiRA trigger was used in conjunction with the energy loss 
in this foil to reject most of the beam contaminants.

The energy calibration of the Si detectors was obtained with a $^{228}$Th 
$\alpha$-particle source. The particle-dependent energy calibrations of the CsI
detectors were achieved with cocktail beams selected with the A1900 separator.
These include protons ($E/A$=60 and 80 MeV), $\alpha$-particles ($E/A$=60 and 80 MeV),
deuterons ($E/A$=21 and 59 MeV), tritons ($E/A$=27 and 36~MeV), $^3$He ($E/A$=37
and 103 MeV) and $^6$Li ($E/A$=60 and 80~MeV) beams.

The HiRA telescopes have excellent isotope separation for the all of the light fragments 
of interest in this work and multiple fragments within a single telescope can be identified 
\cite{Charity07}. The gains for the Si amplifiers were setup for the detection of hydrogen 
and helium isotopes. Some information on lithium isotopes is used in this work,
 but lower-energy Li fragments saturated the Si shaping amplifiers and 
could not be identified (see later). 

\section{Simulations}
\label{sec:simul}
The effect of the detector 
acceptance and resolution is quite important when extracting decay widths, 
branching ratios and correlations between the fragments. These effects have 
been extensively studied using Monte Carlo simulations that include the 
angular acceptance and the angular and energy resolutions of the detectors,
the energy loss \cite{Ziegler85} and small-angle scattering \cite{Anne88} 
of the fragments as they leave the target, and the beam-spot size. 
These simulations have proven quite reliable 
in past experiments with HiRA \cite{Charity07,Charity08}.  

A number of different types of simulations are described in this work
including both sequential and prompt decay processes. However, the predicted 
resolution of the excitation energy of a level was found to be 
mostly insensitive to the 
details of the decay process, but is largely 
determined from the detector properties.

In simulating sequential 3-body decays, where the width of the 
intermediate state 
is significant, we have used the R-matrix formalism \cite{Lane58,Barker99}
to predict
 the distribution of $E_T$, 
the total kinetic energy released, and $E_x$=$E_T$-$E_k^1$, the kinetic energy 
released in the second step; 
\begin{equation}
N(E_T,E_x) \sim \frac{\Gamma_1(E_T,E_x)}{[E_T-Q_{1}-\Delta_{tot}(E_T)]^2+(1/4)\Gamma_{tot}^2(E_T)}
\end{equation}
where 
\begin{gather}
\Gamma_{tot}(E_T) = \int_0^{E_T} \Gamma_1(E_T,E_x)dE_x, \\ 
\Delta_{tot}(E_T) = \int_0^{\infty} \Delta_1(E_T,E_x)dE_x, \\
\Gamma_1(E_T,E_x) = 2 \gamma_1^2 P_{1\ell}(E_T-E_x) \rho(E_x), \\
\Delta_1(E_T,E_x) = - \gamma_1^2 [S_{1\ell}(E_T-E_x)-S_{1\ell}(Q_{1}-E_x)]\rho(E_x), \\
\rho(E_x) = c \frac{\Gamma_2(E_x)}{[E_x-Q_{2}-\Delta_2(E_x)]^2+(1/4)\Gamma_2^2(E_x)},\label{eq:rho} \\
\int_0^{\infty} \rho(E_x) dE_x = 1, \label{eq:c}\\
\Gamma_2(E_x) = 2\gamma_2^2 P_{2\ell}(E_x), \\
\Delta_2(E_x) = -\gamma_2^2 [S_{2\ell}(E_x) - S_{2\ell}(Q_{2})], \\
P_{\ell}(E) = \frac{ka}{F_{\ell}(ka)^2+G_{\ell}(ka)^2},\\
 S_{\ell}(E) = \frac {F'_{\ell}(k a) F_{\ell}(ka) + G'_{\ell}(k a) G_{\ell}(ka)}{F_{\ell}(k a)^2+G_{\ell}(k a)^2} , \\
a = 1.45 \rm{fm} (A_1^{1/3}+A_2^{1/3}),
\end{gather}
$F$ and $G$ are regular and irregulat Coulomb wavefunctions, 
$k$ is the wave number, $c$ is a normalization constant 
determined from Eq.~(\ref{eq:c}), 
$\gamma_1^2$ and $\gamma_2^2$ are the reduced widths
 associated with the first and second decay steps, and $Q_{1}$ and $Q_2$ 
are the 
centroids associated with $E_{T}$ and $E_x$, respectively.
Angular correlations in sequential decay are calculated from 
Refs.~\cite{Biedenharn53,Frauenfelder53}. 

The reduced width can be expressed as 
\begin{equation}
\gamma^2 = S \theta^2_{sp} \frac{\hbar^2}{\mu a^2}
\label{eq:theta2}
\end{equation}
where $S$ is the spectroscopic factor and $\theta^2_{sp}$, the single-particle
dimensionless reduced width, is 
\begin{equation}
\theta^2_{sp} = \frac{a}{2} \frac {u^2(a)}{\int_0^a u^2(r)dr}.
\label{eq:spdrw}
\end{equation}
Here  $u(r)/r$ is the single-particle radial  wave function calculated 
with a Coulomb plus Wood-Saxon potential with standard parameters for 
radii and diffuseness ($r_{0}$=1.25 fm,  $r_C$=1.3 fm, and $a$=0.65 fm) 
and the depth adjusted to fit the resonance energy.

\section{RESULTS}

\label{sec:results}

The deduced properties of the levels investigated in this study
are summarized in Table~\ref{Tbl:states} including their centroids, 
widths, and decay modes. More detailed discussion for each case are contained 
in the rest of this section including comparison with the evaluated 
quantities from the ENSDF database \cite{ENSDF}. Excitation energies are 
determined from the invariant mass method, i.e., the total kinetic energy 
of the fragments in their center-of-mass reference frame minus the decay 
Q value.

\begin{table*}
\caption{The excitation energy, width, spin and isospin of states for which
new information is determined.}
\label{Tbl:states}%
\begin{ruledtabular}
\begin{tabular}{cccccccc}
nucleus& decay   & branching                 &   exit     &$E^*$         & $\Gamma$  & $J^{\pi}$ & $T$ \\
       &         & ratio                     &   channel  &[MeV]           & [keV]       \\
\hline
$^{7}$B & \textit{p}+$^6$Be$_{g.s.}$&81$\pm$10\% & 3\textit{p}+$\alpha$ & 0.0 & 801$\pm$20 & & 3/2 \\
$^{8}$B & \textit{p}+$^{7}$Be$_{4.57}$ &&  \textit{p}+$^{3}$He+$\alpha$ &5.93$\pm$0.02 & 850$\pm$260 & 3, 4 & 1 \\
$^{8}$B & \textit{p}+$^{7}$Be$_{6.73}$ &&  \textit{p}+$^{3}$He+$\alpha$ &8.15$\pm$0.20 & 950$\pm$320 &                         & 1 \\
$^{8}$B & 2\textit{p}+$^{6}$Li$_{IAS}$ &$\sim$97.5\% &  2\textit{p}+$^{6}$Li         & 10.619$\pm$0.009\footnotemark[2]      & $<$60\footnotemark[2]      & 0$^{+}$              & 2 \\
        & 2\textit{p}+$^{6}$Li$_{2.18}$& $\sim$1.0\% & 2\textit{p}+\textit{d}+$\alpha$ \\
        & 2\textit{p}+\textit{d}+$\alpha$& $\sim$1.5\%  \\
$^{8}$Be & \textit{p}+$^{7}$Li$_{4.63}$ && \textit{p}+\textit{t}+$\alpha$ & 22.96$\pm$.02 &  680$\pm$146     & 3, 4 & 1 \\
$^8$C    & 2\textit{p}+$^6$Be$_{g.s.}$ & 100\% & 4\textit{p}+$\alpha$ & 0.0 & 130$\pm$50 & 0$^{+}$ & 2 \\
$^{9}$B & $\alpha$+$^{5}$Li& 97.2$\pm$0.5\%& \textit{p}+2$\alpha$ & 11.70$\pm$0.02 & 880$\pm$80 &  3/2$^{-}$,5/2$^{+}$,7/2$^{-}$   &  1/2 \\
        & \textit{p}+$^8$Be$_{3.03}$& 2.8$\pm$0.5\%& \textit{p}+2$\alpha$ \\
$^{9}$B & \textit{p}+$^{8}$Be$_{T=0+1}$& & \textit{p}+2$\alpha$ & 16.99$\pm$0.03 & 22$\pm$5\footnotemark[2] & 1/2$^-$ & 3/2 \\
$^{9}$B & \textit{p}+$^{8}$Be$_{19.069}$  && 2\textit{p}+$^{7}$Li & 20.64$\pm$0.10\footnotemark[1] & 450$\pm$250 &   \\
$^{10}$C & \textit{p}+$^9$B$_{2.345}$ &&   2\textit{p}+2$\alpha$ & 8.54$\pm$0.02 & $<$200 & \\
$^{10}$C & 2\textit{p}+2$\alpha$ &48\% & 2\textit{p}+2$\alpha$ & 9.69 & 490 &  & 1 \\
         & $\alpha$+$^{6}$Be &35\% & 2\textit{p}+2$\alpha$ \\
         & \textit{p}+$^9$B$_{2.34}$& 17\% & 2\textit{p}+2$\alpha$ \\
$^{10}$C &                            &     & 2\textit{p}+2$\alpha$ & 10.48$\pm$0.2 & $<$200 \\
$^{10}$C &                             &    & 2\textit{p}+2$\alpha$ & 11.44$\pm$0.2 & $<$200 \\
\end{tabular}
\footnotetext[1]{Assumed decay is to the ground state of $^7$Li.}
\footnotetext[2]{value from tabulations}
\end{ruledtabular}
\end{table*}

\subsection{$^8$C ground state}
\label{sec:8C}
\begin{figure}[tbp]
\includegraphics*[ scale=.4]{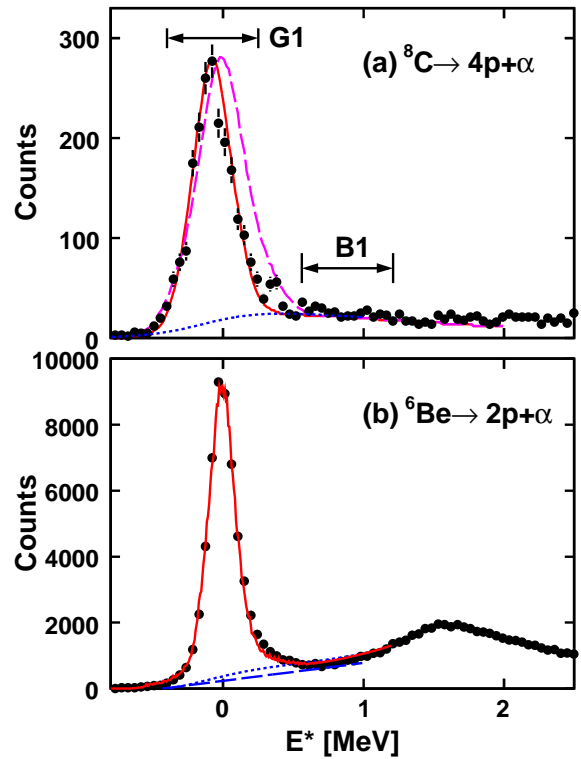}
\caption{(Color online) Excitation-energy spectra for (a) $^8$C fragments
 produce in neutron knockout from the $^9$C beam and (b) $^6$Be fragments 
produced in neutron knockout from the $^7$Be beam. In both cases, the curves 
show simulated ground-state distributions (see text). The solid curve in (a) 
is a fit to the data with a background indicated by the dotted curve. The
dashed curve is the simulated response based on the tabulated peak location 
and width. In dotted and dashed curves in (b) are the two possible background 
distribution considered in the text.  
}
\label{fig:Ex_8C}
\end{figure}

The ground state of $^8$C is unstable to disintegration into four protons and 
an $\alpha$ particle. The distribution of $^8$C excitation energy
 reconstructed from the 4\textit{p}+$\alpha$ channel following neutron 
knockout 
of $^9$C beam particles is displayed 
in Fig.~\ref{fig:Ex_8C}(a).  The excitation energy was determined assuming 
the $^8$C mass excess of 35.094~MeV obtained from the 2003 mass evaluation
of Ref.~\cite{Audi03}. The mass excess is needed to calculate the decay 
Q value. The excitation-energy spectrum shows a strong peak located near 
zero excitation energy indicating that we have populated the ground state 
of $^8$C. The long-dashed curve shows the simulated shape, 
including detector response, based on the evaluated mass excess and the listed
$^8$C decay width of 230~keV. This curve also contains a background
 contribution which is indicated by short-dashed curve.
The simulated shape is both wider and 
shifted up in energy compared to the experimental peak suggesting that 
the evaluated mass excess and width used in the simulations are incorrect.

In order gain a better understanding of the magnitude of the 
experimental uncertainties associated with this measurement, we have 
investigated the line shape of the $^6$Be ground state formed 
in neutron knockout reactions from the $^7$Be beam. In the $^9$C and $^7$Be
 neutron knockout reactions, 
the beam velocities, target thickness, and detector apparatus were identical and
 thus systematic errors are expected to be the same.
The reconstructed $^6$Be excitation-energy spectrum from the detected 
2\textit{p}+$\alpha$ events is displayed in Fig.~\ref{fig:Ex_8C}(b).
 This spectrum contains a peak at zero energy associated with 
the ground state and a wider peak at $\sim$1.7~MeV associated with the 
first excited state. For the ground-state peak, 
the solid curve in this figure shows the simulated spectrum, again including a
background contribution (including a contribution from the first excited state) 
which is indicated by the short-dashed curve. The tabulated $^6$Be mass excess 
and decay width were used in this simulation which reproduces the 
experimental result very well, strongly suggesting that the disagreement of
 the simulation with the data for $^8$C is not an experimental artifact. 
The experimental FWHM of the $^6$Be ground-state peak is 214$\pm$10~keV, 
significantly 
greater than its intrinsic width of $\Gamma$=92$\pm$6~keV. Thus the agreement 
between the $^6$Be$_{g.s.}$ data and simulation indicates that we can correctly account 
for the experimental resolution.

 The solid curve in Fig.~\ref{fig:Ex_8C}(a) shows the results of a 
 simulation where both the $^8$C mass excess and its decay width 
were adjusted to best reproduce the experimental peak. 
The fitted mass excess is 35.030$\pm$0.030~MeV 
and the fitted width is 130$\pm$50~keV.     

Previous measurements of the mass excess and decay width were made 
in Refs.~\cite{Robertson74,Robertson76,Tribble76} using transfer reactions. 
Of these, only the 1976 work of Tribble \textit{et al.} \cite{Tribble76} 
has more than a handful of $^8$C events. The reaction studied was 
$^{12}$C($^4$He,$^8$He)$^8$C and the $^8$He fragments were detected. 
The extracted mass excess was 35.10$\pm$0.03 and the decay width was 
230$\pm$50~keV and 183$\pm$56 keV assuming Breit-Wigner and 
Gaussian intrinsic lines shapes, respectively. As in the present work, 
the experimental resolution was significant. 

\begin{figure}[tpb]
\includegraphics*[ scale=.4]{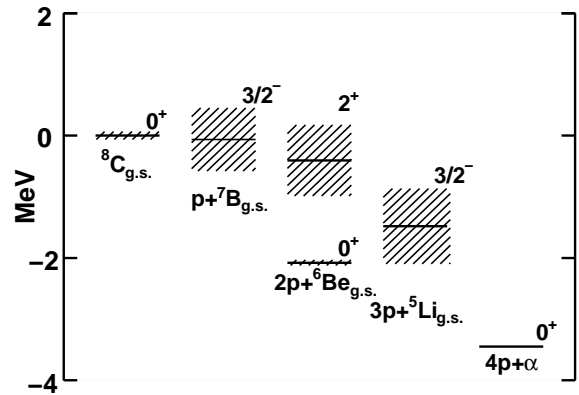}
\caption{ Levels of interest in the decay of the $^8$C 
and $^7$B ground states.}
\label{fig:c8Level}
\end{figure}

\begin{figure}[tpb]
\includegraphics*[ scale=.4]{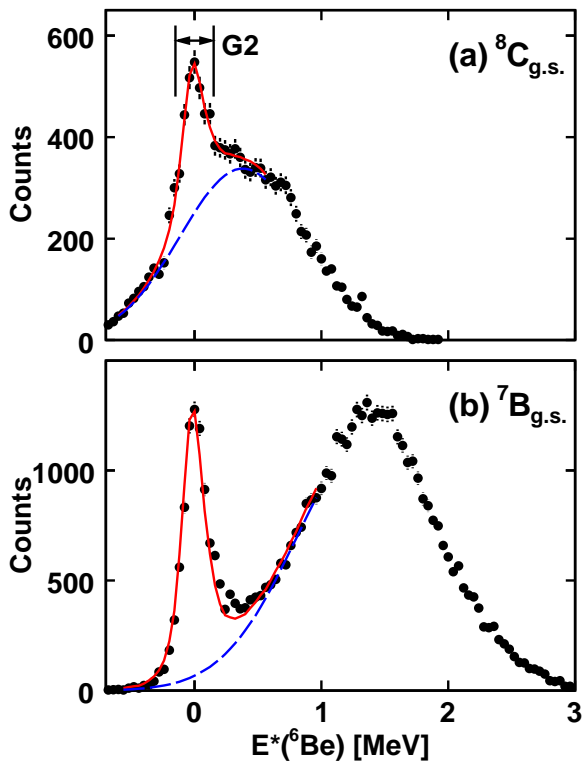}
\caption{(Color online) The spectrum of all possible reconstructed 
$^6$Be excitation energies for events in the ground states of (a) $^8$C and
(b) $^7$B.  The data points are the experimental data, while
the solid curves show  fits to the $^6$Be ground-state region with the dashed
curves showing the fitted background.}
\label{fig:c8b7be6}
\end{figure}

The $^8$C ground state is quite unusual in having a five-particle exit channel.
It is of interest to determine whether this state disintegrates 
directly into five pieces or proceeds in a sequential process 
of two or more steps through intermediate states. Figure~\ref{fig:c8Level}
shows the levels of possible interest in the decay.
Of the possible intermediate states, the $^6$Be ground state is the 
narrowest (longest lived)  and easiest to  detect experimentally.
In each detected 4\textit{p}+$\alpha$ event, one can find 6 ways to 
reconstruct a $^6$Be decay as a 2\textit{p}+$\alpha$ subset. For each of 
these ways, a $^6$Be excitation energy is determined.
We use gate $G1$ in Fig.~\ref{fig:Ex_8C}(a) to select out the ground-state events. 
 The resulting excitation-energy spectrum
of all these potential $^6$Be candidates is displayed in 
Fig.~\ref{fig:c8b7be6}(a) as data points. 
As there can be at most only one $^6$Be 
ground-state fragment per event, at least 5/6 th of the spectrum 
corresponds to incorrectly reconstructed $^6$Be fragments 
and these form a background. The experimental spectrum clearly displays a 
peak centered near zero energy associated with the $^6$Be ground state 
 on a background of approximately Gaussian shape.

To determine what fraction of the $^8$C events  decay through
 $^6$Be$_{g.s.}$, we have fitted the spectrum in Fig.~\ref{fig:c8b7be6}(a) 
with two components;
a $^{6}$Be$_{g.s.}$ component with shape taken from the experimental peak 
in Fig.~\ref{fig:Ex_8C}(b) obtained with the $^7$Be beam 
and a background component taken to be 
Gaussian in the immediate vicinity of the $^{6}$B$_{g.s.}$ peak.
In order to use the experimental $^6$Be$_{g.s.}$ line shape from 
Fig.~\ref{fig:Ex_8C}(b), 
one has to first subtract the small background under this peak. The short and 
long-dashed curves in this figure show two possible backgrounds which were 
considered. The short-dashed curve is associated with the solid curve 
obtained using a symmetric Breit-Wigner line shape. 
The use of the  long-dashed 
background curve would imply a slightly asymmetric $^6$Be$_{g.s.}$ line shape 
which has an enhanced higher-energy tail. A small degree of asymmetry 
is not unreasonable.

The resulting fit (using the original $^6$Be background) is shown by the solid 
curve in Fig.~\ref{fig:c8b7be6}(a) where the Gaussian background is 
indicated by the dashed curve. Taking into account both possible $^6$Be 
backgrounds, these fits imply that, on average, 1.01$\pm$0.05 $^{6}$Be$_{g.s.}$ fragments are created in each 
$^8$C$_{g.s.}$ decay, i.e. essentially all decays pass through $^6$Be$_{g.s.}$.This number was corrected for the small background under the 
$^8$C$_{g.s.}$ peak itself [dotted curve in Fig.~\ref{fig:Ex_8C}(a)], 
using the the region $B1$ to estimate the $^6$Be probability associated with
this background.

In many ways, the decay of $^8$C$_{g.s}$ to 2\textit{p}+$^6$Be$_{g.s.}$ 
is similar to the decay of $^6$Be$_{g.s.}$ to 2\textit{p}+$\alpha$. 
In each case, both the initial and final states are $J^{\pi}$=0$^+$ and 
the possible 
intermediate states ($^7$B$_{g.s.}$ and $^5$Li$_{g.s}$) are $J^{\pi}$=3/2$^-$. 
These intermediates states are both wide, but only their low-energy tails are 
energetically accessible. Both decays are democratic; $R_{E}$=1.2 and 1.9 
and $d_{E}$=8.8 and 6.5~fm for $^8$C and $^6$Be, respectively. 
Thus the disintegration of $^8$C$_{g.s.}$  is novel having 
two sequential steps of democratic two-proton decay. 
To investigate this possibility, 
it is  useful to separate out experimentally the protons from 
the first and second decay steps. However, it is not possible to do this with 
100\% certainty for all decays.

For a fraction of the events it is possible to make this separation with 
 reasonable precision. To this end, we have selected events where one, 
and only one, of the 6 possible reconstructed $^6$Be fragments has an excitation 
energy associated with the ground-state peak indicated by the region $G2$ in 
Fig.~\ref{fig:c8b7be6}(a). For the events which survive this criteria, 
the identities of the $^6$Be$_{g.s.}$ decay products are obvious.

Some events from the tails of the $^6$Be$_{g.s.}$ peak lie outside 
the $G2$ region and will be rejected. Also, and more importantly
 one can
 have the situation that the correct $^6$Be fragment is in the high or 
low-energy tail outside of the gate and an incorrectly identified $^6$Be 
fragment lies within the $G2$ gate. 
Such an event will not be rejected, but  the identities of the 
protons from the first and second steps will be incorrect.   
Monte Carlo simulations were used to optimize the width of the $G2$ region so as to obtain the best compromise between event-rejection and mis-identification.

With the chosen gate width we select only 30\% of the $^8$C$_{g.s.}$ events
and, of these, we estimate that a selected event is incorrectly 
identified 30\% of the time. While this is significant, these misidentifications
involve cases were the two protons assigned to one of the decay steps are 
actually from the two separate steps. As the two steps are separated by a 
$^6$Be$_{g.s.}$ fragment with $J^{\pi}$=0$^+$, there is no angular correlations 
between the protons from the two steps and these misidentified events 
will give rise only to smooth backgrounds in the correlation plots.

\begin{figure}[tbp]
\includegraphics[scale=0.4]{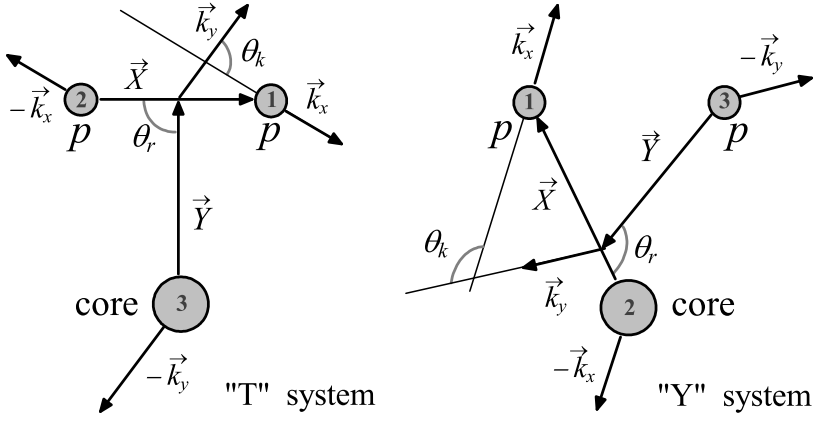}
\caption{Independent ``T'' and ``Y'' Jacobi systems for the core+$N$+$N$
three-body system in coordinate and momentum spaces.}
\label{fig:jacobi}
\end{figure}

The energy and angular correlations between the particles
produced in two-proton decay
can be described by the hyperspherical Jacobi vectors 
$\mathbf{X}$ and $\mathbf{Y}$ and their conjugate momenta $\mathbf{k}_x$ 
and $\mathbf{k}_y$. There are two 
independent ways of defining the coordinates 
which are referred to as the ``T'' and 
``Y'' systems. These are illustrated in Fig.~\ref{fig:jacobi} where the core
(fragment 3 in the ``T'' system or fragment 2 in the ``Y'' system)
is the $^6$Be$_{g.s.}$ fragment in the first 2\textit{p} decay and the $\alpha$ particle in the second 2\textit{p} decay. 
In terms of the position vectors $\mathbf{r}_i$, momentum vectors 
$\mathbf{k}_i$ and masses $m_i$ ($i$=1, 2, and 3)  
the Jacobi coordinates are 
\begin{subequations}
\begin{eqnarray}
\mathbf{X} &=& \mathbf{r}_{1} - \mathbf{r}_{2} \, , \\
\mathbf{Y} &=& \frac{m_1 \mathbf{r}_{1} + m_2\mathbf{r}_{2}}{m_1+m_2} 
- \mathbf{r}_{3} \, ,\\
\mathbf{k}_x &=& \frac {m_2 \mathbf{k}_{1}-m_1 \mathbf{k}_{2}}{m_1+m_2}\, , \\
\mathbf{k}_y &=& \frac{m_3(\mathbf{k}_{1}+\mathbf{k}_{2}) -
(m_1+m_2)\mathbf{k}_{3}}{m_1+m_2+m_3}. 
\end{eqnarray}
\end{subequations}

Of the six degrees of freedom required to define the $\mathbf{k}_x$ and 
$\mathbf{k}_y$ distributions, three describe the Euler rotation of the 
decay plane and one is constrained from energy conservation. 
Thus the complete correlation information can be described by two variables 
which we take as $E_x/E_T$ and $\theta_k$, where $E_x$ is the 
energy associated 
with the $X$ coordinate,
\begin{equation}
E_x = \frac{(m_1+m_2) k_x^2}{2 m_1 m_2},
\end{equation}  
$E_T$ is the total three-body energy
and $\theta_k$ is the angle between the Jacobi momenta,
\begin{equation}
\theta_k = \frac{\mathbf{k}_x \cdot \mathbf{k}_y}{k_x k_y}.
\end{equation}

For each event, there are two ways of labeling the two protons
and thus there are two possible values 
of the [$E_x/E_T$,$\cos(\theta_k)$] coordinate, both of which are used to 
increment the correlation histograms. 
For the ``T'' system, this 
produces a symmetrization of the angular distributions 
about $\cos(\theta_k)$=0. 

Distributions constructed in the two Jacobi systems are just different representations 
of the same physical picture.
Each Jacobi system can reveal different aspects of the correlations. The Jacobi ``Y'' system 
is particularly useful if the two protons are emitted sequentially through an intermediate state.
In this case, $E_x$ is the kinetic energy released in the second step and $\theta_k$ is the angle
between the two decay axes. The Jacobi ``T'' system can have a similar interpretation for a diproton decay 
with an intermediate $^2$He state.

\begin{figure}[tbp]
\includegraphics[scale=0.48]{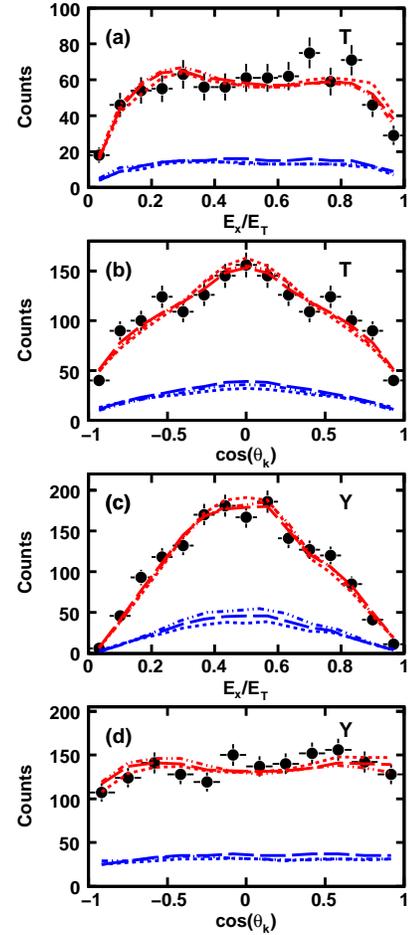}
\caption{(Color online) Projected correlations in  
(a,b) ``T'' and (c,d) ``Y'' 
Jacobi systems 
for the three-body decay of $^6$Be formed in the second step of 
 $^8$C$_{g.s.}$ decay.
Angular correlations are shown in (b) and (d) and energy correlations 
are shown in (a) and (c). The data points are the experimental data, 
while simulated results are indicated by the  curves which pass 
through the data points.  The lower curves are estimates of the 
``backgrounds'' of
 incorrectly identified events. The short and long-dashed and 
solid curves are associated with simulations (A), (B) , and (C), 
respectively (see text). }
\label{fig:C8_Jacobi_Be6}
\end{figure}

\begin{figure}[tbp]
\includegraphics[scale=0.48]{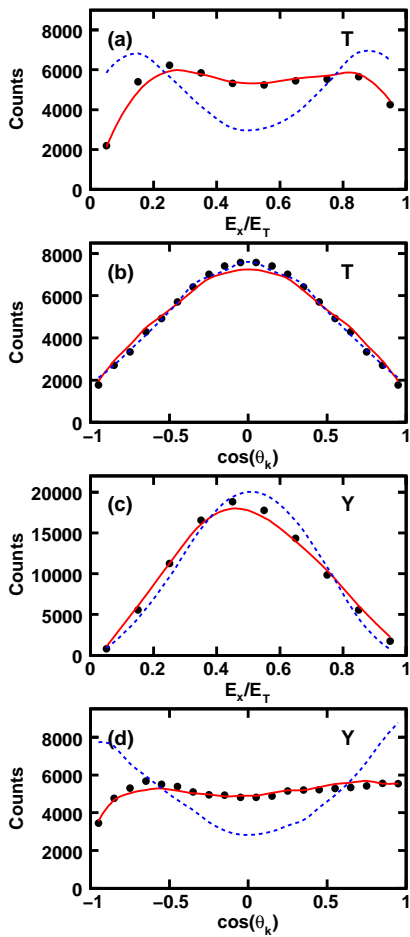}
\caption{As for Fig.~\ref{fig:C8_Jacobi_Be6}, but now for $^6$Be$_{g.s.}$
events produced via neutron knockout from the $^7$Be beam.
The solid and dashed curves show predictions of the 3-body cluster model 
and a sequential-decay calculation, respectively.}
\label{fig:Be6_Jacobi}
\end{figure}

The projected correlations in both the ``T'' and ``Y'' systems 
are plotted  in 
Fig.~\ref{fig:C8_Jacobi_Be6} for the second decay step of $^8$C$_{g.s.}$, i.e., the 2\textit{p} decay of the 
$^6$Be$_{g.s.}$ intermediate state.
These can be compared to results from the $^7$Be beam in 
Fig.~\ref{fig:Be6_Jacobi}, where  
$^6$Be$_{g.s.}$ is formed more directly following a neutron-knockout reaction.
In this case, the statistical uncertainties are significantly smaller and there 
are no misidentified events. A consistency between the projected 
correlations in Fig.~\ref{fig:C8_Jacobi_Be6} and \ref{fig:Be6_Jacobi} is 
necessary if the $^8$C$_{g.s.}$ level disintegrates through $^6$Be$_{g.s.}$
and indeed the corresponding projected correlations are quite similar in 
the two figures. 

In order to make more detailed comparisons of the experimental 
$^6$Be correlations in 
Figs.~\ref{fig:C8_Jacobi_Be6} and \ref{fig:Be6_Jacobi},
  we have made extensive Monte-Carlo simulations 
including the effect of the detector response and the event selection 
requirements.  
The results from the $^7$Be beam in Fig.~\ref{fig:Be6_Jacobi} are compared to 
simulations where the effect of the detector resolution and acceptance is 
considered with no misidentification or selection bias. The detector effects induce only small modifications to the 
projected correlations in this case.
The solid curves are
predicted distributions from the Quantum-Mechanical 3-body cluster  
model of Ref.~\cite{Grigorenko09} (Calculation P2 in that reference). 
This model reproduces
the experimental total energy $E_T$, $\Gamma$, and the  experimental correlations in Fig.~\ref{fig:Be6_Jacobi}.
These predicted 
correlations will therefore be used in the simulations in the disintegration of 
$^8$C.

The dashed curves in Fig.~\ref{fig:Be6_Jacobi} are from an R-matrix 
approximation of sequential 2-proton emission from $^6$Be$_{g.s.}$ 
through the $^5$Li$_{g.s.}$
intermediate state as described in Sec.~\ref{sec:simul}. 
The angular correlations in this simulation are particularly  strong 
and favor decays where the two 
sequential decay axes are collinear, i.e. $\cos(\theta_k)=\pm$1 in the 
Jacobi ``Y'' system [see Fig.~\ref{fig:Be6_Jacobi}(d)]. This angular correlation
is also responsible for the predicted broad minimum 
in the energy distribution in the 
Jacobi ``T'' system [Fig.~\ref{fig:Be6_Jacobi}(a)]. However, these features 
associated with the predicted angular correlation are lacking in the 
experimental data. This indicates that the decay of $^6$Be$_{g.s.}$ does not 
pass through an intermediate state of well defined angular 
momentum ($J^{\pi}$=3/2$^-$). Similar conclusions have been made in 
other studies of $^6$Be$_{g.s.}$ \cite{Geesaman77,Bochkarev89}.

Now let us return to $^6$Be$_{g.s.}$ decay in the process of 
$^8$C$_{g.s.}$ disintegration. In order to include the bias from 
the event selection and the contribution from misidentified events, it is 
necessary to simulate the first step of $^8$C$_{g.s.}$ disintegration.
For this first decay step, $^8$C$\rightarrow$2\textit{p}+$^6$Be$_{g.s.}$,
we have considered three quite different sets of 
simulated [$E_x/E_t$, $\cos(\theta_k)$] correlations:
\begin{enumerate}[(A)]
\item  The momenta of the three fragments are chosen according to 
available phase-space volume.
\item The [$E_x/E_t$, $\cos(\theta_k)$] correlations for the Jacobi ``T'' system 
is taken to be the same as for $^6$Be$_{g.s.}$ decay. Note, as the core mass 
is different in the $^8$C$_{g.s.}$ 2-proton decay, 
the Jacobi-``Y'' distribution will not be 
exactly the same as for $^6$Be$_{g.s.}$ decay.
\item The decay is treated in the R-matrix approximation as a sequential 
2-proton
decay through the $^7$B intermediate state in a similar manner as just 
described for the sequential-2 proton decay of $^6$Be$_{g.s.}$. The R-matrix 
parameters for $^7$B$_{g.s.}$ are taken from Sec.~\ref{sec:7B}. 
\end{enumerate}

The short-dashed, long-dashed, and the solid curves in Fig.~\ref{fig:C8_Jacobi_Be6} are 
the results for $^6$Be$_{g.s.}$ decay in the second step of $^8$C$_{g.s.}$ 
disintegration for simulations (A), (B), and (C), respectively. 
The curves, which pass 
through the data points in each panel, include the detector response, 
the selection bias and 
the contribution from the incorrectly identified events. The latter 
contributions are indicated by the lower curves in each panel. In all cases, 
the three curves from the different simulations almost overlap. 
Thus the simulated 
results for $^6$Be$_{g.s.}$ decay show very little sensitively to the decay 
correlations in the first $^8$C$\rightarrow$2\textit{p}+$^6$Be$_{g.s.}$ step.
Our assertion that the distributions for the misidentified events  show 
no strong structure is borne out and the simulations for all selected events 
reproduce the experimental data. Therefore within 
the statistical errors on the experimental data and uncertainty 
associated with extracting the data, that correlations in the second step 
of $^8$C disintegration is consistent with $^6$Be$_{g.s.}$ decay.

\begin{figure}[tbp]
\includegraphics[scale=0.48]{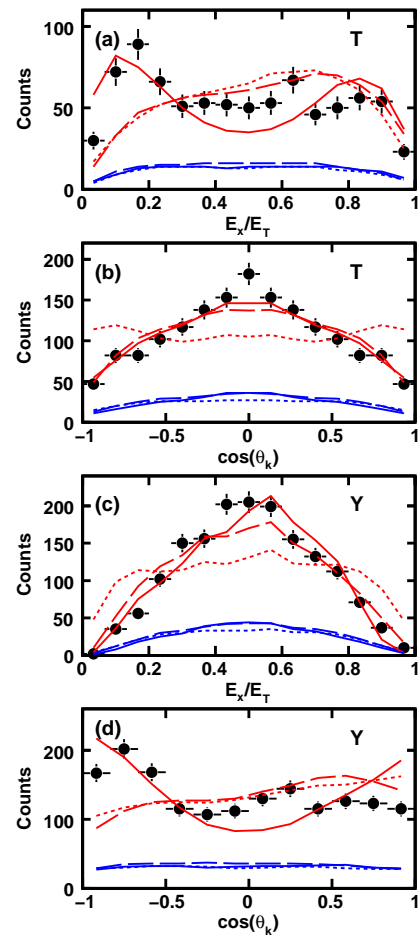}
\caption{As for Fig.~\ref{fig:C8_Jacobi_Be6}, but now the correlations 
are associated with the first step of $^8$C$_{g.s.}$ decay 
forming 2\textit{p}+$^6$Be$_{g.s.}$.}
\label{fig:C8_Jacobi}
\end{figure}

For the first step (2\textit{p}+$^6$Be$_{g.s.}$) of $^8$C$_{g.s.}$ decay,
the projected $E_x/E_t$ and $\cos(\theta_k)$ distributions for 
both the Jacobi ``T'' and ``Y'' systems are plotted in Fig.~\ref{fig:C8_Jacobi}.
The curves in each panel show the simulated distributions and backgrounds 
from the three simulations discussed above. Again these backgrounds are 
similar 
in all the simulations and show relatively smooth behaviors.
Clearly, the phase-space simulation (short-dashed curve) is inconsistent 
with all the experimental projected correlations. 
Also the simulations using the 
theoretical $^6$Be correlations (long-dashed curves) does not fit the data
in Figs.~\ref{fig:C8_Jacobi}(a) and \ref{fig:C8_Jacobi}(d) 
indicating that two-proton correlations are 
different in the first and second steps.
 
The sequential decay simulations gives the best description 
to all the experimental distributions, however,  it is clearly not perfect.
The predicted angular correlations indicated by the solid curve in 
Fig.~\ref{fig:C8_Jacobi}(d) show strong  enhancements near 
$\cos(\theta_k)$=$\pm$1. These angular correlations also cause the 
large structure predicted 
for the Jacobi ``T'' energy distribution of Fig.~\ref{fig:C8_Jacobi}(a). 
The experimental data also show an enhancement near $\cos(\theta_k)$=-1, 
sometimes called the diproton region, however, the corresponding enhancement 
near $\cos(\theta)$=1 is not evident. In the sequential scenario, 
$\cos(\theta)$=1 corresponds to the $^6$Be core from the second 
step being directed towards the proton from the first step. Coulomb 
final-state interactions may deflect the trajectories and thus suppress the yield
near $\cos(\theta)$=1. Alternatively the enhancement near  $\cos(\theta_k)$=-1
may not be related to  a sequential angular correlation but rather might be 
a signature of some
 ``diproton'' character of the decay similar to that observed for the 
6.57-MeV state in $^{10}$C \cite{Charity09}.

The total decay width in the sequential scenario can be estimated as  
\cite{Lane58, Barker99}
\begin{equation}
\Gamma^0 = \frac{\Gamma_{tot}(Q_1)}{1+\gamma_1^2 \int_0^{\infty} 
\frac{dS_{1\ell}(E_T-E_x}{dEx}\big)_{Ex=Q_1}\rho(E_x)dE_x}.
\label{eq:gamma0}
\end{equation}
The \textit{p}+$^7$B$_{g.s.}$ spectroscopic factor was calculated with CKI 
Hamiltonian \cite{Cohen65,*Cohen67} as $S$=3.20 (8/7) where (8/7) is 
the center of mass correction. With $\theta^2_{sp}$=0.44 from Eq.~(\ref{eq:spdrw}) we 
estimate $\gamma^2_1$=4.28 MeV and thus  $\Gamma^0$=12~keV  
which is significantly less than the experimental value of 130$\pm$50~keV 
and thus is clear that this sequential decay scenario cannot provide a 
total description of the 2-proton decay of $^8$C$_{g.s.}$.

Given the enhanced ``diproton'' character of the decay it is useful 
to consider the  diproton decay model which 
can be treated in the R-matrix formulation of Barker \cite{Brown67} 
together with
the diproton cluster decay model developed in Ref.~\cite{Barker02,*Barker03}.
For the wavefunctions we use the \textit{p}-shell basis with the
CKI Hamiltonian \cite{Cohen65,*Cohen67}.

The diproton decay  spectroscopic factor is given as a product
of the three terms as in Eq. (7) of \cite{Brown67}. For the $^{8}$C decay
these are (8/6)$^{2}$ = 1.78 (the center of mass correction),
$G^{2}(p)$=1/2 (the cluster overlap factor) and $C$=1.21 (the
\textit{p}-shell spectroscopic factor for $L$=$S$=0), to give
$S_{62}$=1.08. The related single-particle dimensionless reduced width
of $\theta^{2}_{sp}$=1.00 is also taken from \cite{Barker02,*Barker03}. 
The diproton
decay width obtained for the experimental decay energy of $Q_1$=2.077~MeV 
is $\Gamma$=88~keV. If we use
the older $Q_1$ value of 2.147~MeV based on the tabulated masses,
 the width would be 100~keV.
We have assumed a narrow $^{6}$Be final state. The change coming from
a folding with its actual width is small. Our spectroscopic factor $S_{62}$
is a factor of five larger than given by Barker
following Eq. (13) of Ref.~\cite{Barker02,*Barker03}. 
We do not know why, but assume
that there was an error in Barker's calculation.

The calculated diproton decay width of $\Gamma$=88~keV is consistent with the
experimental value of 130$\pm$50~keV, supporting the interpretation of the 
enhancement at low $E_x$/$E_T$ in the Jacobi ``T'' system [Fig.~\ref{fig:C8_Jacobi}(a)] 
as a diproton feature.  However, the diproton model does not provide 
a complete description of the experimental correlations. For example in the diproton model, the 
Jacobi ``T'' angular distribution is just the distribution of the diproton decay angle, 
which should be isotropic as the diproton has $J^\pi$=0$^+$. However the experimental distribution 
in Fig.~\ref{fig:C8_Jacobi}(b) is not uniform and thus inconsistent with this notion.  

We have also considered the diproton model for $^6$Be$_{g.s.}$ decay.
For the spectroscopic factor we obtain
(6/4)$^{2}$ = 2.25 (the center of mass correction),
$G^{2}(p)$=1/2 (the cluster overlap factor) and $C$=1.00 (the
\textit{p}-shell spectroscopic factor for $L$=$S$=0), to give
$S_{42}$=1.12. With the single-particle dimensionless reduced width
of $\theta^{2}_{sp}$=1.13 \cite{Barker02,*Barker03}, the diproton
decay width obtained for $Q_1$=1.371~MeV is $\Gamma$=98~keV again consistent with the
tabulated value of 92$\pm$6~keV. However in this case the diproton nature of the experimental
correlations (Fig.~\ref{fig:Be6_Jacobi}) is even less apparent.
Clearly a complete understanding of $^6$Be$_{g.s.}$ and $^8$C$_{g.s.}$ 
requires a full three-body decay calculation.

\subsection{$^{7}$B levels}
\label{sec:7B}
\begin{figure}[tbp]
\includegraphics*[ scale=.4]{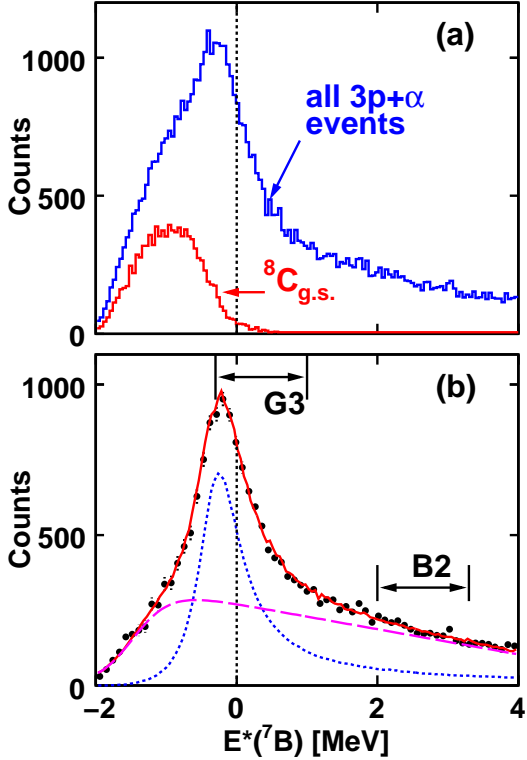}
\caption{(Color online) (a) The excitation-energy spectra for 
$^7$B events determined from all detected 3\textit{p}+$\alpha$ events and from
$^8$C$_{g.s.}$ 4\textit{p}+$\alpha$ events where one of the protons was 
ignored. (b) The data points are the $^7$B excitation-energy spectrum 
after the $^8$C contamination was subtracted. The solid curve shows a fit 
to this data assuming an R-matrix line shape and including the detector 
response. The long-dashed curves indicate the background obtained from the fit and the raw R-matrix line shape is indicated by the dotted curve.}   
\label{fig:Ex_7B}
\end{figure}

The $^7$B excitation-energy spectrum determined from all detected 
3\textit{p}+$\alpha$ 
events is displayed in Fig.~\ref{fig:Ex_7B}(a). 
The peak near zero excitation energy
indicates that the ground state of $^7$B was produced in the reaction. There 
is a significant contamination in this spectrum from $^8$C$_{g.s.}$ 
disintegrations 
where only 3 of the 4 protons were detected. To determine the contribution 
from this process, we have taken detected 4\textit{p}+$\alpha$ events 
associated with $^8$C$_{g.s.}$ and generated pseudo 
3\textit{p}+$\alpha$ events by alternatively removing one of the protons 
and determined the $^7$B excitation energy. The distribution of 
these excitation energies 
is labeled as $^8$C$_{g.s.}$ in 
Fig.~\ref{fig:Ex_7B}(a) and indicates that the incompletely detected 
$^8$C$_{g.s.}$ events lead to an enhancement of the low-energy side of 
the $^7$B$_{g.s.}$ peak. The Monte Carlo simulations were used to normalize 
the $^8$C$_{g.s.}$ distribution to the expected number of 
incomplete detected events.
The data points in Fig.~\ref{fig:Ex_7B}(b) show the results for 
the $^7$B excitation-energy spectrum after 
this $^8$C$_{g.s.}$ contamination was subtracted.  

The subtracted distribution peaks at $E^*\sim$-100~keV rather than zero,
suggesting that 
the evaluated mass excess from Ref.~\cite{Audi03} used to calculate the 
Q value for this figure is incorrect. This evaluated mass 
excess of 27.94$\pm$0.10 MeV was 
derived from an average of two measurements; 27.80$\pm$0.10 MeV from the
 $^7$Li($\pi^+$,$\pi^-$)$^7$B reaction and 27.94$\pm$0.10~MeV from the 
$^{10}$B($^3$He,$^6$He)$^7$B reaction \cite{McGrath67}.

As this  is a rather wide level near threshold, the use of a
 Breit-Wigner line shape may not be appropriate. We have used R-matrix theory 
\cite{Lane58} 
assuming \textit{p}+$^6$Be$_{g.s.}$ as the only open decay channel.
The line shape is given by Eq.~(\ref{eq:rho}) where now $E_x$ is the 
energy above the \textit{p}+$^6$Be$_{g.s.}$ threshold, and
$Q_{2}$ is the resonance energy.
Using $J^{\pi}$=3/2$^-$ listed in  the ENSDF database and $\ell$=1,
the experimental spectrum was fit by adjusting 
$E_2$ and $\gamma_2^2$ and other parameters defining a 
smooth background under the peak. The effects of the detector acceptance 
and resolution was taken into account via the Monte Carlo simulations.
The resulting fit is displayed as the 
solid curve in Fig.~\ref{fig:Ex_7B}(b) and the fitted background is shown as 
the dashed curve. The dotted curve shows the fitted peak shape without the 
effects of the detector acceptance and resolution. 

The fitted resonance energy is
$Q_2$=2.013$\pm$0.025~MeV which implies a mass excess of 27.677$\pm$0.025~MeV.
The error includes the effect in a 20\% uncertainty in the magnitude of
 the subtracted $^8$C contamination.
 This mass excess is 
consistent with the previous measurement from the  
$^7$Li($\pi^+$,$\pi^-$)$^7$B reaction, but is inconsistent with that 
from the $^{10}$B($^3$He,$^6$He)$^7$B reaction.
The fitted reduced width is $\gamma_2^2$=1.32$\pm$0.02 MeV and 
thus the level width\cite{Lane58},
\begin{equation}
\Gamma^0 = \frac {\Gamma(Q_2)}{1+\gamma_2^2 \, \frac {dS_{\ell}}{dE_x}\big)_{E_x=Q_2}},
\label{eq:gamma2}
\end{equation} 
 is 0.80$\pm$0.02 MeV. 
The evaluated width in the ENDSF database is 1.4$\pm$0.2~MeV 
derived from the $^{10}$B($^3$He,$^6$He)$^7$B reaction \cite{McGrath67}. 
Again our results are inconsistent with this measurement.

In the above analysis of this level, we assumed 
$^7$B$_{g.s.}\rightarrow$ \textit{p}+$^6$Be$_{g.s.}$ sequential decay. 
We can check this by looking for a $^6$Be$_{g.s.}$ fragment as described 
 in Sec.~\ref{sec:8C}.
Here, there are only 
three ways to 
reconstruct a $^6$Be fragment from 3 protons and an $\alpha$ particle. 
The distribution of all three possible $^6$Be excitation energies is shown 
in Fig.~\ref{fig:c8b7be6}(b) for the gate $G3$ in Fig.~\ref{fig:Ex_7B}(b) 
on the $^7$B
 excitation energy. This gate does not cover the whole of the $^7$B$_{g.s.}$ 
peak width, but avoids the lower energies where the $^8$C contamination is present.
The  $^6$Be excitation-energy spectrum displays a prominent peak at zero 
excitation energy indicating that $^6$Be$_{g.s.}$ fragments were produced 
in the decay. The solid curve shows a fit to the data using the experimental 
$^6$Be$_{g.s.}$ line shape [Fig.~\ref{fig:Ex_8C}(b)] and a smooth Gaussian background (dashed curve) 
under this peak. From this fit we determine that, on average, 
there is a 54$\pm$6\% probability of finding a $^6$Be$_{g.s.}$ fragment in the 
$G3$ gate. However it is clear from Fig.~\ref{fig:Ex_7B}(b) that 
there is a very significant background under this peak. Using gate $B2$ in 
Fig.~\ref{fig:Ex_7B}(b) we estimate that the background 
has a 19$\pm$2\% probability of containing a $^6$Be$_{g.s.}$ fragment. 
Using this, we find that 81$\pm$10\% of the $^7$B$_{g.s.}$ events 
decay via \textit{p}+$^6$Be$_{g.s.}$.

Shell-Model calculations for $^7$B$_{g.s.}$ with the CKI Hamiltonian \cite{Cohen65,*Cohen67}
give the spectroscopic factor for the \textit{p}+$^{6}$Be configuration as 
$S$=0.59(7/6) = 0.688. The \textit{p}+$^6$Be$_{1.67}$(2$^+$) configuration is 
predicted to be $\sim$3 times stronger but decay to this channel is suppressed 
due to its significantly smaller barrier penetration factor. 
With $\theta^2_{sp}$=0.688 from Eq.~(\ref{eq:theta2}), this  gives us a 
predicted reduced width of 1.42~MeV [Eq.~(\ref{eq:spdrw})]. 
Taking into account the 81\% branching ratio, we estimate 
the experimental value to be $\gamma^2$=1.07$\pm$0.15~MeV which is quite 
similar to the predicted value confirming that the \textit{p}+$^6$B$_{g.s.}$ 
is not the strongest configuration in $^7$B$_{g.s.}$.  
 
 Although it is possible to consider a proton
 decay through the 1.670-MeV ($J^{\pi}$=2$^+$) first-excited state of $^6$Be 
(see Fig.~\ref{fig:c8Level}), this $^6$Be intermediate state is
sufficiently wide that the disintegration is democratic. 
The average energy released in the first step is $E_k$=343~keV 
(determined from the differences in centroids of the two levels) which is very
 small compared 
to the width of the intermediate state, $\Gamma$=1.16~MeV. Alternatively, 
the distance traveled by the proton on average before the $^6$Be fragment 
decays is 2.0~fm which is small compared to the nuclear diameter. Thus this 
decay strength is probably best described as a 4-body decay.

\subsection{$^{8}$B levels}

\label{sec:8B} 

\begin{figure}[tbp]
\includegraphics*[ scale=.42]{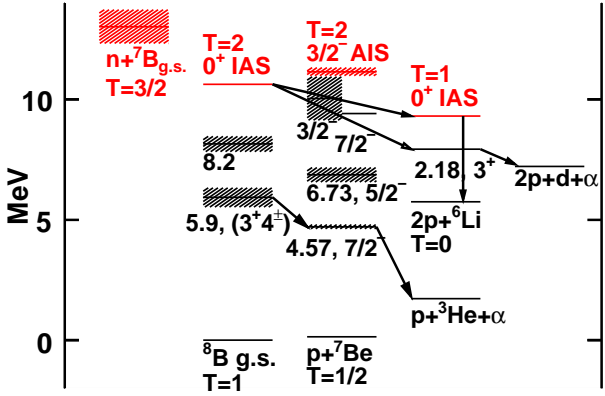}
\caption{(Color online) Level scheme showing the levels of $^{8}$B discussed in this work and their decay paths.}
\label{fig:levelB8}
\end{figure}

The $^8$B excited states studied in this work and their observed decay 
paths are illustrated
 in the level diagram of Fig.~\ref{fig:levelB8}. We observed excited states in 
three detected exit channels; \textit{p}+$^3$He+$\alpha$, 2\textit{p}+$^6$Li,
 and 2\textit{p}+\textit{d}+$\alpha$. The \textit{p}+$^7$Be channel, which is probably 
the most important exit channel for many states, was not accessible 
in this work as Be fragments saturate the Si shaping amplifiers.

\subsubsection{\textit{p}+$^3$He+$\alpha$ exit channel}
\label{sec:p3a}

\begin{figure}[tbp]
\includegraphics*[ scale=.6]{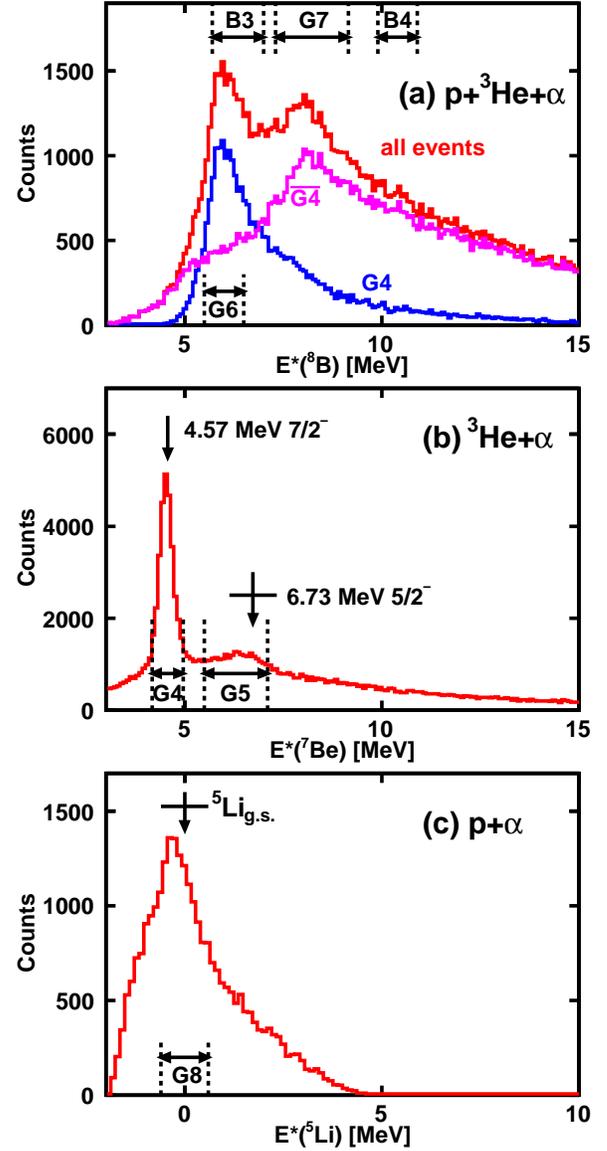}
\caption{(Color online) Excitation-energy spectra determined from 
\textit{p}+$^{3}$He+$\alpha$ events.  
(a) The $^8$B excitation spectrum from all  
\textit{p}+$^{3}$He+$\alpha$ events and its decomposition 
into those associated with the 
4.57-MeV $^7$Be intermediate state and those not. (b) The $^7$Be 
excitation spectrum obtained from the $^3$He-$\alpha$ pairs of all  
\textit{p}+$^{3}$He+$\alpha$ events. (c) Background-subtracted $^5$Li 
excitation-energy spectrum associated with the 8.15-MeV state in $^8$B. 
Vertical arrows
 locate known levels with their widths indicated by the error bars. 
The gates used in this work are also indicated by the dashed lines.}
\label{fig:Ex_8B}
\end{figure}

The $^8$B excitation-energy spectrum derived from all detected
 \textit{p}+$^3$He+$\alpha$ events, shown in Fig.~\ref{fig:Ex_8B}(a), 
displays wide peaks at 5.93 and 8.15 MeV. The $^{7}$Be excitation-energy 
distribution obtained from the $^{3}$He+$\alpha$ pairs associated with each 
of these events is shown in Fig.~\ref{fig:Ex_8B}(b). A peak associated 
with the 4.57-MeV 7/2$^{-}$ level is prominent. The total 
$^8$B excitation spectra is subdivided in Fig.~\ref{fig:Ex_8B}(a) 
into those in coincidence with the $^{7}$Be peak 
[gate $G4$ in Fig.~\ref{fig:Ex_8B}(b)] and those that are not. 
This subdivision separates
 the two $^8$B peaks. The 5.93-MeV peak is associated with proton decay 
to the 4.57-MeV state of $^7$Be and 8.15-MeV state does not decay by this 
path. Let us first concentrate on the 5.93-MeV peak.

\begin{figure}[tbp]
\includegraphics*[ scale=.4]{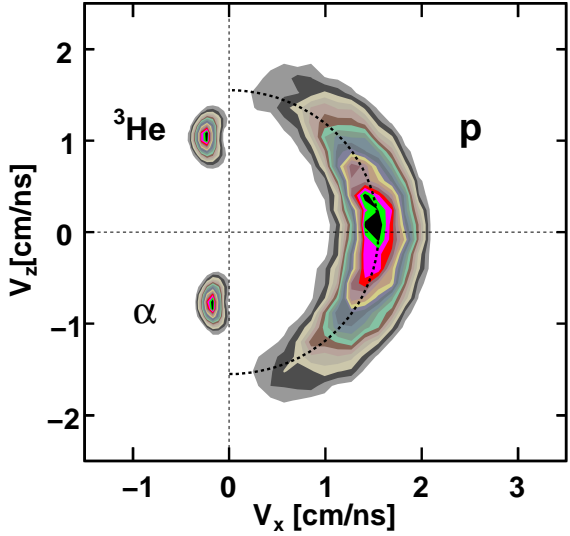}
\caption{(Color online) Distributions of velocities of the three fragments 
for the \textit{p}+$^{3}$He+$\alpha$ exit channel associated 
with the 5.93-MeV $^8$B state. Events were selected using the $G4$ and $G6$ 
gates in Fig.~\ref{fig:Ex_8B}. The distributions are projected onto the 
plane of the decay, with all 
$^3$He  and all $\alpha$-particles locations forced to be 
approximately coincident.}
\label{fig:b8map}
\end{figure}

As there are only three fragments in the exit channel, the velocity vectors
 of the decay products are located in a plane in the $^8$B center-of-mass 
frame. 
To display the correlations between the decay fragments, we have projected 
these vectors onto that plane, and then within this plane, rotated all the 
vectors such that the locations of all the $\alpha$-particle velocities 
approximately coincide and at the same time the locations of  all the 
$^{3}$He fragments 
velocities approximately coincide. 
This was achieved by requiring that the relative
 $^3$He-$\alpha$ velocity was parallel to the $V_z$ axis. A contour plot of 
the resulting distribution of \textit{p}, $^{3}$He, and $\alpha$ velocities 
is displayed in Fig.~\ref{fig:b8map}. The $^{3}$He-$\alpha$ separation is 
constant as these fragments come from the decay of the  4.57-MeV $^7$Be state.
The protons lie approximately on an arc centered at the origin 
($^8$B center of mass) as indicated by the dashed curve. Clearly the protons were emitted in the first step and the
 magnitude of their velocity is independent of their emission direction and 
the subsequent decay axis of the $^7$Be intermediate state.

\begin{figure}[tbp]
\includegraphics*[ scale=.4]{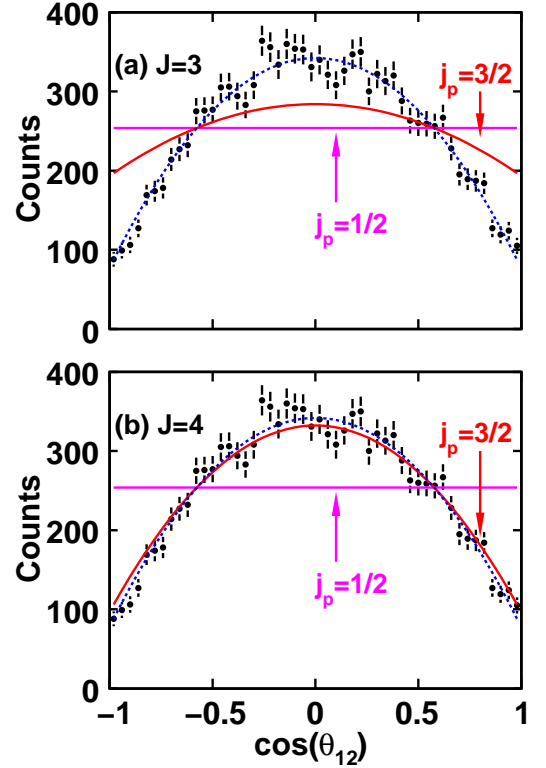}
\caption{(Color online) Angular correlations for the 5.93-MeV state of 
$^{8}$B. Events were selected using both the $G4$ and $G6$ gates in 
Fig.~\ref{fig:Ex_8B}. The angle $\theta_{12}$ in the relative angle between 
the initial proton decay axis and the subsequent $^3$He-$\alpha$ decay 
axis in the $^8$B center-of-mass frame. The proton and $\alpha$ particle 
are emitted in the same direction when $\theta_{12}$=0. The solid curves are predictions for pure values of $j_p$, while the dashed curves are for mixed values where the degree of mixing was chosen to best fit the data.}
\label{fig:b8cos_level1}
\end{figure}

The decay of this state is expected to be  sequential 
with $R_{E}$=0.11 and $d_{E}$=60~fm, and thus the two decay steps are largely 
independent, apart from consideration of conservation laws. Angular momentum 
conservation gives rise to the angular correlations.
Let us consider the relative angle $\theta_{12}$ between the two decay steps
which is related to the angle $\theta_k$ in the Jacobi ``Y'' system.
A $\theta_{12}$=0 corresponds to an event where the proton in the first step
 and the $\alpha$ in the second step are emitted in the same direction. 
The distribution of $\theta_{12}$ is shown as the data points in both 
panels of Fig.~\ref{fig:b8cos_level1}. The experimental distribution is 
symmetric about $\cos(\theta_{12})$=0 consistent with the expectation for a 
sequential decay scenario. The experimental distribution is clearly not 
isotropic (flat). The predicted 
correlations in a sequential decay depend of the initial 
spin $J$ and the total angular momentum 
removed by the proton $j_p$ in the first decay step. 
The predicted correlations do not depend on the 
parity of the initial state for these proton decays, but 
 mixing between $j_p$ values should also  considered, i.e., the 
angular correlation becomes
\begin{eqnarray}
w(\theta_{12}) &=& \left| \alpha_1 \right|^2 w_1(\theta_{12}) +
\left| \alpha_2 \right|^2 w_2(\theta_{12}) \nonumber \\
&& + (\alpha_1 \alpha_2^* +
 \alpha_1^* \alpha_2) w_{inter}(\theta_{12}) , 
\end{eqnarray}
where $w_1$ and $w_2$ are the correlations associated with two pure values 
of $j_p$, $\alpha_1$ and $\alpha_2$ are their complex amplitudes and 
$w_{inter}$ is the interference term.

With this mixing, only two possible $J$ values were consistent with
the experimental distributions. Figure \ref{fig:b8cos_level1}(a) shows 
the results for $J$=3. The two solid curves show the predictions for two
 pure values of $j_p$; 1/2 and 3/2. 
These predictions have been normalized to the same number of events 
as in the experimental distribution and Monte Carlo
simulations indicate that the
distortions due to the angular acceptance and energy resolution of 
the detectors are minimal.  Neither of these predictions fit 
the data at all. However,  a mixed solution with relative amplitudes 
of $\alpha_1$=1 and 
$\alpha_2$=0.49$\pm$0.14 for the two $j_p$ values, respectively, 
fits the data very well [dashed curve in Fig.~\ref{fig:b8cos_level1}(a)]. 

The other solution was obtained for $J$=4 and is shown in 
Fig.~\ref{fig:b8cos_level1}(b). Again, pure $j_p$=1/2 and 3/2 
predictions are shown by the solid curves. In this case the $j_p$=3/2 
predictions fits the data reasonably well. If we allow for mixing, the 
best fit is obtained with relative amplitudes of $\alpha_2$=1 and 
$\alpha_1$=-0.07$\pm$0.10, however, the level of mixing is minimal.
All told, the angular correlations indicate that the 5.91-MeV state is either 
$J$=3 or 4.


For the 8.15~MeV state in $^8$B, which does not
 proton decay to the 4.57-MeV $^7$Be state [Fig.~\ref{fig:Ex_8B}(a)], 
it is difficult to clearly define its decay path due to the 
larger background under this peak and also because there are no possible 
narrow intermediate states through which it can decay. 
Sequential decay
 through a higher-lying $^7$Be state and through the $^5$Li ground state are possible. 
Apart from the narrow 4.57-MeV state, the $^7$Be excitation spectra in 
Fig.~\ref{fig:Ex_8B}(b) shows a broad peak at $\sim$6.5~MeV,
 close to the location of the $E^*$=6.73-MeV,
$J^{\pi}$=5/2$^{-}$, $\Gamma$=1.2-MeV state in $^7$Be, indicated
 by the arrow in the Fig.~\ref{fig:Ex_8B}(b). 
For a wide intermediate state, 
the full width of the state is not always populated by sequential decay as the 
Coulomb barrier of the first step can suppress the higher excitation region.
The peak at 6.5 MeV may be associated with the 
6.73-MeV state and gating of this peak enhances the 8.15-MeV peak. 
However for all the \textit{p}+$^3$He+$\alpha$ events, 
only 30$\pm$6\% of the kinetic energy, in the reconstructed $^8$B frame, is
 associated with the proton. Thus the $^3$He+$\alpha$ pair 
accounts for the remaining 70$\pm$6\% and thus the $^{3}$He-$\alpha$ 
reconstructed excitation energy is 
strongly correlated to \textit{p}+$^3$He+$\alpha$ excitation energy suggesting
that the 
6.5-MeV peak in $^7$Be spectra in Fig.~\ref{fig:Ex_8B}(b) is 
a reflection of the peak at 8.15 MeV in Fig.~\ref{fig:Ex_8B}(a) due to this 
strong correlation and not associated with an intermediate state.  

\begin{figure}[tbp]
\includegraphics*[ scale=.4]{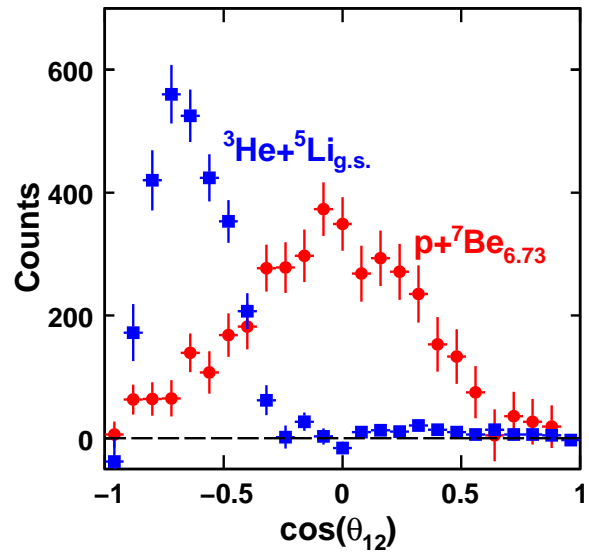}
\caption{(Color online) Background-subtracted angular correlations 
associated with possible 
sequential-decay scenarios for the 8.15-MeV state of $^{8}$B. 
The 8-15-MeV peak was selected with gate $G7$ in Fig.~\ref{fig:Ex_8B} 
and the background was estimated from $B3\cup B4$.
For  proton decay to the 6.73-MeV state of $^7$Be, the extra gate $G5$ was 
applied and for $^3$He decay to the ground state of $^5$Li, 
the extra gate $G8$ was applied.}
\label{fig:b8cos_level2}
\end{figure}

Also possible is the decay $^8$B$^*\rightarrow^3$He+$^5$Li$_{g.s.}$
Figure~\ref{fig:Ex_8B}(c) shows the 
reconstructed $^5$Li excitation-energy  spectrum determined for the 
\textit{p}-$\alpha$ pairs associated with the 8.15-MeV peak [gate $G7$ in 
Fig.~\ref{fig:Ex_8B}(a)]).
A rough background correction was made using gates $B3$ and $B4$ on either side of the 8.15-MeV peak. A broad peak at zero
 excitation energy is observed overlapping strongly with the ground state
 of $^5$Li. Angular correlations were constructed for both the two sequential 
decay scenarios, and after background subtractions, are shown in 
Fig.~\ref{fig:b8cos_level2}. In this figure $\theta_{12}$=0 for 
$^3$He+$^5$Li$_{g.s.}$ 
decay corresponds to the $^3$He and $\alpha$ being emitted in the 
same direction. The $^3$He-$^5$Li$_{g.s.}$ results is clearly asymmetric about 
$\cos(\theta_{12})$=0 and cannot be explained by sequential decay. For the 
\textit{p}+$^7$Be$_{6.73}$ case, although the distribution is symmetric about 
$\cos(\theta_{12})$=0, we were unable to fit it with 
the range of possible correlations expected for a 5/2$^{-}$ intermediate state
and thus there is no evidence for a sequential decay of this state which 
suggets the decay is 3-body in nature.
 
\subsubsection{2\textit{p}+$^6$Li exit channel}
\label{sec:2p6Li}
\begin{figure}[tbp]
\includegraphics*[ scale=.4]{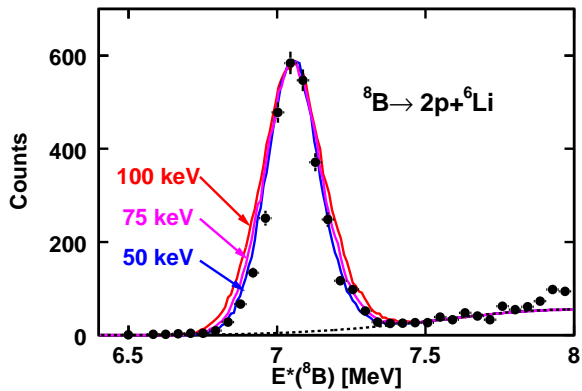}
\caption{(Color online) Excitation-energy spectra from 2\textit{p}+$^{6}$Li events. The curves show simulated distributions for Breit-Wigner line shapes with
the indicated decay widths. The lower curve below the data, 
indicates the background which was included in the simulated distributions.}
\label{fig:Ex_8B_fit}
\end{figure}

The excitation-energy spectrum derived from 2\textit{p}+$^6$Li events in
Fig.~\ref{fig:Ex_8B_fit} displays a very prominent peak at 7.06$\pm$0.02~MeV. 
The excitation energy in Fig.~\ref{fig:Ex_8B_fit} was calculated 
 based on the assumption 
that the $^6$Li fragments were created in their ground states and did not 
$\gamma$ decay. Only one $^6$Li excited state has a significant 
$\gamma$-decay branch, the 3.562-MeV, $J^{\pi}$=0$^+$, $T$=1 level whose $\gamma$ 
branching ratio is practically 100\%. Therefore the peak observed in 
Fig.~\ref{fig:Ex_8B_fit} can correspond to a level of excitation energy of 
either 7.06$\pm$0.02 or 10.62$\pm$0.02 MeV. 

The intrinsic width of this level is quite narrow as 
the experimental width is similar to the predicted instrumental 
resolution. If we assume
a Breit-Wigner line shape, we can further constrain the width.
The curves in Fig.~\ref{fig:Ex_8B_fit} show Monte Carlo predictions
including the detector response for intrinsic widths of 50, 75, 
and 100 keV normalized to the experimental peak height. 
The behavior  in the region of 
the low-energy tail, 
where the uncertainty due to the background contribution is minimal,
suggests an intrinsic width of less than 75~keV.

There are no previously known levels at $E^*$=7.06~MeV. From the mirror nucleus
$^8$Li, we expect only one narrow state near this energy, the mirror of the 
6.53-MeV $J^{\pi}$=4$^+$, $\Gamma$=35~keV level. However, this $^8$Li state has a 
significant branch to the \textit{n}+\textit{t}+$\alpha$ exit channel 
passing through the 4.63-MeV $^7$Li \cite{Grigorenko02}. 
As the 7.06-MeV state is not observed is the mirror exit channel \textit{p}+$^3$He+$\alpha$ 
(see Fig.~\ref{fig:Ex_8B}) is seems unlikely 
that it can be this $J^{\pi}$=4$^+$ state.

On the other hand, if the peak in Fig.~\ref{fig:Ex_8B_fit} corresponds to a 
10.62-MeV state, then we can immediately  identify it with the known 
10.619$\pm$0.009~MeV $J^{\pi}$=0$^+$ $T$=2 state in $^8$B with $\Gamma<$60~keV
\cite{Robertson75}.
 The almost exact matching of the excitation energy and a consistent limit 
to the width  strongly support this assignment.

With this assignment, the $^8$B state is thus the isobaric analog of $^8$C$_{g.s.}$ 
discussed in Sec.~\ref{sec:8C} and its structure should 
be similar. It decays by the emission of two protons  to the isobaric analog 
of $^6$Be$_{g.s.}$ in $^6$Li. Such a decay can conserve isospin only 
if the two protons are emitted promptly in a $T$=1 configuration. 
Sequential two-proton decay does not conserve isospin as there
are no energetically accessible $T$=3/2 states available in the $^7$Be 
intermediate nucleus (see Fig.~\ref{fig:levelB8}).
This is an 
isospin equivalent to the Goldansky two-proton decay, as the isospin allowed 
intermediate state is energetically forbidden.

Further elucidation of the nature of this two-proton decay 
may be obtained from the correlations between the fragments. 
Unfortunately in this experiment, most 
$^6$Li fragments were not identified due to a saturation of the Si 
amplifiers, thus most of the 2\textit{p}+$^6$Li events associated with 
this state were not observed except where the $^6$Li fragment had 
high kinetic energy (low $\Delta E$) strongly biasing any correlation measurement.
A future experiment is planned to study this correlation.

\subsubsection{2\textit{p}+\textit{d}+$\alpha$ exit channel}

\begin{figure}[tbp]
\includegraphics*[ scale=.4]{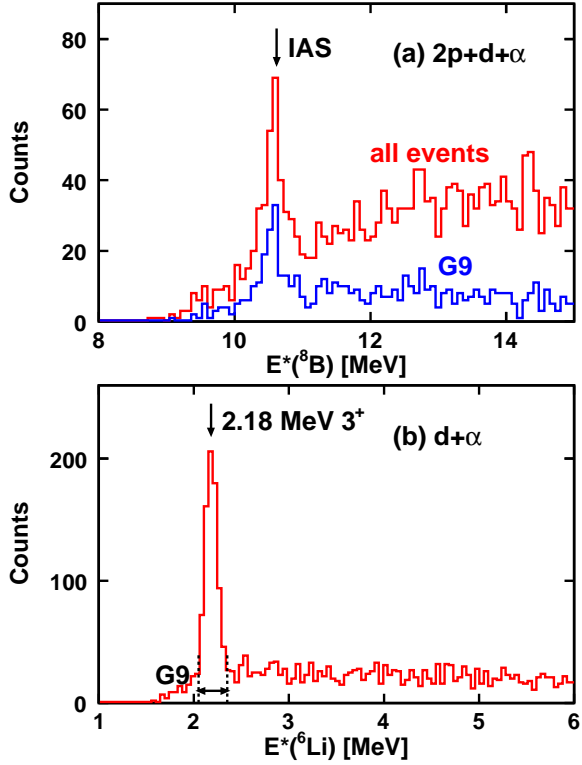}
\caption{(Color online) (a) $^8$B excitation-energy spectra obtained 
from 2\textit{p}+\textit{d}+$\alpha$ events. (b) $^6$Li excitation-energy 
spectrum obtained from each \textit{d}-$\alpha$ pair of each  
2\textit{p}+\textit{d}+$\alpha$ event. The arrows indicate the locations of 
known states. The gate $G9$ in (b) is used to gate the spectrum in (a). }
\label{fig:Ex_8B2}
\end{figure}

The $^8$B isobaric analog state  also decays to the 
2\textit{p}+\textit{d}+$\alpha$ exit channel. Here, the decay is 
 isospin forbidden and the yield is much smaller than that obtained for
the 2\textit{p}+$^6$Li channel (Table~\ref{Tbl:states}), further suggesting that the latter 
is an isospin-allowed decay. The excitation-energy spectrum extracted from 
these events is plotted in Fig.~\ref{fig:Ex_8B2}(a) and shows a 
narrow peak at the energy of the isobaric analog state (arrow). Possible 
narrow intermediate states are the $^6$Be ground state and the 2.18-MeV 
$J^{\pi}$=3$^+$ $^6$Li excited state, although we found no evidence that the  former is
 associated with this peak and in fact used the $^6$Be ground state as a veto 
to reduce the background.  The background was also reduced by gating on the 
reconstructed $^8$B velocity distribution associated with the IAS peak for 
the 2\textit{p}+$^6$Li events. The spectrum labeled ``all events'' 
contained these two background reducing gates. For ``all events'', 
the reconstructed $^6$Li excitation spectra from the \textit{d}-$\alpha$ 
pairs is shown in Fig.~\ref{fig:Ex_8B2}(b) where the 2.18-MeV state is 
quite prominent. Gating on this state [gate $G4$ in Fig.~\ref{fig:Ex_8B}(b)], 
produces the excitation-energy spectrum  shown in 
Fig.~\ref{fig:Ex_8B2}(a). The peak in this case contains only $\sim$50\% of 
the strength of the ``all events'' peak. The origin of the remaining 
strength is not clear, possibly from 3 or 4-body decays, 
but certainly is not associated with sequential decay through a narrow state.

For the strength associated with the 2.18-MeV $^6$Li state, we were not able 
to identify any $^7$Be intermediate levels. Either the two protons 
are emitted in an initial 3-body decay or the $^7$Be intermediate state(s) are 
too wide to isolate experimentally. Because of the low yields, the statistical errors associated with the background-subtracted correlations are too 
large to extract any further information.  

\subsection{$^{8}$Be levels}
\label{sec:8Be} 

\begin{figure}[tbp]
\includegraphics*[ scale=.4]{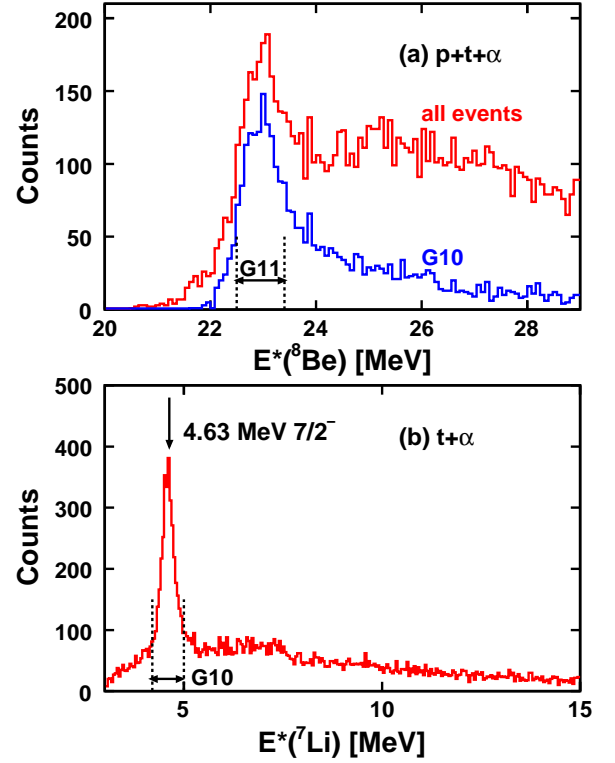}
\caption{(Color online) (a) $^8$Be excitation-energy spectra from 
\textit{p}+\textit{t}+$\alpha$ events. (b)$^7$Li excitation-energy spectrum
from the \textit{t}-$\alpha$ pairs in each \textit{p}+\textit{t}+$\alpha$ 
event.}
\label{fig:Ex_8Be}
\end{figure}

A high lying $^8$Be state is seen in the \textit{p}+\textit{t}+$\alpha$ 
excitation spectrum. The distribution from all detected events is displayed 
in Fig.~\ref{fig:Ex_8Be}(a)(``all events''). 
A broad peak is observed at 22.96 MeV. Such a peak
 was also observed previously following the fragmentation 
of a $^{12}$Be beam and 
was associated with proton decay to the 4.63-MeV $J^{\pi}$=7/2$^{-}$ $^7$Li 
excited state \cite{Charity08}. This is confirmed in the present work. 
The excitation spectrum obtained from the \textit{t}-$\alpha$ pairs is shown 
in Fig.~\ref{fig:Ex_8Be}(b) where the 4.63-MeV $^7$Li state is quite 
prominent. Gating on this peak [gate $G10$ in Fig.~\ref{fig:Ex_8Be}(b)] 
produced the ``$G10$'' spectrum in Fig.~\ref{fig:Ex_8Be}(a). 
The gate significantly
 reduced the background, but did not reduce the yield in the peak indicating 
that proton decay to the 4.63-MeV state is the dominant decay mode. 

\begin{figure}[tbp]
\includegraphics*[ scale=.4]{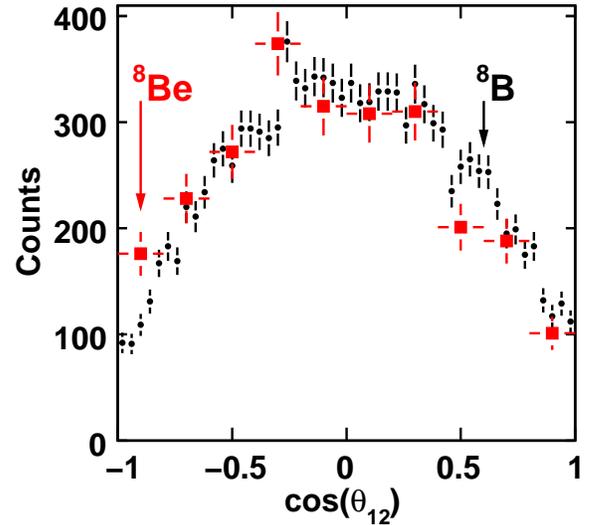}
\caption{(Color online) Comparison of angular correlations obtained for the 5.92-MeV state in $^{8}$B and the 
22.96-MeV state in $^{8}$Be. The correlations for the $^8$Be state 
were obtained using gates $G10$ and $G11$ in Fig.~\ref{fig:Ex_8Be}.}
\label{fig:be8cos}
\end{figure}

This state has similarities to the 5.93-MeV state in $^8$B which proton decays 
to the 4.57-MeV state in $^7$Be, i.e., the mirror state to the 4.63-MeV in 
$^7$Li. Figure~\ref{fig:be8cos} compares the angular distributions of 
$\theta_{12}$ in the decay of these two states. Here the $^7$B data were scaled 
to the same number of total counts as the $^7$Be results. 
Within the statistical errors,
the correlations are identical suggesting that the 22.96-MeV state in 
$^8$Be is the analog of the 5.93-MeV $^8$B state and thus would have the same 
possible spin assignments ($J$=3 or 4 ). 
The intrinsic widths of these two states are also
comparable; 850$\pm$260 and 680$\pm$146~keV for $^8$B and $^8$Be, respectively.

\begin{figure}[tbp]
\includegraphics*[ scale=.4]{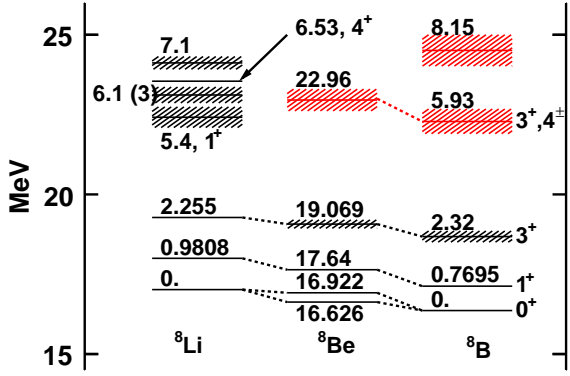}
\caption{(Color online) Isobar level diagram of $T$=1 states for $A$=8.
The levels discussed in this work are displayed in red. The energy scale on 
the \textit{y} axis is with respect to the $^8$Be ground state.}
\label{fig:A8}
\end{figure}

In Fig.~\ref{fig:A8} we show these levels in an isobar diagram of $T$=1 levels.
The relative excitation energies of these two levels is also consistent 
with their assignment to the same isospin triplet. 
The corresponding analog in the $^8$Li nucleus is not entirely clear.
A 6.1-MeV $^8$Li state with $\Gamma\sim$1~MeV is a possibility;
it has been tentatively assigned $J$=3.

\subsection{$^9$B}

\begin{figure}[tbp]
\includegraphics*[ scale=.45]{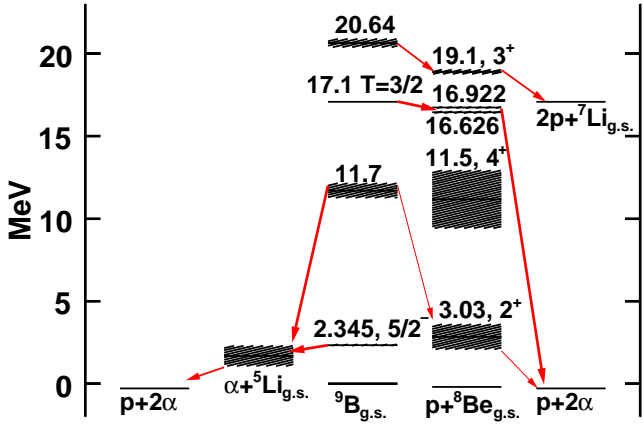}
\caption{(Color online) Level diagram for $^9$B showing the levels discussed in this
work and their decay paths. The energy scale on the \textit{y} axis is with respect to the $^9$B ground state.}
\label{fig:b9}
\end{figure}

We observed $^9$B levels in the \textit{p}+2$\alpha$ and 2\textit{p}+$^{7}$Li 
exit channels. 
Figure~\ref{fig:b9} shows a level diagram with the levels of interest 
and their decay paths.

\subsubsection{\textit{p}+2$\alpha$ exit channel}

\begin{figure}[tbp]
\includegraphics*[ scale=.7]{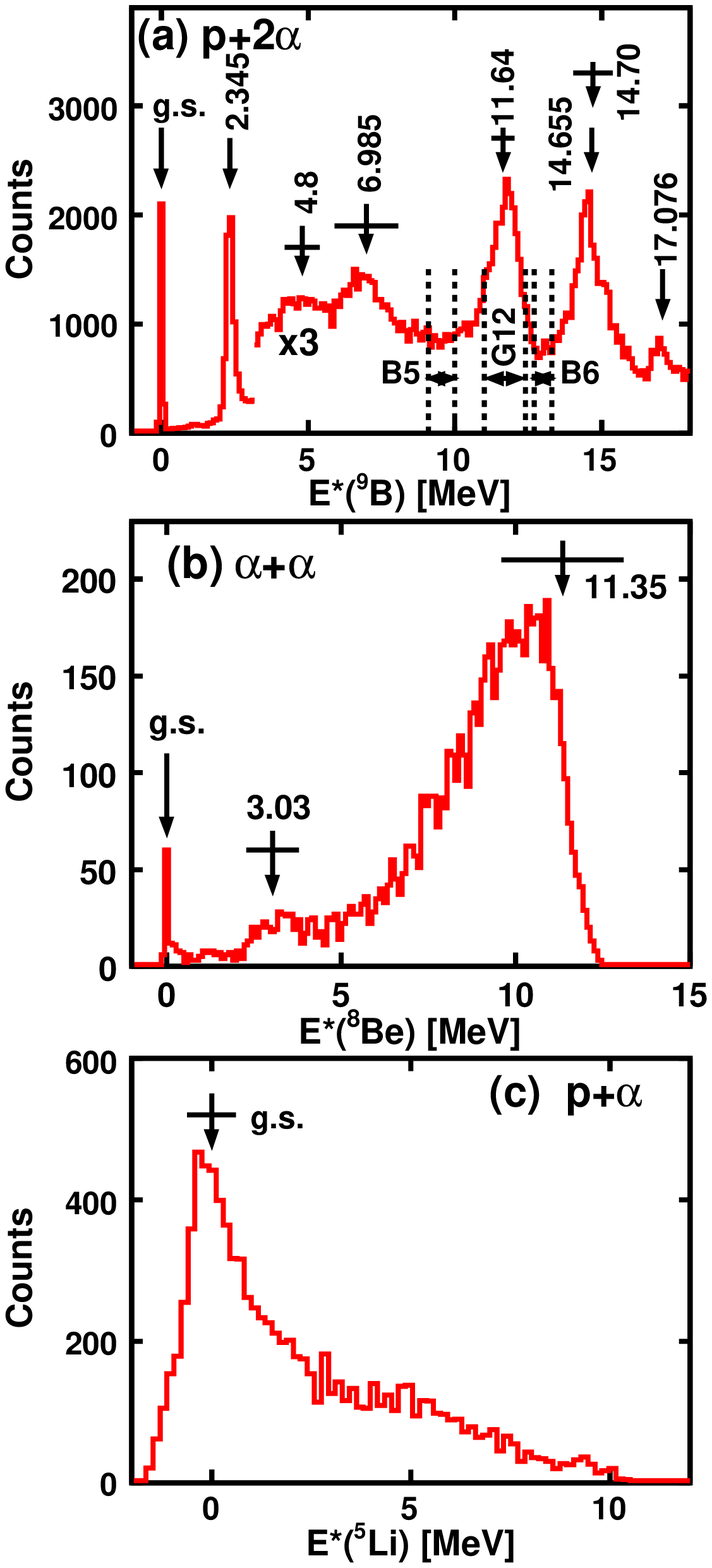}
\caption{(Color online) (a) Excitation-energy spectra of $^{9}$B derived from 
\textit{p}+2$\alpha$ events. (b) The $^8$Be excitation-energy spectra 
obtained from the $\alpha$-$\alpha$ pairs for events in the 11.7-MeV $^9$B 
($G12$ gate), 
(c) Background subtracted $^5$Li excitation-energy spectra for events in the 
11.7-MeV $^9$B peak. The arrows indicate the locations of known levels.}
\label{fig:Ex_9B}
\end{figure}

The $^9$B excitation-energy spectrum derived from \textit{p}+2$\alpha$ events, 
displayed in Fig.~\ref{fig:Ex_9B}(a), shows many peaks. The most prominent of 
these are the ground state and the 2.345-MeV $J^{\pi}$=5/2$^-$ level. The 
spectrum above
 these two peaks has been scaled by a factor of three to show additional 
detail. Prominent peaks are also observed at 11.7 
and 14.7 MeV. For the latter, two listed levels can contribute, 
 the narrow 14.655-MeV isobaric analog state and the wider 14.70-MeV 
($\Gamma$=1.35 MeV) state. Judging by the shape of the 14.7-MeV peak, it is 
likely that both levels are present. As it is impossible to separate 
the decay modes of these two levels, we will not report on any further 
analysis of this peak. 

The peak at 11.7 MeV is close to a listed 11.640-MeV level with 
 a tentative spin assignment of $J^{\pi}$=7/2$^{-}$.
The width of this listed level (780$\pm$45~keV) is also close to the value of 
880$\pm$80keV extracted from the peak after correction for the experimental resolution.
For further analysis of the correlations associated with this peak, we use the gate $G12$ in 
Fig.~\ref{fig:Ex_9B}(a) with the regions $B5$ and $B6$ on either side to 
estimate the background under the peak. The $^8$Be excitation-energy 
spectra from the $\alpha$-$\alpha$ pairs associated with each event in the
$G12$ gate is plotted in Fig.~\ref{fig:Ex_9B}(b). The ``background'' was not 
subtracted from this spectrum as the reconstructed $^8$Be ($\alpha$+$\alpha$)
 and $^9$B (\textit{p}+$\alpha$+$\alpha$)
excitation energies are highly correlated. This is similar to the
 strong correlation between the $^3$He+$\alpha$ and 
\textit{p}+$^3$He+$\alpha$ reconstructed excitation energies found in 
Sec.~\ref{sec:p3a}. Due to this correlation, the spectra from gates $B5$
and $B6$ are not good approximations to the background associated with gate $G12$
for the $^8$Be excitation energy.
In Fig.~\ref{fig:Ex_9B}(b), contributions from the $^8$Be ground state
 and the 3.03-MeV first excited state are clearly present. However, the 
ground-state contribution is essentially all background because when 
we gate on this peak 
and project the $^9$B excitation-energy distribution, no indication of the 
11.7-MeV peak is seen. On the other hand, the 3.03~MeV peak is not all 
background. 
With a similar gating arrangement, we find $\sim$2.8\% of the strength at 
11.7 MeV is be associated with this $^8$Be intermediate state. 

Most of the 11.7-MeV $^9$B peak is thus associated with the 
broad structure above 
$E^*(^{8}Be)$=5~MeV in Fig.~\ref{fig:Ex_9B}(b). 
This structure overlaps with the wide 11.35-MeV $J^{\pi}$=4$^+$ 
$^8$Be level. The situation here is similar to the 8.15-MeV $^8$B state.  
The  centroid of the structure in Fig.~\ref{fig:Ex_9B}(b) is below the centroid for this level, 
possibly a consequence of sequential feeding. However the $^8$Be and $^9$B 
excitation energies are well correlated and thus the structure could just be a 
consequence of this correlation. The alternative intermediate state is $^5$Li. 
Fig.~\ref{fig:Ex_9B}(c) shows the background-subtracted $^5$Li excitation-energy 
spectrum associated with the 11.7-MeV peak obtained from the \textit{p}+$\alpha$ pairs. 
There are two possible  \textit{p}+$\alpha$ pairs from each \textit{p}+2$\alpha$ 
event and both are included in this spectrum. A peak corresponding to the 
ground state of $^5$Li is clearly present indicating that $\alpha$+$^5$Li$_{g.s.}$
is an important decay branch.

\begin{figure}[tbp]
\includegraphics*[ scale=.4]{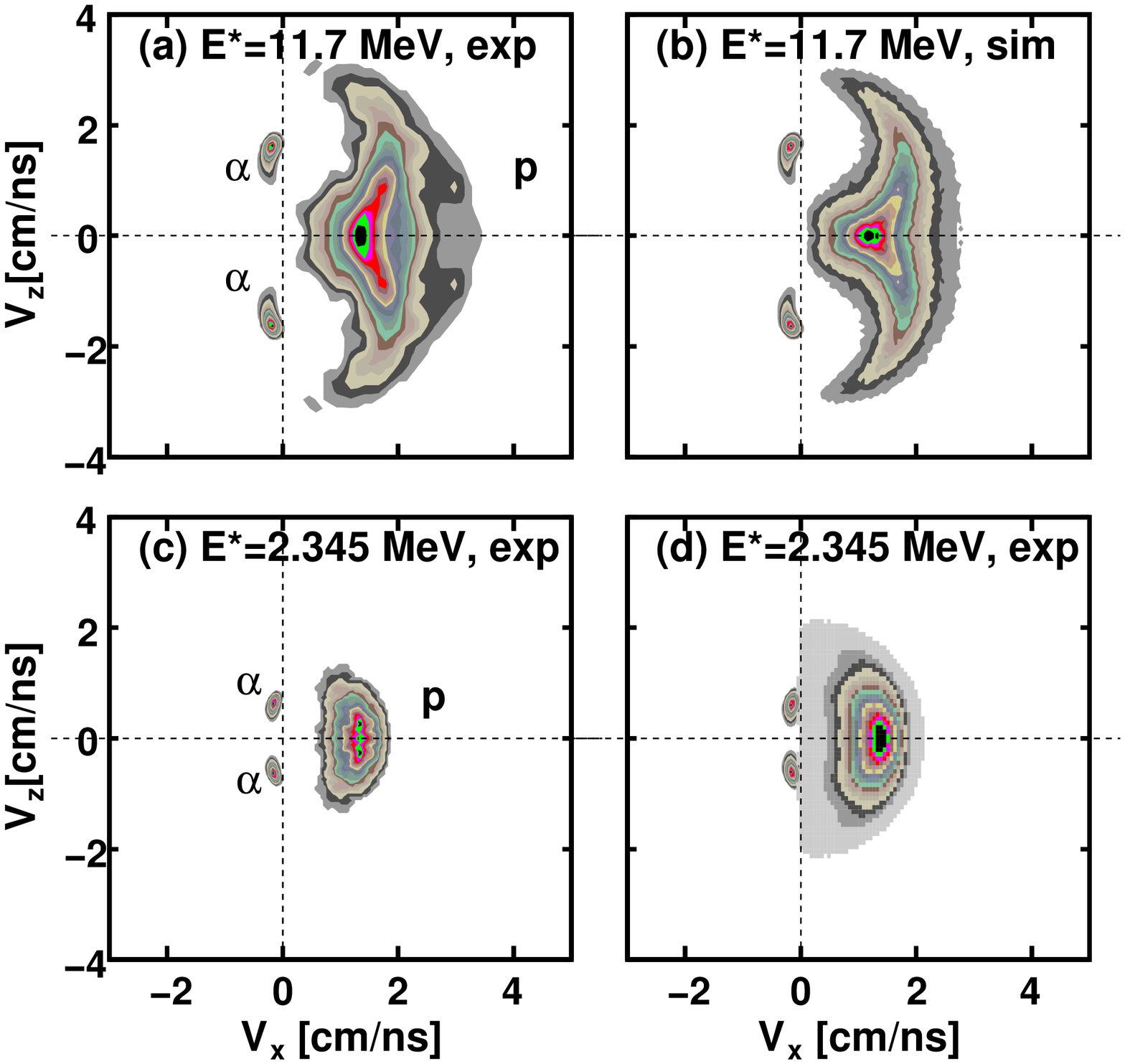}
\caption{(Color online) Distributions of the velocities of the three 
fragments for the \textit{p}+2$\alpha$ exit channel associated with (a,b) 
the 11.7-MeV and (c,d) the 2.345-MeV states of $^9$B. The distributions are projected
 on the plane of the decay with the two $\alpha$-particle velocities localized.
The experimental distributions in (a,c) are compared to simulated 
distributions for $\alpha$-$^5$Li$_{g.s.}$ decay in (b,d).} 
\label{fig:b9_map}
\end{figure}

The correlations associated with the decay of the 11.7-MeV state can be displayed 
in a manner similar to Fig.~\ref{fig:b8map} where we projected the velocity vectors onto 
the decay plane. To localized the distribution of $\alpha$ particles, the 
$\alpha$-$\alpha$ relative velocity was made parallel to the $V_z$ axis, and as we
cannot distinguish the two $\alpha$ particles, we have symmetrized the distributions
about $V_z$=0. The velocity plot for the 11.7-MeV state is displayed in 
Fig.~\ref{fig:b9_map}(a) and can be compared to the corresponding plot 
for the 2.345-MeV state in Fig.~\ref{fig:b9_map}(c). Unlike  Fig.~\ref{fig:b8map}
where the protons lie on an arc centered at the origin, the protons in Fig.~\ref{fig:b9_map}(a)
lie on two overlapping arcs each centered approximately near an $\alpha$ particle.
These two arcs indicate that the protons are not emitted from the $^9$B center of mass, as expected 
for direct proton decay, but come predominantly from the decay of a $^5$Li intermediate fragment.
Thus the broad peak in the $\alpha$-$\alpha$ excitation spectrum in Fig.~\ref{fig:Ex_9B}(b) 
should not be interpreted as an $^8$Be excited state.

The distributions in Figs.~\ref{fig:b9_map}(b) and \ref{fig:b9_map}(d) 
show the results of simulations of a sequential disintegration initiated 
by an $\alpha$+$^5$Li$_{g.s.}$ decay. 
The 2.345-MeV level was previously 
identified as having strong  $\alpha$+$^5$Li$_{g.s.}$  decay character \cite{Charity07}, 
however the two arcs are not separated as the two $\alpha$ particles are located too 
close together. In both simulations we have determined the emission energies in the two steps
using the R-matrix formalism as described in Sec.~\ref{sec:simul}. Both simulations reproduce 
the experimental distribution reasonably well.

\begin{figure}[tbp]
\includegraphics*[ scale=.5]{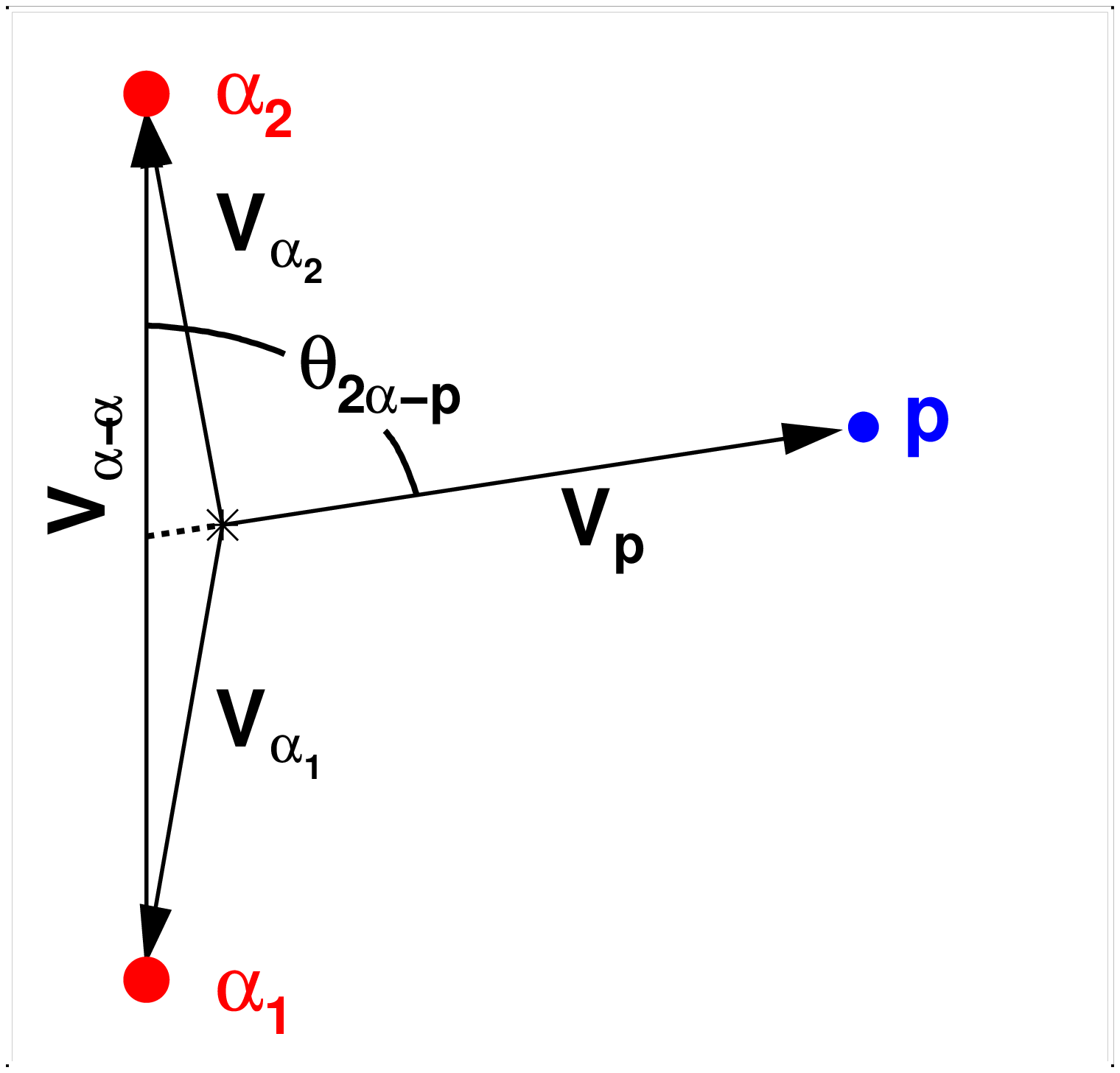}
\caption{(Color online) Schematic showing the definition of 
the angle $\theta_{2\alpha-p}$.}
\label{fig:b9angle}
\end{figure}

\begin{figure}[tbp]
\includegraphics*[ scale=.5]{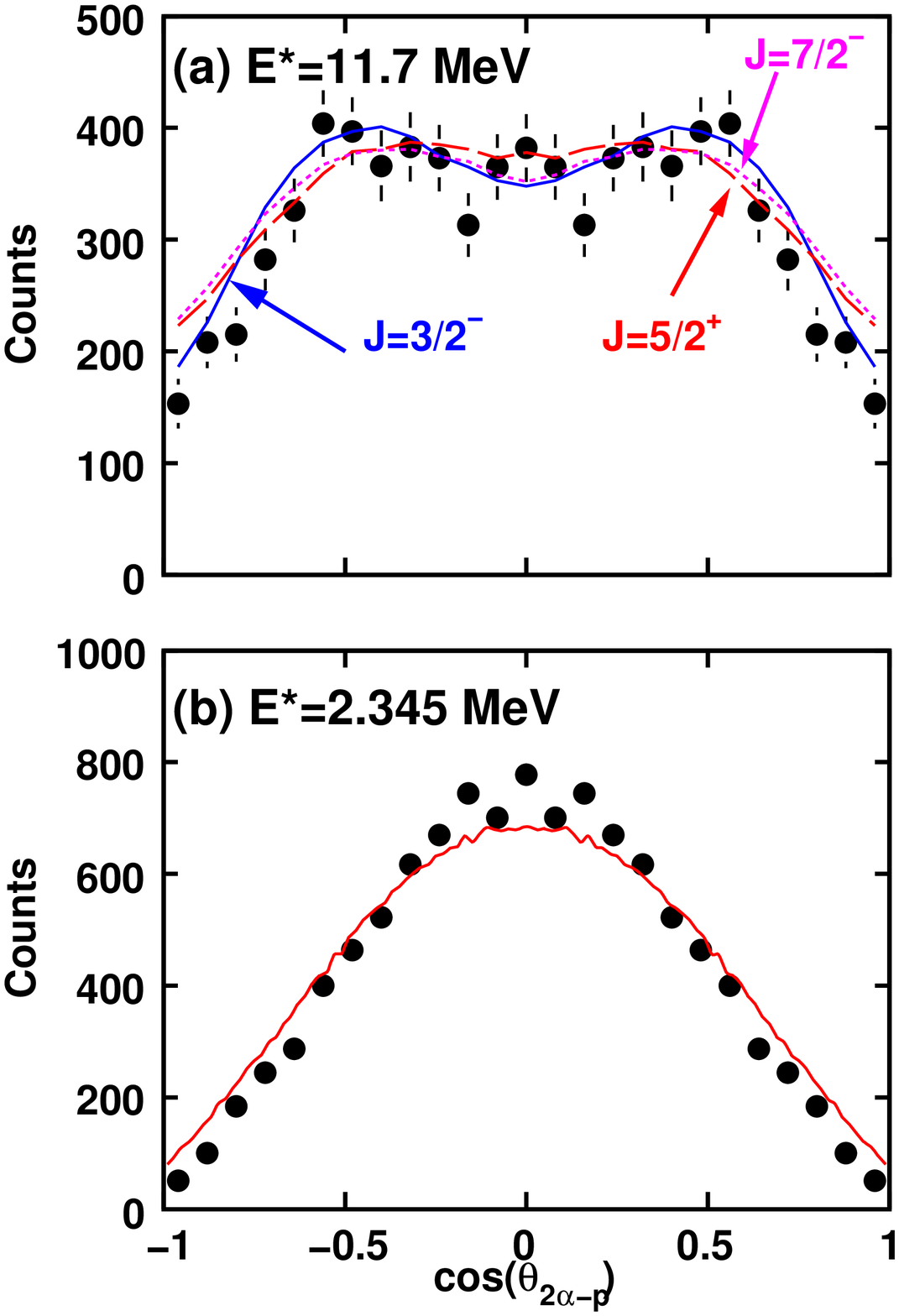}
\caption{(Color online) Background-subtracted angular corrections measured 
in the decay of the (a) 11.7-MeV and (b) the 2.345-MeV states of $^9$B. 
The curves indicate calculated distributions assuming an initial 
$\alpha$-$^5$Li$_{g.s.}$ decay. In (a), the calculated results are given for 
a number of assumed spins for the 11.7-MeV state.} 
\label{fig:b9_cos}
\end{figure}

Information about the spin of the 11.7-MeV state can be determined from the angular correlations 
of the two decay steps. Unfortunately as the two arcs overlap, we cannot always identify which $\alpha$ 
particle is emitted in the first and second steps and thus we cannot uniquely determine $\theta_{12}$.
Instead we have looked at the angle $\theta_{2\alpha-p}$ defined in Fig.~\ref{fig:b9angle}. This is the angle 
between the relative velocity vectors of the two $\alpha$ particles ($V_{\alpha-\alpha}$)
and the proton velocity ($V_p$) in the $^9$B 
center-of-mass frame. Background-subtracted distributions of this quantity are shown for both the 
11.7 and 2.345-MeV states in Fig.~\ref{fig:b9_cos}. For an initial 
$\alpha$-particle decay, unlike proton decay,
 the angular correlations are sensitive to the parity of the initial level. Furthermore if we constrain ourselves 
to $\alpha$ decays with the lowest possible $\ell$ waves, 
there is no mixing to consider and each $J^\pi$ value has a unique 
correlation. The prediction obtained for the 2.345-MeV level with its known spin of 5/2$^-$ 
is compared to the experimental data in Fig.~\ref{fig:b9_cos}(b). 
It fits the data extremely well and thus again is consistent with sequential decay initiated by an 
 $\alpha$-$^5$Li$_{g.s.}$ decay.  In Fig.~\ref{fig:b9_cos}(a) for the 11.7-MeV
state, we show three predictions which fit the experimental data with $J^\pi$=3/2$^{-}$, 5/2$^{+}$, and 7/2$^-$. 
Of these, the 3/2$^-$ predictions gives the best fit with a $\chi^2/\nu$=1.7. The other two cases have 
$\chi^2/\nu$=2.6 and 3.0 respectively, so $J^\pi$=3/2$^{-}$ is slightly preferred. We note, that 
the listed 11.64-MeV state in $^9$B has a tentative assignment as $J^\pi$=7/2$^-$ which is the least likely 
of our three possible fits.        

\begin{figure}[tbp]
\includegraphics*[ scale=.5]{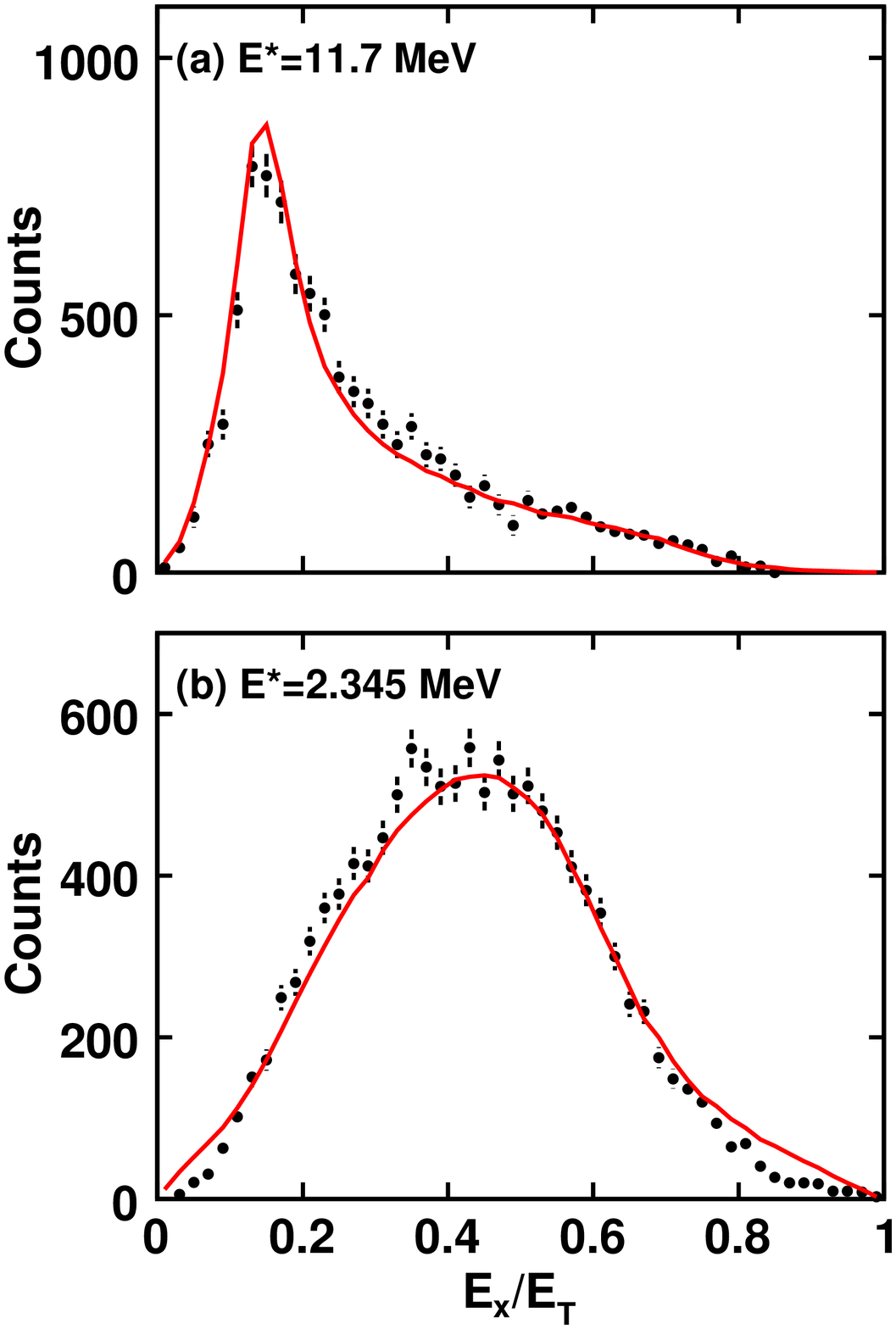}
\caption{(Color online) Background-subtracted energy corrections 
in the Jacobi ``Y'' system measured 
in the decay of the (a) 11.7-MeV and (b) the 2.345-MeV states of $^9$B. 
The curves indicate calculated distributions assuming 
$\alpha$-$^5$Li$_{g.s.}$ decay in the R-matrix approximation.} 
\label{fig:Ex_9B_J}
\end{figure}

Figure~\ref{fig:Ex_9B_J} shows the energy correlations for the 11.7 and 2.345-MeV states in $^{9}$B.
The correlations are presented in the Jacobi ``Y'' system 
where the ``core'' in Fig.~\ref{fig:jacobi} is now the proton and the 
``protons'' in that figure are the $\alpha$ particles. The quantity $E_x$ is now the relative kinetic energy in the 
\textit{p}-$\alpha$ subsystem and, for each event, the spectra are incremented
 twice, once for each the two 
\textit{p}-$\alpha$ pairs. The curves are the R-matrix predictions using
 parameters for the 
$^5$Li$_{g.s.}$ resonance from Ref.~\cite{Woods88}. These sequential calculations reproduce the data 
quite well, further indicating a sequential character to these decays. The strong peak associated with 
the $^5$Li ground state is clearly resolved in Fig.~\ref{fig:Ex_9B_J}(a) for the 11.7~MeV state.

The discussion of these $^9$B decays has ignored interactions between the 
two decay steps. For the  11.7-MeV level, the short life time of the 
$^5$Li$_{g.s.}$ intermediate state is partially ameliorated by the large 
kinetic energy released in the first step ($E_k^{1}$=10~MeV). With this value 
we  obtain 
$R_{E}$= 0.12 and $d_{E}$=16~fm for this decay. The products from the first 
step thus are separated by $\sim$6 nuclear radii, on average, before
 the second steps occurs.  It is clear that in this case we would expect 
the sequential character of the decay to largely survive final-state 
interactions 
between the final fragments.   

The sequential character of the 2.345-MeV state is  more surprising. Here 
$E_k^1$ is much smaller and $R_{E}$=1.88 and $d_{E}$=4.0~fm and this should be 
a democratic decay. The value of 
$d_{E}$ is in fact one of the smallest values found in this work, 
even smaller than the value of
6.5~fm obtained for $^6$Be$_{g.s.}$ decay.

Why does this state have such 
a strong 
sequential character, when $^6$Be$_{g.s.}$ decay displays very little?
Specifically we find that the angular correlations associated 
with sequential decay 
are found for the 2.345-MeV $^9$B state, but not for the $^6$Be$_{g.s.}$. 
A possible reason is that the $^6$Be$_{g.s.}$ decay is approaching a 
Goldansky decay; most of the strength of the intermediate state is energetically 
inaccessible, whereas all the $^5$Li$_{g.s.}$ strength is accessible for the 
2.345-MeV state. 
Another consideration is that final state interactions between the products 
would be largest for $\cos(\theta_{2\alpha-p})$=$\pm$1, i.e., when one of 
the fragments from the second step is directed directly towards the $\alpha$ 
particle from the first step.  However, the angular corrections are minimal 
for these cases and so the effect of final state interactions will also be 
minimized. Although the correlations do maintain a strong sequential character,
the magnitude of the decay width cannot be explained from a R-matrix 
calculation (Sec.~\ref{sec:simul}). The R-matrix prediction by Barker 
was a factor of 4 too small \cite{Barker03a} indicating the need for a 
three-body calculation.

\begin{figure}[tbp]
\includegraphics*[ scale=.4]{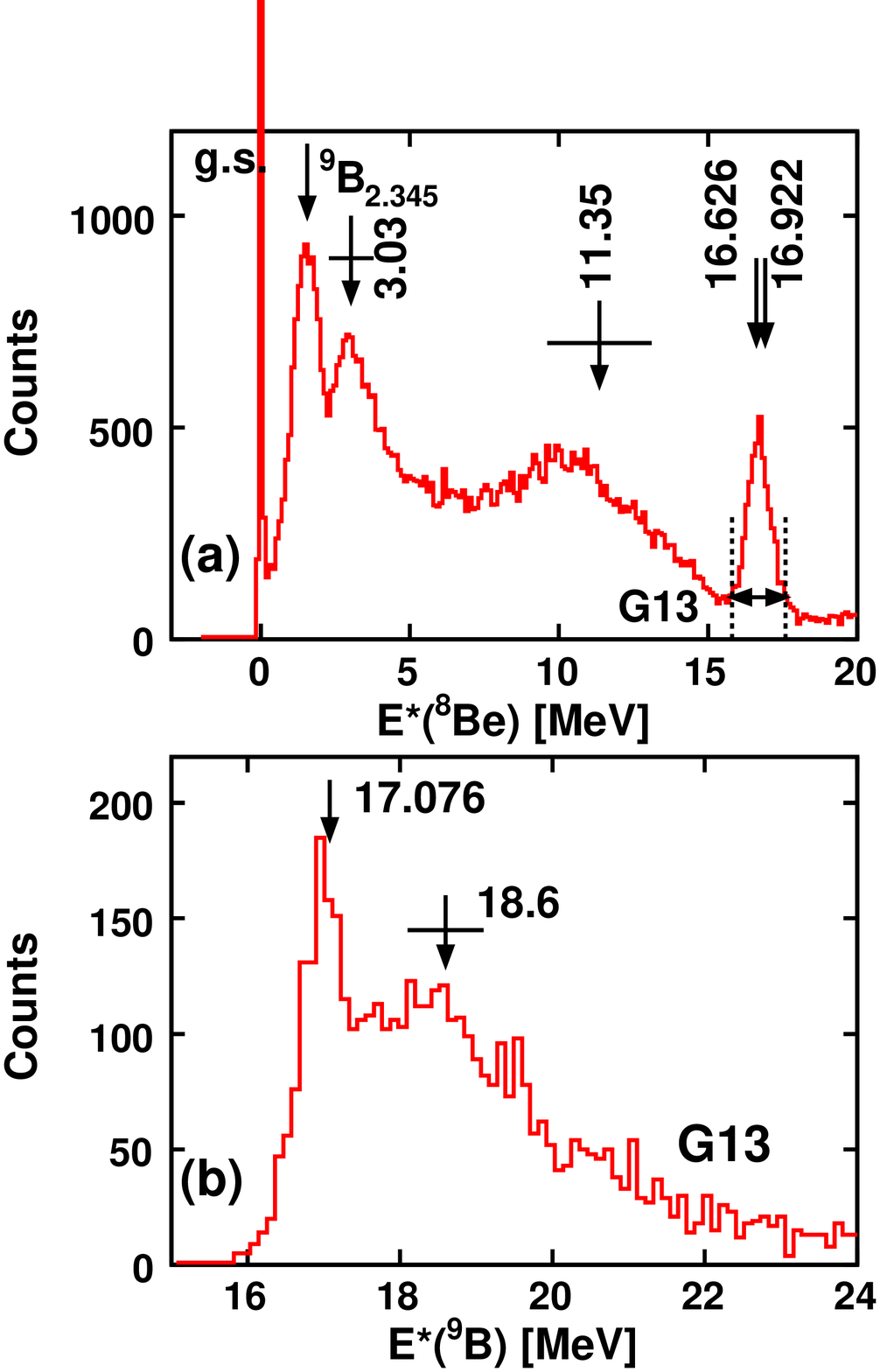}
\caption{(Color online) (a) Excitation-energy distribution for $^8$Be fragments reconstructed from the $\alpha$-$\alpha$ pairs of all \textit{p}+2$\alpha$ events.
(b) The $^{9}$B excitation-energy distributions obtained with the $G13$ gate on the $^{8}$Be excitation energy.}
\label{fig:Ex_9B_IAS}
\end{figure}

One can also look for $^9$B levels which decay predominantly to $^8$Be states. The 
reconstructed $^8$Be excitation-energy spectrum from the $\alpha$-$\alpha$ 
pairs of each \textit{p}+2$\alpha$ event is shown in 
Fig.~\ref{fig:Ex_9B_IAS}(a). The $^8$Be ground state and a number of excited states
indicated by the arrows are clearly present. The peak at $\sim$1.5 MeV is not 
a $^8$Be excited state, but rather is associated with the decay of the 
2.345-MeV $^{9}$B excited state. Of particular interest is the strong 
peak associated
with the 16.626 and 16.922-MeV mixed $T$=0+1 $J^\pi$=2$^+$ states. 
These are narrow states, 
but the experimental resolution is not sufficient for their separation. 
The peak thus can contain contributions from both of these mixed-isospin 
states.
 
The $^9$B excitation-energy spectrum obtained by gating of the 
$E^*\sim$17-MeV peak 
[gate $G13$ in Fig.~\ref{fig:Ex_9B_IAS}(a)] is shown in 
Fig.~\ref{fig:Ex_9B_IAS}(b). The peak at 16.99$\pm$0.04 MeV is clearly 
isolated although it was also present in the ungated spectrum of 
Fig.~\ref{fig:Ex_9B}(a). This peak could correspond to a previously known level 
at 16.71$\pm$0.10~MeV or the 17.076$\pm$0.04-MeV $T$=3/2 level. Both decays 
satisfy isospin conservation and the former, if it has a significant 
\textit{d}+$^7$Be branching ratio, may be important in big-bang 
nucleosynthesis \cite{Chakraborty10}.
The magnitude of the yield in the observed peak is consistent with
 that from the ungated spectrum in Fig.~\ref{fig:Ex_9B}(a)  
indicating that this state predominantly decays to 
one or both of the $^{8}$Be  $T$=0+1 levels. Presumably most of the decay is to
the 16.626-MeV state as only the low-energy tail of the 16.922~MeV 
state is energetically accessible.

The 16.71-MeV level has only been observed before in one experimental 
study \cite{Dixit91} 
where it was suggested to be the mirror of the 16.67~MeV $J^\pi$=5/2$^+$ 
state in $^9$Be. Experimental angular distributions in that work 
were consistent
with this spin assignment. The spin of the 17.076-MeV 
state is not listed in the ENSDF database, however, it is probably the 
mirror of the 16.977-MeV $T$=3/2 $^9$Be state which has a 
spin of  $J^\pi$=1/2$^{-}$. In the two cases we would thus expect the 
first step in the sequential decay would be dominated by $\ell$=0 and 
1 decay, respectively, giving rise to different 
angular correlations. 

\begin{figure}[tbp]
\includegraphics*[ scale=.4]{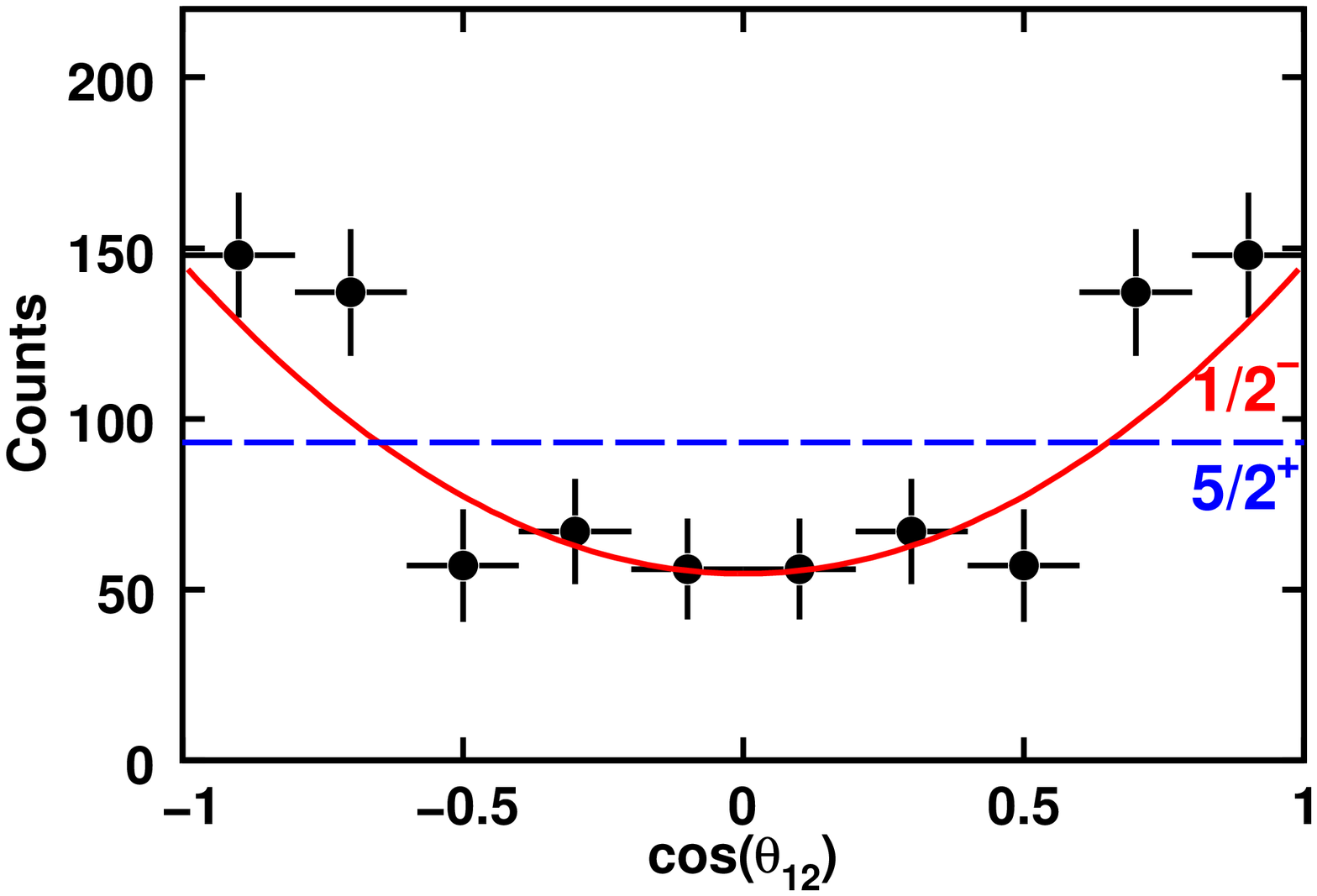}
\caption{(Color online) Angular correlation for the 16.99~MeV state in $^9$B.
Data points indicate the background-subtracted experimental results and the 
curves shows the results expected for the indicated level spins.
} 
\label{fig:angle9B17}
\end{figure}

We have extracted the angular 
correlation between 
the first proton emission step and the second 2$\alpha$ decay step for the 
events in the peak located at $\sim$17 MeV in Fig.~\ref{fig:Ex_9B_IAS}(b). 
The higher-energy region 
directly adjacent in excitation energy was used for the background subtraction.
The background-subtracted correlation is plotted in Fig.~\ref{fig:angle9B17}
and is compared to calculations for $J^\pi$=1/2$^{-}$ ($\ell$=1) and 
$J^\pi$=5/2$^+$ ($\ell$=0).  Of these calculations, the data are in much better
agreement with the  $J^\pi$=1/2$^{-}$ case, suggesting that the state observed
is the $T$=3/2 state.
The decay to the 16.626~MeV state is $R_{E}$=0.41 and $d_{E}$=45~fm, and the 
large value of the latter number suggests this is approaching a truly 
sequential decay.

\subsubsection{2\textit{p}+$^7$Li exit channel}

\begin{figure}[tbp]
\includegraphics*[ scale=.5]{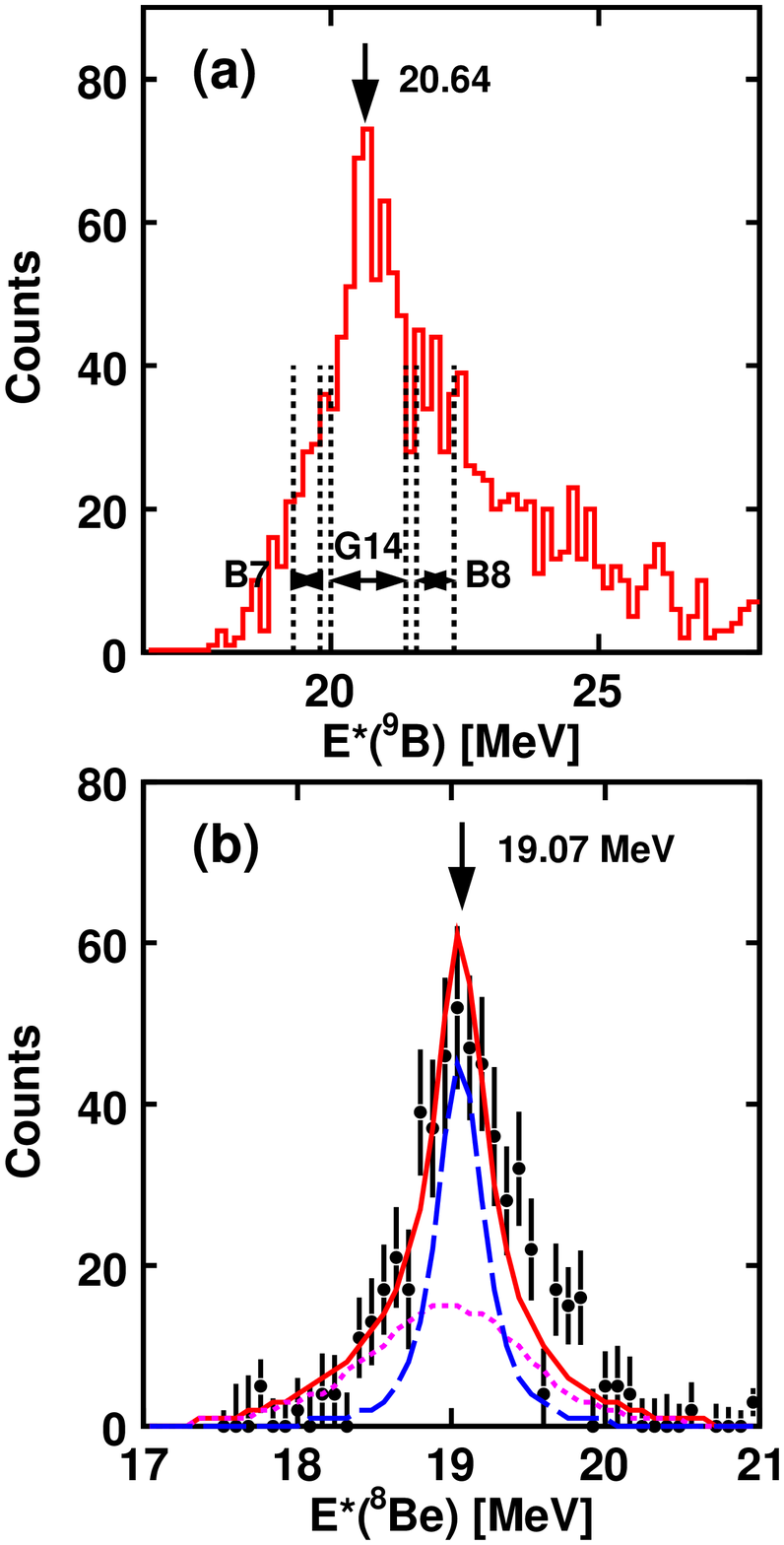}
\caption{(Color online) (a) Excitation-energy distribution for $^9$B fragments 
reconstructed from  2\textit{p}+$^{7}$Li events.
(b) The $^{8}$Be excitation-energy distributions obtained with the $G1$ 
gate on the $^{9}$Be excitation energy. The distributions contains contributions from both possible \textit{p}-$^7$Li combinations.}
\label{fig:Ex_9B20}
\end{figure}

Due to the problem with the saturation of the amplifiers, 
only a small fraction of the $^7$Li fragments could be identified.  
However, there were sufficient statistics for the 2\textit{p}+$^7$Li 
exit channel to isolate a new level in $^9$B.
The excitation-energy spectrum determined from 2\textit{p}+$^7$Li events is 
displayed in Fig.~\ref{fig:Ex_9B20}(a) and a peak at 20.64$\pm$0.10 MeV is evident with a width of 
$\Gamma$=450$\pm$250~keV. 
The large error is a consequence of an uncertainty of 
how to define 
the background when fitting this peak. We do not know whether the 
$^7$Li fragment was created in the ground state or the gamma-emitting 
0.477-MeV excited state. The energy of the level could thus be either 20.64 or 
21.12~MeV. We have attempted to see if the decay of this level could be 
associated with a sequential two-proton decay through a $^8$Be excited state. 
As there are two protons, we can reconstruct the two possible $^8$Be excitation 
energies from each 2\textit{p}+$^7$Li event. The background-subtracted 
distribution of both of these 
gated on the 20.64-MeV peak [gate G14 in Fig.~\ref{fig:Ex_9B20}(a) 
with background gates $B7$ and $B8$] is shown in Fig.~\ref{fig:Ex_9B20}(b).
Again there is some uncertainty in the background subtraction, but the location 
of the peak was not affected by this. The peak can be identified with the
19.069~MeV, $\Gamma$=271~keV, $J^\pi$=3$^+$ in $^8$Be. The curves in 
Fig.~\ref{fig:Ex_9B20}(b) show Monte Carlo simulations of the decay through this state.
The long and short-dashed curves show the contributions from the correctly and 
incorrectly identified proton associated with $^8$Be decay, and the solid curve 
is the sum of these two. Both the correct and incorrect distributions have almost the same 
average energy, but the incorrect distribution is quite wide and the peak position 
is defined by the correctly identified proton which has a much narrower distribution. 
The simulation was performed 
assuming the second decay is to the ground state of $^7$Li. If we assumed the decay 
was to 0.477-MeV state, then the reconstructed $^8$Be peak energy would be at $\sim$19.5 
MeV. No previously known $^8$Be excited state could be found at this energy but it is 
possible that decays occurs through an unknown $^8$Be excited state, however it seems 
more likely that the decay is to the $^7$Li ground state through the known 19.07-MeV state.

\subsection{$^{10}$C levels}
\begin{figure}[tbp]
\includegraphics*[ scale=.5]{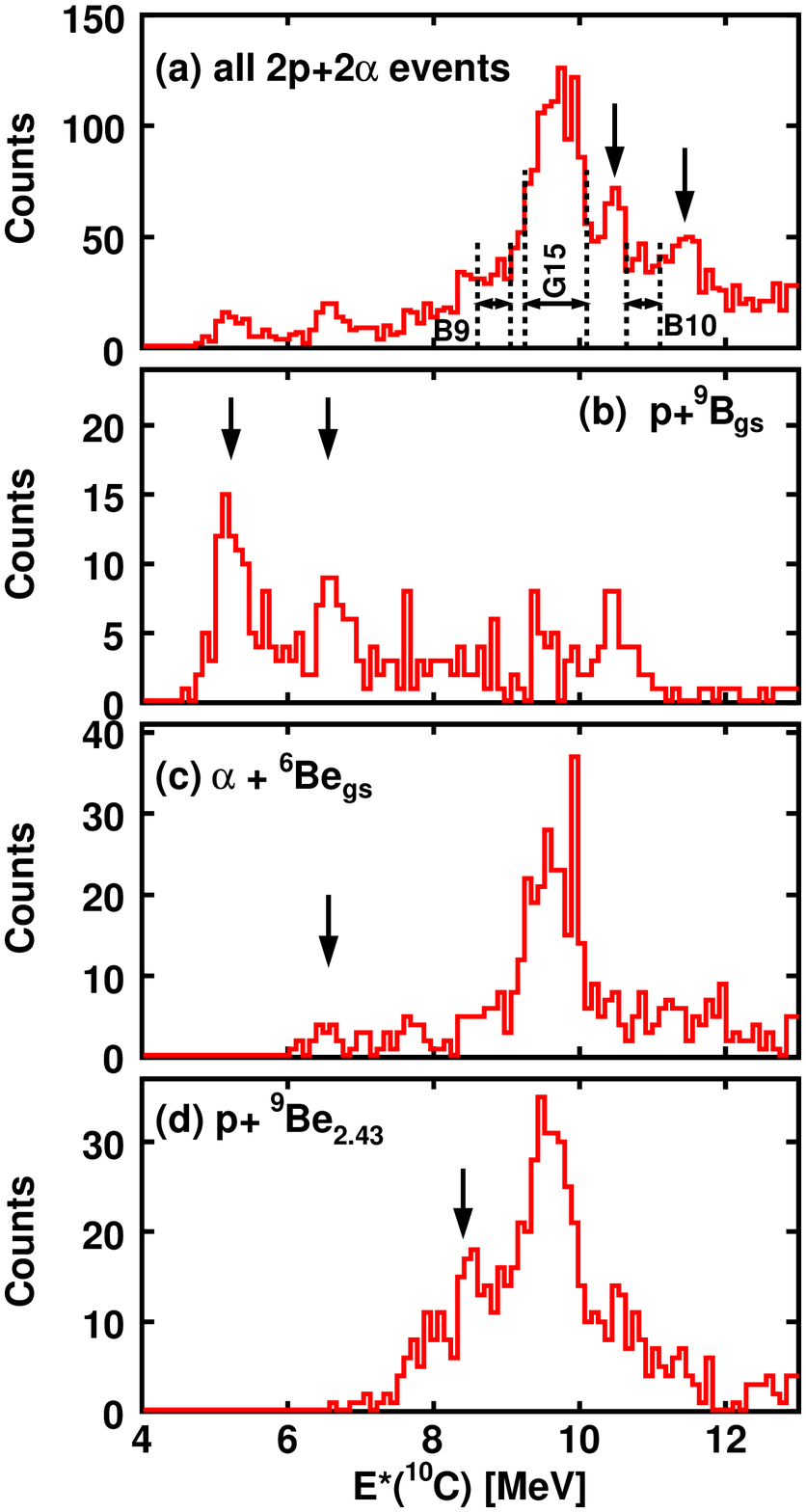}
\caption{(Color online)  Excitation-energy distributions for $^{10}$C 
fragments 
reconstructed from  2\textit{p}+2$\alpha$  events.
Panel (a) is for all events, while (b), (c) and (d) are gated on the presence 
of  $^9$B$_{g.s.}$, $^6$Be$_{g.s.}$, and $^9$B$_{2.345}$ intermediate states.
Arrows show the locations of the peaks discussed in this work. 
The gate $G15$ used to select the 9.69-MeV peak and the 
gates $B9$ and $B10$ used for background subtraction are indicted. } 
\label{fig:Ex_10C}
\end{figure}

 Information on $^{10}$C states produced by neutron pickup and more 
complicated reactions was obtained from the 
2\textit{p}+2$\alpha$ exit channel. This is the only channel available for 
particle-decaying states with $E^*<$ 15 MeV. The excitation-energy spectrum 
from all of these events is displayed in Fig.~\ref{fig:Ex_10C}(a) 
and  is dominated 
by a peak at 9.69 MeV, although other smaller peaks are evident. 
In our previous work, levels for this exit channel 
were revealed by gating on the presence of various narrow 
intermediate states \cite{Charity09}. Figures~\ref{fig:Ex_10C}(b), \ref{fig:Ex_10C}(c), and 
\ref{fig:Ex_10C}(d) show the $^{10}$C excitation-energy distribution gated on 
the $^9$B ground state, the $^6$Be ground state, and the 2.345-MeV $^9$B 
excited state, respectively. 

The arrows in Fig.~\ref{fig:Ex_10C}(b) 
show the locations of two states (5.222 and 6.553 MeV) that were excited via 
the inelastic scattering of a $^{10}$C beam and which decay 
by proton emission to the $^9$B ground state  \cite{Charity09}. 
These states dominated the total $^{10}$C excitation 
energy spectrum in inelastic scattering, 
but here they are weakly populated
and require a gate by the $^9$B$_{g.s.}$ to enhance them 
[see Fig.~\ref{fig:Ex_10C}(b)]. As 
these states are 
expected to have 2$\alpha$ cluster structure \cite{Charity09}, 
then it is not surprising that a neutron-pickup by 
$^9$C (which does not have any 2$\alpha$ cluster structure) 
does not populate these states strongly.

The widths of these states also appear to be wider (by 50-70\%) 
than we would expect based on the intrinsic widths given in 
Ref.~\cite{Charity09} and the simulated experimental resolution. 
Possibly these peaks are actually 
doublets which would not be surprising as other levels near these energies 
are expected based on the mirror nucleus $^9$Be.  At higher excitation
 energies in Fig.~\ref{fig:Ex_10C}(b), there may also be indications 
of other states, though the statistical errors are large. We note the absence 
of any significant contribution from the 9.69-MeV level in this spectrum.
 
A 6.56-MeV state which decayed via the $\alpha$+$^6$Be$_{g.s.}$ channel 
was also observed in the inelastic excitation of a $^{10}$C beam 
\cite{Curtis08,*Curtis10,Charity09}. Its location is indicated by 
the arrow in Fig.~\ref{fig:Ex_10C}(c). This state is also expected to 
have strong cluster structure and is clearly only weakly populated 
in the present experiment.  In Fig.~\ref{fig:Ex_10C}(c), the spectrum is 
again dominated by the 9.69-MeV peak, indicating that $\alpha$+$^6$Be$_{g.s.}$ 
is an important decay branch for this level.

The spectrum in Fig.~\ref{fig:Ex_10C}(d) is also dominated by the 
9.69-MeV level indicating that this level has a second decay branch 
(\textit{p}+$^9$B$_{2.345}$). Also a smaller peak appears at 
8.5~MeV. The arrow in this figure shows the location of an 8.4~MeV state that
 was previously shown to decay through the 2.345-MeV $^9$B intermediate level.
However, the intrinsic width of that state was $\sim$1~MeV, significantly 
wider than the present peak. 
The experimental width of the peak in this work ($\sim$300 keV) 
is consistent with the simulated resolution. Within the statistical uncertainties we estimate 
$\Gamma<$200~keV.

Other small peaks are also suggested in the total distribution in 
Fig.~\ref{fig:Ex_10C}(a). The peaks at 10.48 and 11.44 MeV, indicated by the 
arrows in this figure, are previously unknown. Their widths are 
consistent with the experimental resolution. The decay of these states is not known; 
they do not feature prominently in the distributions gated by the long-live 
intermediates.

\subsubsection{9.69-MeV state}
The strong 
9.69-MeV state is wider than the other observed $^{10}$C states; from the simulation we estimate its intrinsic width is 490~keV.
It was not observed via $^{10}$C inelastic scattering in 
Refs.~\cite{Curtis08,*Curtis10,Charity09} suggesting that it does not
 have a strong cluster structure and maybe is more single-particle in nature.
In any case we will show its decay modes are quite interesting. 
In the following we give an analysis of correlations associated with this peak 
using gate $G15$ and background  gates $B9$ and $B10$ in 
Fig.~\ref{fig:Ex_10C}(a). All the  distributions discussed in the remainding of this section are background subtracted.

\begin{figure}[tbp]
\includegraphics*[ scale=.4]{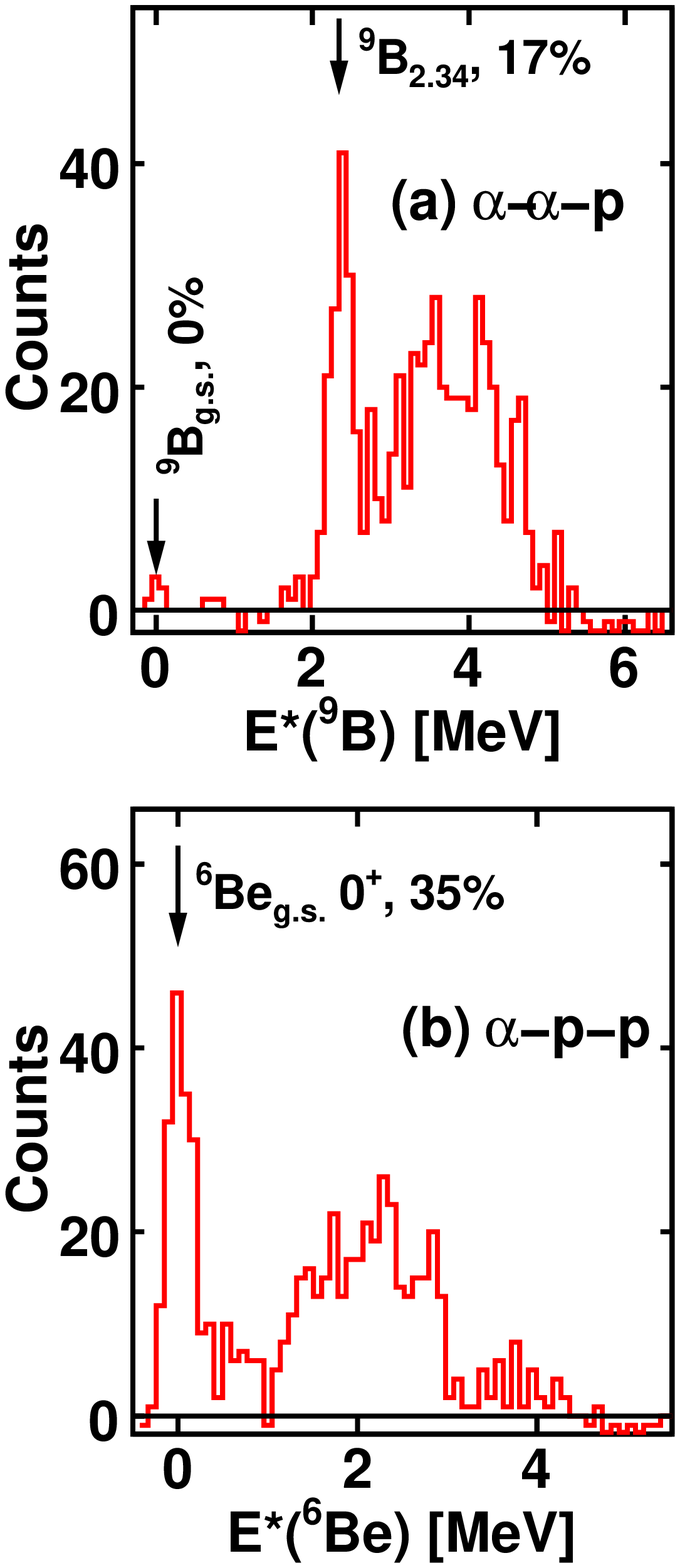}
\caption{(Color online) Background-subtracted excitation-energy 
distributions for (a) $^9$B and (b) 
$^6$Be fragments produced in the decay of the 9.69-MeV level of $^{10}$C 
determined from \textit{p}+2$\alpha$ and 2\textit{p}+$\alpha$ triplets.}
\label{fig:c10_be6b9}
\end{figure}

As we have observed in Fig.~\ref{fig:Ex_10C}(c) and \ref{fig:Ex_10C}(d), 
the 9.69~MeV level has strength for decays into the $\alpha$+$^6$Be$_{g.s.}$ 
and \textit{p}+$^9$B$_{2.345}$ channels. 
Estimates of the branching ratios for these two channels are based on the
 $^6$Be and $^9$B excitation-energy spectra gated on the 
9.69-MeV state displayed in 
Fig.~\ref{fig:c10_be6b9}. One can clearly see peaks 
associated with $^6$Be ground state [Fig.~\ref{fig:c10_be6b9}(b)] 
and the 2.345-MeV state [Fig.~\ref{fig:c10_be6b9}(a)] in $^9$B. The yield associated with the ground state of $^9$B 
is clearly minimal. The yield in the $^6$Be and $^9$B$_{2.345}$ peaks 
were fit with Gaussians with smooth backgrounds. From these yields and correcting for the detector efficiency, we determine branching ratios of  35\% and 17\%  for $\alpha$+$^6$Be$_{g.s.}$ and \textit{p}+$^9$B$_{2.345}$ decay paths, respectively. Both of these decay branches 
are safely sequential with $d_{E}$=136 and 200~fm, respectively.
The determination of the decay mechanism of the remaining 
$\sim$50\% of the yield is more difficult and, to isolate this component, 
we have vetoed events associated
with the $^6$Be$_{g.s.}$ and $^9$B$_{2.435}$ peaks in the following analysis.

\begin{figure}[tbp]
\includegraphics*[ scale=.4]{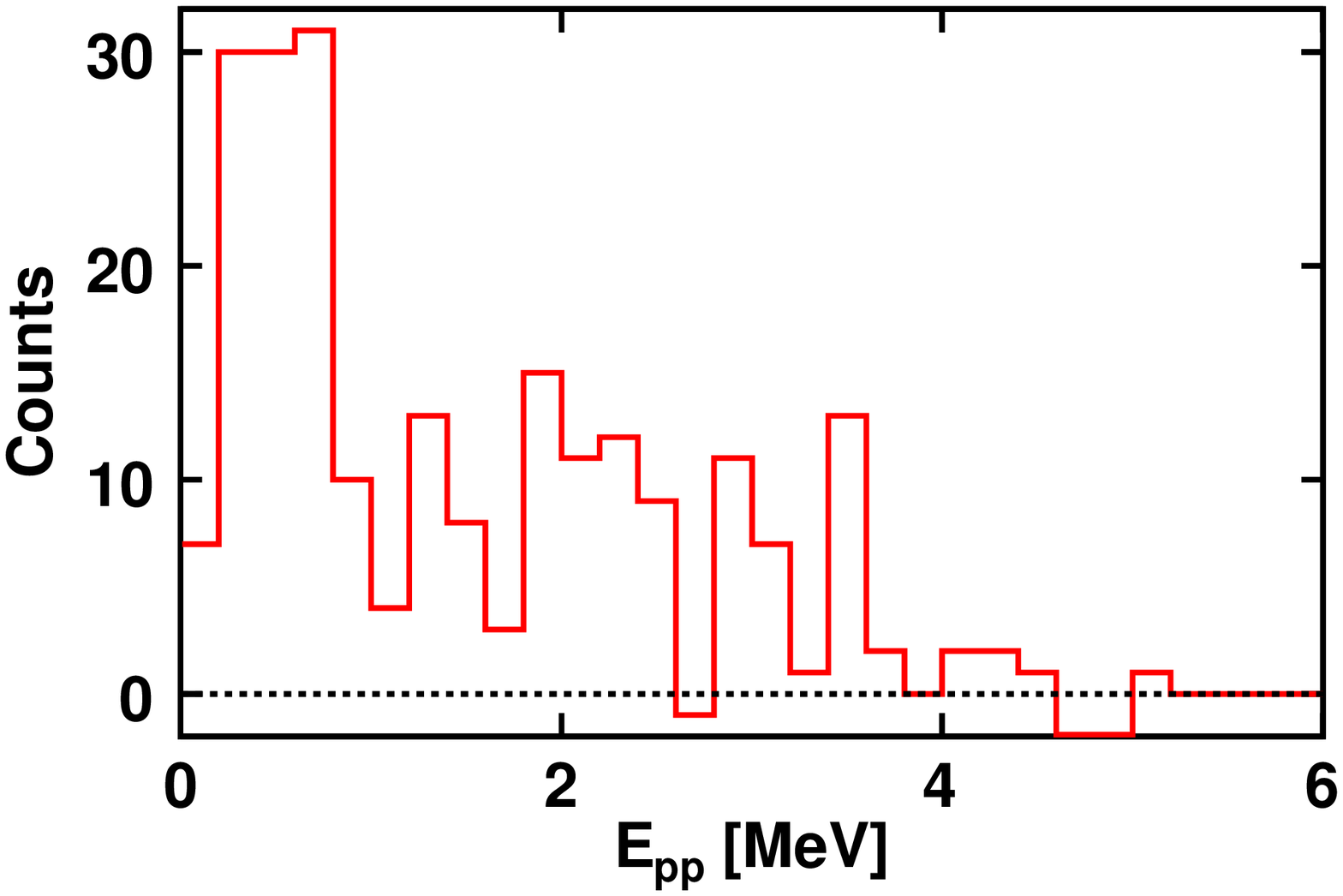}
\caption{(Color online) Distribution of relative kinetic energy
 between the two 
protons for the 9.69-MeV level in $^{10}$C.}
\label{fig:c10_pp}
\end{figure}

\begin{figure}[tbp]
\includegraphics*[ scale=.4]{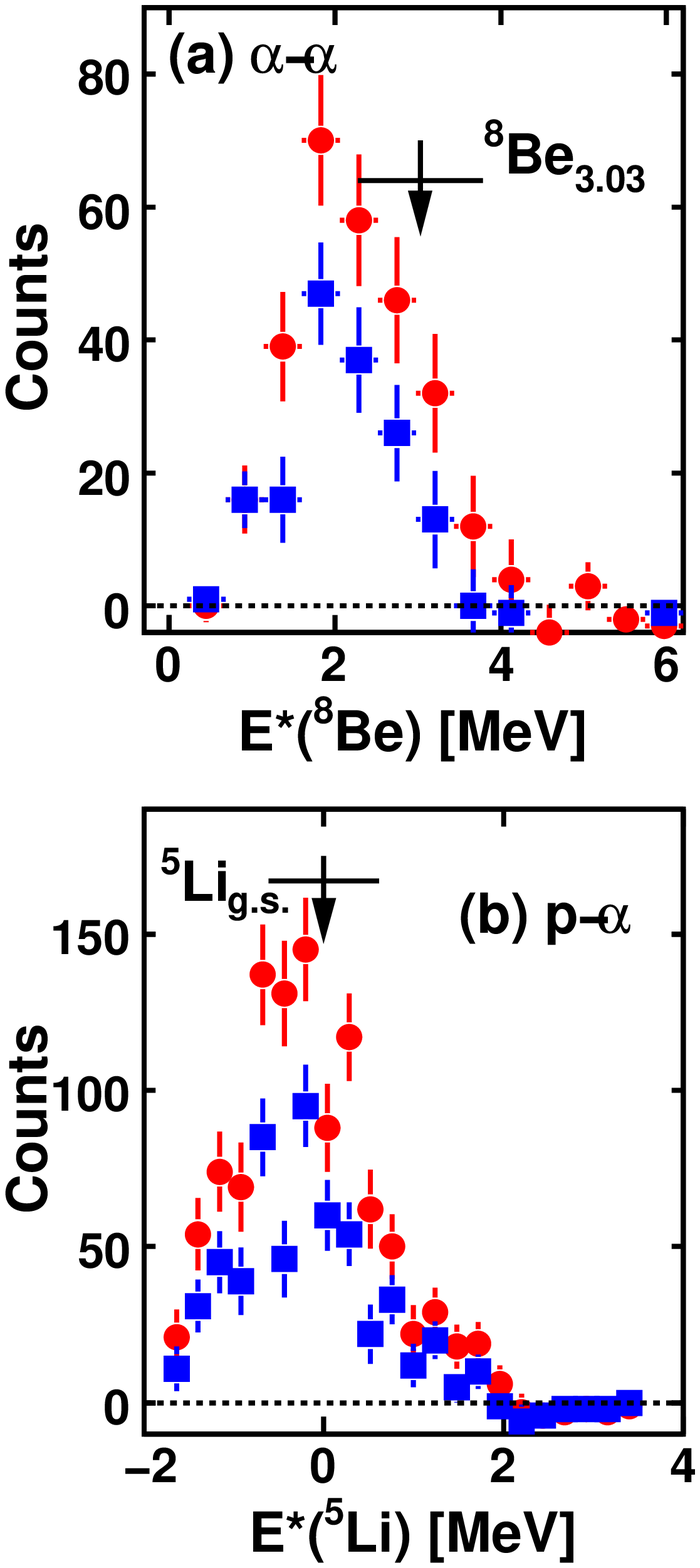}
\caption{(Color online) Background-subtracted excitation-energy distributions
 for (a) $^8$Be and (b) 
$^5$Li fragments produced in the decay of the 9.69-MeV level of $^{10}$C 
determined from $\alpha$-$\alpha$ and \textit{p}-$\alpha$ pairs.
For the latter case, the distribution contains all four possible 
\textit{p}-$\alpha$
pairs associated with each detected 2\textit{p}+2$\alpha$ event.
The distributions indicated by the circular data points have 
no restriction on the \textit{p}-\textit{p} relative energy, 
while for the squared-shaped data points $E_{pp} < $ 1.0 MeV.
The arrows with error bars show the mean energy and width of the 
indicated resonances.
} 
\label{fig:c10_be8li5}
\end{figure}

The distribution of $E_{pp}$, the relative kinetic energy between the 
two protons, is plotted in Fig.~\ref{fig:c10_pp}. This distribution shows 
a strong enhancement at small relative energies ($E_{pp} <$1.0 MeV), 
the diproton region, but also displays a tail to significantly higher values.
Figures \ref{fig:c10_be8li5}(a) and \ref{fig:c10_be8li5}(b) show the 
distributions of $^8$Be and $^5$Li excitation energy determined from the 
detected $\alpha$-$\alpha$ pairs and all 4 combinations of 
\textit{p}-$\alpha$ pairs. The total distributions (circular data points)
 have strong overlaps with the wide 3.03-MeV $^8$Be and the $^5$Li ground-state 
resonances, respectively. The distributions are rather insensitive to 
whether the events 
are associated with the diproton peak in Fig.~\ref{fig:c10_pp} or associated 
with the higher-energy tail. To illustrate this, the squared data points in 
Fig.~\ref{fig:c10_be8li5}(a) and \ref{fig:c10_be8li5}(b) show the 
distributions gated on the diproton peak ($E_{pp}<$1.0 MeV). 
Within the statistical uncertainties, they are the same shape as 
the ungated distributions.

\begin{figure}[tbp]
\includegraphics*[ scale=.4]{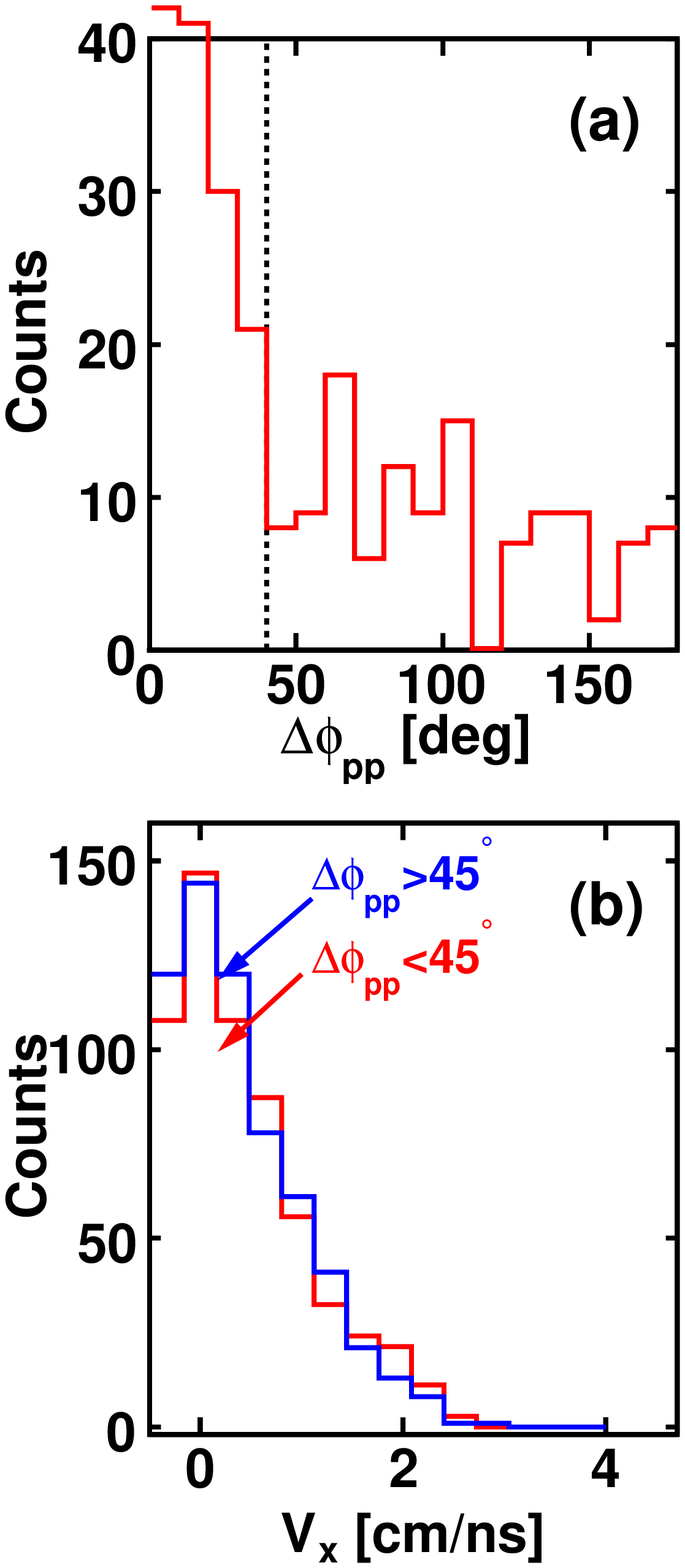}
\caption{(Color online) Background-subtracted distributions associated with the
decay of the 9.69-MeV state in $^{10}$C. (a) Distribution of the angle 
$\Delta\phi_{pp}$. This is the angle between the two planes in velocity space
defined by the two $\alpha$ particles and each of the protons.
(b) For the two indicated gates on $\Delta\phi_{pp}$, the distributions of the 
\textit{x} component of the proton velocity (see text). 
For comparison purposes, the results for $\Delta\phi_{pp}>40^{\circ}$ 
have been normalized to the same number of counts as in the 
$\Delta\phi_{pp}<40^{\circ}$ distribution.} 
\label{fig:c10_deltaPhi}
\end{figure}

The strong overlap with the $^5$Li$_{g.s.}$ resonance 
may initially suggest that the
$^{10}$C fragment decays via the $^5$Li$_{g.s.}$+$^{5}$Li$_{g.s.}$ channel. 
However in such a case, the excitation energy reconstructed from an 
$\alpha$ particle and a proton originating from 
different $^5$Li fragments should 
populate the $^5$Li excitation energies significantly higher than 
$^5$Li$_{g.s.}$ resonance region. However, 
the experimental data indicated that all four
 $\alpha$-\textit{p} pairs are consistent with the $^5$Li$_{g.s.}$ resonance.
The decay thus appears to be four-body in nature and cannot be 
described by a sequential process. 
A possible interpretation is that all four \textit{p}-$\alpha$ pairs are 
simultaneously in $^5$Li$_{g.s.}$ resonances and some fraction of the 
\textit{p}-\textit{p} pairs are tightly correlated in a ''diproton''. 
Possibly, the 
$\alpha$-$\alpha$ pair may also be in a 3.03-MeV $^8$Be resonance. 
On the other 
hand, the overlap with the $^5$Li$_{g.s.}$ and other resonances 
may just be coincidental. If all of these associations with resonances are correct,
it is not clear that a configuration exists which would satisfy all of the implied angular 
momentum couplings.
To display the correlations graphically, we note that each \textit{p}+2$\alpha$
triplet in each 2\textit{p}+2$\alpha$ event defines a plane in velocity space. 
With the two protons, we define two such planes in each event. The distribution 
of $\Delta\phi_{pp}$, the angle between these two planes is plotted in 
Fig.~\ref{fig:c10_deltaPhi}(a). This distribution resembles the 
relative-proton-energy distribution in Fig.~\ref{fig:c10_pp} with a diproton region and a high-energy tail, and in fact the quantities $E_{pp}$ and 
$\Delta\phi_{pp}$ are strongly correlated. 

\begin{figure}[tbp]
\includegraphics*[ scale=.4]{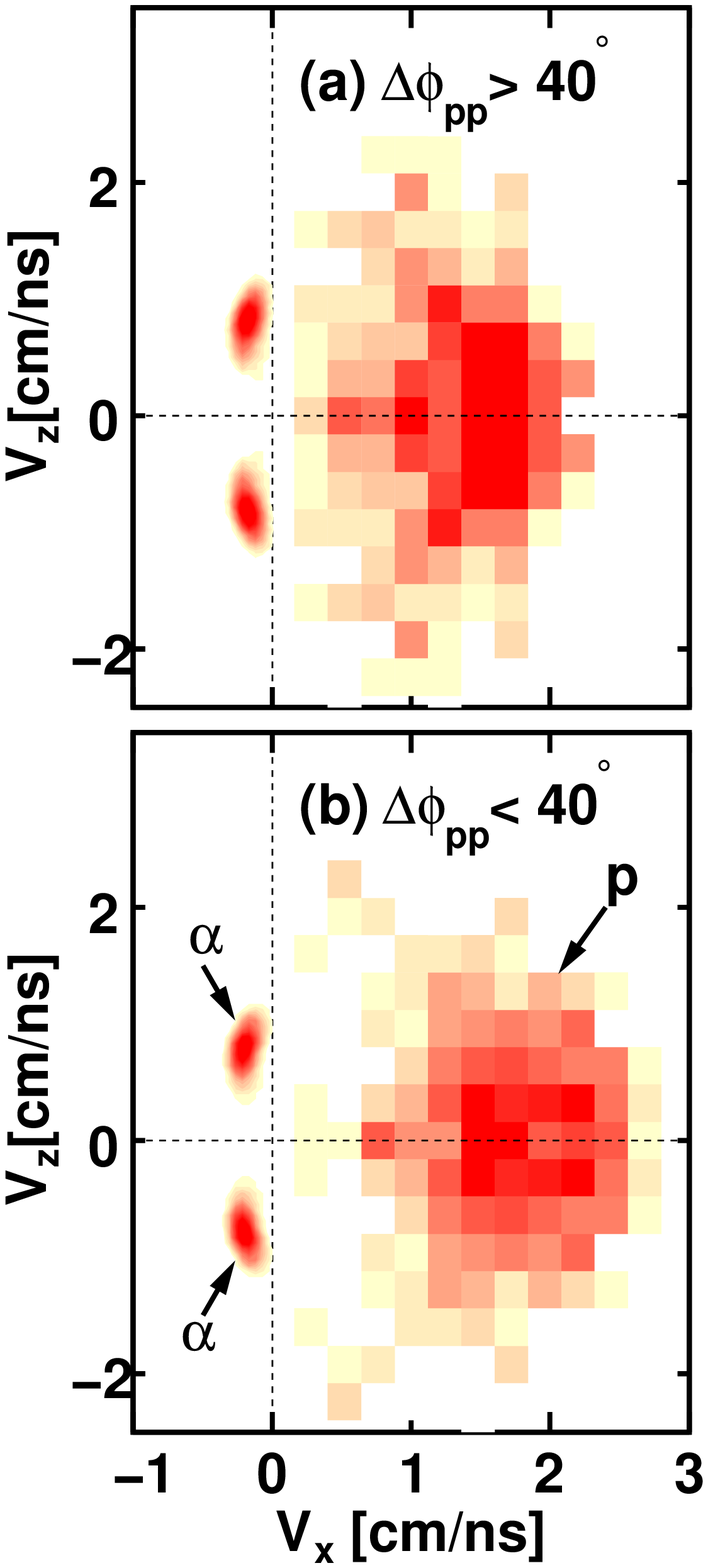}
\caption{(Color online) Velocity distributions of $\alpha$ particles and 
protons produced in the decay of the 9.69-MeV state in $^{10}$C.
For each 2$\alpha$+\textit{p} triplet in each detected 2$\alpha$+2\textit{p}
event, the velocities have been projected onto the plane of the 
three fragments 
and rotated in this plane to align the locations of the $\alpha$ particles.
The results are shown for (a) $\Delta\phi_{pp}<40^{\circ}$ 
and (b) $\Delta\phi_{pp}>40^{\circ}$ where the $\Delta\phi_{pp}$ is the 
angle between the two planes defined by the two protons.}     
\label{fig:c10_plane}
\end{figure}

In order to help visualize the correlations between each proton 
and the two alpha particles, we have employed the procedure
used to generate Fig.~\ref{fig:b8map}.
For each \textit{p}+2$\alpha$ triplet, the $\alpha$-particle and proton
 velocity 
vectors are projected into their decay plane. Subsequently they are 
rotated in this plane to align
the locations of the $\alpha$ particles. The 
2-dimensional velocity distributions are plotted in Fig.~\ref{fig:c10_plane}  
for gates on the diproton region ($\Delta\phi_{pp}<40^{\circ}$) 
and the high-energy tail ($\Delta\phi_{pp}>40^{\circ}$).

Both distributions are quite similar and thus 
again there is little difference in the 
\textit{p}-$\alpha$ correlations for the diproton and high-energy tail 
regions of the \textit{p}-\textit{p} relative energy. This is further 
illustrated in Fig.~\ref{fig:c10_deltaPhi}(b) where the projected $V_{x}$ 
distributions for these two regions are compared. 
Again we see very little difference in the results for the two gates.
The average location of the  two $\alpha$ particles and the proton form 
an isosceles triangle in their defining plane; i.e., 
the two \textit{p}-$\alpha$ average 
relative velocities are identical and thus all \textit{p}-$\alpha$ pairs 
can have the same reconstructed $^5$Li excitation energy.

\section{DISCUSSION AND CONCLUSIONS}
\label{conclusion}
This work highlights the wide range of many-particle decays of ground  
and excited states in light nuclei. It is not clear that there is a 
simple way of the predicting when the decays should be sequential or prompt.
Certainly the presence of possible narrow intermediate states does not guarantee
that the system will take advantage of them. For example 
the 9.69-MeV state in $^{10}$C has $^6$Be$_{g.s.}$ and $^9$B$_{2.34}$ 
intermediate states with $\Gamma\sim$90~keV; yet 50\% of the time it 
bypasses these
and undergoes a 4-body decay. The 6.56~MeV state $^{10}$C also has competing 
prompt and sequential decay branches \cite{Charity09}. 
The nature of the decay is certainly 
related to the structure of the state and, in principle with suitable 
models, this information can be used to improve structure calculations. 
Presently there 
are only a few cases where prompt 3-body decay has been modeled in a 
sophisticated manner, as for example the quantum-mechanical cluster 
calculation \cite{Grigorenko09}, where comparisons to experimental 
correlations are useful. The prompt 4-body decay poses an even more difficult 
theoretical problem.

In this work we have studied many-particle exit channels associated with states
formed following the interactions of an $E/A$=70~MeV $^9$C beam with a $^9$Be 
target. The decay products were detected in the HiRA array and the states 
were identified from the invariant mass of the detected fragments. 
Correlations between these fragments were used to deduce the nature of the 
decay and search for long-lived intermediate states associated with 
sequential decay processes. For such decays, angular correlations between the 
fragments provide information on the spin of the level.

The ground state of $^8$C disintegrates into four protons and an 
$\alpha$-particle. This decay was found to proceed through the 
$^6$Be$_{g.s.}$ 
intermediate state, with two steps of 2-proton decay. 
The correlations between the protons in the first step exhibit a 
significant enhancement in a region of phase space  where the two 
protons have small relative energy. The correlations in the second step 
were found to be consistent with $^6$Be$_{g.s.}$ 2-proton decay obtained from 
 an independent measurement using a  $^7$Be beam.

Evidence suggests that
the isobaric analog of this state in $^8$B,  undergoes 2-proton decay to the 
isobaric analog of $^6$Be$_{g.s.}$  in $^6$Li. This is the first case of 
two-proton decay between two isobaric analog states.
The exact nature of the 2-proton decay 
awaits future measurements, but it may constitute a Goldansky-type decay with 
a single proton emission being energetically allowed but isospin forbidden.

A 9.69-MeV state of $^{10}$C was found, which $\sim$50\% of the time, 
undergoes a prompt 
4-body decay into the 
2\textit{p}+2$\alpha$ channel. There is a strong diproton-like correlation 
between the two protons, at the same time, all four possible 
\textit{p}-$\alpha$ pairs have energies consistent with a $^5$Li$_{g.s.}$ resonance.  
No sequential decay scenario is possible to produce 4 simultaneous 
$^5$Li$_{g.s.}$ resonances.

Evidence for a prompt 3-body decay of the 8.15-MeV state of $^8$B was found.
On the other hand, a number of states with three-body exit channels 
were observed that are well described as sequential decays through 
intermediate states. These include the isobaric pair, $^8$B$_{5.93}$
and $^8$Be$_{22.9}$, and the 11.7, 16.9 and 20.6~MeV states of  $^9$B.
The correlations in the decay of the 2.345-MeV state of $^9$B to 
the \textit{p} + 2$\alpha$ channel were found to have strong sequential 
character associated with a $^5$Li intermediate, even though the lifetime
of this intermediate state is very short and, at the very least, we would 
expect there to be very important final-state interactions.

Finally, new mass excesses and decay widths were obtained for the ground 
states of $^7$B and $^8$C.

\acknowledgments
 This work
was supported by the U.S. Department of Energy, Division of Nuclear Physics
under grants DE-FG02-87ER-40316 and DE-FG02-04ER41320 and the National
Science Foundation under grants PHY-0606007 and PHY-9977707.


\begin{thebibliography}{10}%
\makeatletter
\providecommand \@ifxundefined [1]{%
 \ifx #1\undefined \expandafter \@firstoftwo
 \else \expandafter \@secondoftwo
\fi
}%
\providecommand \@ifnum [1]{%
 \ifnum #1\expandafter \@firstoftwo
 \else \expandafter \@secondoftwo
\fi
}%
\providecommand \enquote [1]{``#1''}%
\providecommand \bibnamefont  [1]{#1}%
\providecommand \bibfnamefont [1]{#1}%
\providecommand \citenamefont [1]{#1}%
\providecommand\href[0]{\@sanitize\@href}%
\providecommand\@href[1]{\endgroup\@@startlink{#1}\endgroup\@@href}%
\providecommand\@@href[1]{#1\@@endlink}%
\providecommand \@sanitize [0]{\begingroup\catcode`\&12\catcode`\#12\relax}%
\@ifxundefined \pdfoutput {\@firstoftwo}{%
 \@ifnum{\z@=\pdfoutput}{\@firstoftwo}{\@secondoftwo}%
}{%
 \providecommand\@@startlink[1]{\leavevmode}%
 \providecommand\@@endlink[0]{}%
}{%
 \providecommand\@@startlink[1]{%
  \leavevmode
  \pdfstartlink
   attr{/Border[0 0 1 ]/H/I/C[0 1 1]}%
   user{/Subtype/Link/A<</Type/Action/S/URI/URI(#1)>>}%
  \relax
 }%
 \providecommand\@@endlink[0]{\pdfendlink}%
}%
\providecommand \url  [0]{\begingroup\@sanitize \@url }%
\providecommand \@url [1]{\endgroup\@href {#1}{\urlprefix}}%
\providecommand \urlprefix [0]{URL }%
\providecommand \Eprint[0]{\href }%
\@ifxundefined \urlstyle {%
  \providecommand \doi [1]{doi:\discretionary{}{}{}#1}%
}{%
  \providecommand \doi [0]{doi:\discretionary{}{}{}\begingroup
  \urlstyle{rm}\Url }%
}%
\providecommand \doibase [0]{http://dx.doi.org/}%
\providecommand \Doi[1]{\href{\doibase#1}}%
\providecommand \bibAnnote [3]{%
  \BibitemShut{#1}%
  \begin{quotation}\noindent
    \textsc{Key:}\ #2\\\textsc{Annotation:}\ #3%
  \end{quotation}%
}%
\providecommand \bibAnnoteFile [2]{%
  \IfFileExists{#2}{\bibAnnote {#1} {#2} {\input{#2}}}{}%
}%
\providecommand \typeout [0]{\immediate \write \m@ne }%
\providecommand \selectlanguage [0]{\@gobble}%
\providecommand \bibinfo [0]{\@secondoftwo}%
\providecommand \bibfield [0]{\@secondoftwo}%
\providecommand \translation [1]{[#1]}%
\providecommand \BibitemOpen[0]{}%
\providecommand \bibitemStop [0]{}%
\providecommand \bibitemNoStop [0]{.\EOS\space}%
\providecommand \EOS [0]{\spacefactor3000\relax}%
\providecommand \BibitemShut [1]{\csname bibitem#1\endcsname}%
\bibitem{Goldansky60}%
  \BibitemOpen
  \bibfield{author}{%
  \bibinfo {author} {\bibfnamefont{V.~I.}\ \bibnamefont{Goldansky}},\ }%
  \bibfield{journal}{%
  \bibinfo {journal} {Nuclear Physics}\ }%
  \textbf{\bibinfo {volume} {19}},\ \bibinfo {pages} {482 } (\bibinfo {year}
  {1960})%
  \bibAnnoteFile{NoStop}{Goldansky60}%
\bibitem{Miernik07}%
  \BibitemOpen
  \bibfield{author}{%
  \bibinfo {author} {\bibfnamefont{K.}~\bibnamefont{Miernik}}, \bibinfo
  {author} {\bibfnamefont{W.}~\bibnamefont{Dominik}}, \bibinfo {author}
  {\bibfnamefont{Z.}~\bibnamefont{Janas}}, \bibinfo {author}
  {\bibfnamefont{M.}~\bibnamefont{Pf\"utzner}}, \bibinfo {author}
  {\bibfnamefont{L.}~\bibnamefont{Grigorenko}}, \bibinfo {author}
  {\bibfnamefont{C.~R.}\ \bibnamefont{Bingham}}, \bibinfo {author}
  {\bibfnamefont{H.}~\bibnamefont{Czyrkowski}}, \bibinfo {author}
  {\bibfnamefont{M.}~\bibnamefont{Cwiok}}, \bibinfo {author}
  {\bibfnamefont{I.~G.}\ \bibnamefont{Darby}}, \bibinfo {author}
  {\bibfnamefont{R.}~\bibnamefont{Dkabrowski}}, \bibinfo {author}
  {\bibfnamefont{T.}~\bibnamefont{Ginter}}, \bibinfo {author}
  {\bibfnamefont{R.}~\bibnamefont{Grzywacz}}, \bibinfo {author}
  {\bibfnamefont{M.}~\bibnamefont{Karny}}, \bibinfo {author}
  {\bibfnamefont{A.}~\bibnamefont{Korgul}}, \bibinfo {author}
  {\bibfnamefont{W.}~\bibnamefont{Ku\ifmmode~\acute{s}\else \'{s}\fi{}mierz}},
  \bibinfo {author} {\bibfnamefont{S.~N.}\ \bibnamefont{Liddick}}, \bibinfo
  {author} {\bibfnamefont{M.}~\bibnamefont{Rajabali}}, \bibinfo {author}
  {\bibfnamefont{K.}~\bibnamefont{Rykaczewski}},\ and\ \bibinfo {author}
  {\bibfnamefont{A.}~\bibnamefont{Stolz}},\ }%
  \bibfield{journal}{%
  \Doi{10.1103/PhysRevLett.99.192501}{\bibinfo {journal} {Phys. Rev. Lett.}}\
  }%
  \textbf{\bibinfo {volume} {99}},\ \bibinfo {pages} {192501} (\bibinfo {year}
  {2007})%
  \bibAnnoteFile{NoStop}{Miernik07}%
\bibitem{Blank05}%
  \BibitemOpen
  \bibfield{author}{%
  \bibinfo {author} {\bibfnamefont{B.}~\bibnamefont{Blank}}, \bibinfo {author}
  {\bibfnamefont{A.}~\bibnamefont{Bey}}, \bibinfo {author}
  {\bibfnamefont{G.}~\bibnamefont{Canchel}}, \bibinfo {author}
  {\bibfnamefont{C.}~\bibnamefont{Dossat}}, \bibinfo {author}
  {\bibfnamefont{A.}~\bibnamefont{Fleury}}, \bibinfo {author}
  {\bibfnamefont{J.}~\bibnamefont{Giovinazzo}}, \bibinfo {author}
  {\bibfnamefont{I.}~\bibnamefont{Matea}}, \bibinfo {author}
  {\bibfnamefont{N.}~\bibnamefont{Adimi}}, \bibinfo {author}
  {\bibfnamefont{F.}~\bibnamefont{De~Oliveira}}, \bibinfo {author}
  {\bibfnamefont{I.}~\bibnamefont{Stefan}}, \bibinfo {author}
  {\bibfnamefont{G.}~\bibnamefont{Georgiev}}, \bibinfo {author}
  {\bibfnamefont{S.}~\bibnamefont{Gr\'evy}}, \bibinfo {author}
  {\bibfnamefont{J.~C.}\ \bibnamefont{Thomas}}, \bibinfo {author}
  {\bibfnamefont{C.}~\bibnamefont{Borcea}}, \bibinfo {author}
  {\bibfnamefont{D.}~\bibnamefont{Cortina}}, \bibinfo {author}
  {\bibfnamefont{M.}~\bibnamefont{Caamano}}, \bibinfo {author}
  {\bibfnamefont{M.}~\bibnamefont{Stanoiu}}, \bibinfo {author}
  {\bibfnamefont{F.}~\bibnamefont{Aksouh}}, \bibinfo {author}
  {\bibfnamefont{B.~A.}\ \bibnamefont{Brown}}, \bibinfo {author}
  {\bibfnamefont{F.~C.}\ \bibnamefont{Barker}},\ and\ \bibinfo {author}
  {\bibfnamefont{W.~A.}\ \bibnamefont{Richter}},\ }%
  \bibfield{journal}{%
  \Doi{10.1103/PhysRevLett.94.232501}{\bibinfo {journal} {Phys. Rev. Lett.}}\
  }%
  \textbf{\bibinfo {volume} {94}},\ \bibinfo {pages} {232501} (\bibinfo {year}
  {2005})%
  \bibAnnoteFile{NoStop}{Blank05}%
\bibitem{Geesaman77}%
  \BibitemOpen
  \bibfield{author}{%
  \bibinfo {author} {\bibfnamefont{D.~F.}\ \bibnamefont{Geesaman}}, \bibinfo
  {author} {\bibfnamefont{R.~L.}\ \bibnamefont{McGrath}}, \bibinfo {author}
  {\bibfnamefont{P.~M.~S.}\ \bibnamefont{Lesser}}, \bibinfo {author}
  {\bibfnamefont{P.~P.}\ \bibnamefont{Urone}},\ and\ \bibinfo {author}
  {\bibfnamefont{B.}~\bibnamefont{VerWest}},\ }%
  \bibfield{journal}{%
  \Doi{10.1103/PhysRevC.15.1835}{\bibinfo {journal} {Phys. Rev. C}}\ }%
  \textbf{\bibinfo {volume} {15}},\ \bibinfo {pages} {1835} (\bibinfo {year}
  {1977})%
  \bibAnnoteFile{NoStop}{Geesaman77}%
\bibitem{Grigorenko09}%
  \BibitemOpen
  \bibfield{author}{%
  \bibinfo {author} {\bibfnamefont{L.~V.}\ \bibnamefont{Grigorenko}}, \bibinfo
  {author} {\bibfnamefont{T.~D.}\ \bibnamefont{Wiser}}, \bibinfo {author}
  {\bibfnamefont{K.}~\bibnamefont{Mercurio}}, \bibinfo {author}
  {\bibfnamefont{R.~J.}\ \bibnamefont{Charity}}, \bibinfo {author}
  {\bibfnamefont{R.}~\bibnamefont{Shane}}, \bibinfo {author}
  {\bibfnamefont{L.~G.}\ \bibnamefont{Sobotka}}, \bibinfo {author}
  {\bibfnamefont{J.~M.}\ \bibnamefont{Elson}}, \bibinfo {author}
  {\bibfnamefont{A.~H.}\ \bibnamefont{Wuosmaa}}, \bibinfo {author}
  {\bibfnamefont{A.}~\bibnamefont{Banu}}, \bibinfo {author}
  {\bibfnamefont{M.}~\bibnamefont{McCleskey}}, \bibinfo {author}
  {\bibfnamefont{L.}~\bibnamefont{Trache}}, \bibinfo {author}
  {\bibfnamefont{R.~E.}\ \bibnamefont{Tribble}},\ and\ \bibinfo {author}
  {\bibfnamefont{M.~V.}\ \bibnamefont{Zhukov}},\ }%
  \bibfield{journal}{%
  \Doi{10.1103/PhysRevC.80.034602}{\bibinfo {journal} {Phys. Rev. C}}\ }%
  \textbf{\bibinfo {volume} {80}},\ \bibinfo {pages} {034602} (\bibinfo {year}
  {2009})%
  \bibAnnoteFile{NoStop}{Grigorenko09}%
\bibitem{Charity10}%
  \BibitemOpen
  \bibfield{author}{%
  \bibinfo {author} {\bibfnamefont{R.~J.}\ \bibnamefont{Charity}}, \bibinfo
  {author} {\bibfnamefont{J.~M.}\ \bibnamefont{Elson}}, \bibinfo {author}
  {\bibfnamefont{J.}~\bibnamefont{Manfredi}}, \bibinfo {author}
  {\bibfnamefont{R.}~\bibnamefont{Shane}}, \bibinfo {author}
  {\bibfnamefont{L.~G.}\ \bibnamefont{Sobotka}}, \bibinfo {author}
  {\bibfnamefont{Z.}~\bibnamefont{Chajecki}}, \bibinfo {author}
  {\bibfnamefont{D.}~\bibnamefont{Coupland}}, \bibinfo {author}
  {\bibfnamefont{H.}~\bibnamefont{Iwasaki}}, \bibinfo {author}
  {\bibfnamefont{M.}~\bibnamefont{Kilburn}}, \bibinfo {author}
  {\bibfnamefont{J.}~\bibnamefont{Lee}}, \bibinfo {author}
  {\bibfnamefont{W.~G.}\ \bibnamefont{Lynch}}, \bibinfo {author}
  {\bibfnamefont{A.}~\bibnamefont{Sanetullaev}}, \bibinfo {author}
  {\bibfnamefont{M.~B.}\ \bibnamefont{Tsang}}, \bibinfo {author}
  {\bibfnamefont{J.}~\bibnamefont{Winkelbauer}}, \bibinfo {author}
  {\bibfnamefont{M.}~\bibnamefont{Youngs}}, \bibinfo {author}
  {\bibfnamefont{S.~T.}\ \bibnamefont{Marley}}, \bibinfo {author}
  {\bibfnamefont{D.~V.}\ \bibnamefont{Shetty}}, \bibinfo {author}
  {\bibfnamefont{A.~H.}\ \bibnamefont{Wuosmaa}}, \bibinfo {author}
  {\bibfnamefont{T.~K.}\ \bibnamefont{Ghosh}},\ and\ \bibinfo {author}
  {\bibfnamefont{M.~E.}\ \bibnamefont{Howard}},\ }%
  \bibfield{journal}{%
  \Doi{10.1103/PhysRevC.82.041304}{\bibinfo {journal} {Phys. Rev. C}}\ }%
  \textbf{\bibinfo {volume} {82}},\ \bibinfo {pages} {041304} (\bibinfo {year}
  {2010})%
  \bibAnnoteFile{NoStop}{Charity10}%
\bibitem{Wallace07}%
  \BibitemOpen
  \bibfield{author}{%
  \bibinfo {author} {\bibfnamefont{M.~S.}\ \bibnamefont{Wallace}}, \bibinfo
  {author} {\bibfnamefont{M.~A.}\ \bibnamefont{Famiano}}, \bibinfo {author}
  {\bibfnamefont{M.-J.~V.}\ \bibnamefont{Goethem}}, \bibinfo {author}
  {\bibfnamefont{A.~M.}\ \bibnamefont{Rogers}}, \bibinfo {author}
  {\bibfnamefont{W.~G.}\ \bibnamefont{Lynch}}, \bibinfo {author}
  {\bibfnamefont{J.}~\bibnamefont{Clifford}}, \bibinfo {author}
  {\bibfnamefont{J.}~\bibnamefont{Lee}}, \bibinfo {author}
  {\bibfnamefont{S.}~\bibnamefont{Labostov}}, \bibinfo {author}
  {\bibfnamefont{M.}~\bibnamefont{Mocko}}, \bibinfo {author}
  {\bibfnamefont{L.}~\bibnamefont{Morris}}, \bibinfo {author}
  {\bibfnamefont{A.}~\bibnamefont{Moroni}}, \bibinfo {author}
  {\bibfnamefont{B.~E.}\ \bibnamefont{Nett}}, \bibinfo {author}
  {\bibfnamefont{D.~J.}\ \bibnamefont{Oostdyk}}, \bibinfo {author}
  {\bibfnamefont{R.}~\bibnamefont{Krishnasamy}}, \bibinfo {author}
  {\bibfnamefont{M.~B.}\ \bibnamefont{Tsang}}, \bibinfo {author}
  {\bibfnamefont{R.~D.}\ \bibnamefont{de~Souza}}, \bibinfo {author}
  {\bibfnamefont{S.}~\bibnamefont{Hudan}}, \bibinfo {author}
  {\bibfnamefont{L.~G.}\ \bibnamefont{Sobotka}}, \bibinfo {author}
  {\bibfnamefont{R.~J.}\ \bibnamefont{Charity}}, \bibinfo {author}
  {\bibfnamefont{J.}~\bibnamefont{Elson}},\ and\ \bibinfo {author}
  {\bibfnamefont{G.~L.}\ \bibnamefont{Engel}},\ }%
  \bibfield{journal}{%
  \bibinfo {journal} {Nucl. Instrum. Methods A}\ }%
  \textbf{\bibinfo {volume} {583}},\ \bibinfo {pages} {302} (\bibinfo {year}
  {2007})%
  \bibAnnoteFile{NoStop}{Wallace07}%
\bibitem{Engel07}%
  \BibitemOpen
  \bibfield{author}{%
  \bibinfo {author} {\bibfnamefont{G.~L.}\ \bibnamefont{Engel}}, \bibinfo
  {author} {\bibfnamefont{M.}~\bibnamefont{Sadasivam}}, \bibinfo {author}
  {\bibfnamefont{M.}~\bibnamefont{Nethi}}, \bibinfo {author}
  {\bibfnamefont{J.~M.}\ \bibnamefont{Elson}}, \bibinfo {author}
  {\bibfnamefont{L.~G.}\ \bibnamefont{Sobotka}},\ and\ \bibinfo {author}
  {\bibfnamefont{R.~J.}\ \bibnamefont{Charity}},\ }%
  \bibfield{journal}{%
  \bibinfo {journal} {Nucl. Instrum. Methods A}\ }%
  \textbf{\bibinfo {volume} {573}},\ \bibinfo {pages} {418} (\bibinfo {year}
  {2007})%
  \bibAnnoteFile{NoStop}{Engel07}%
\bibitem{Charity07}%
  \BibitemOpen
  \bibfield{author}{%
  \bibinfo {author} {\bibfnamefont{R.~J.}\ \bibnamefont{Charity}}, \bibinfo
  {author} {\bibfnamefont{S.~A.}\ \bibnamefont{Komarov}}, \bibinfo {author}
  {\bibfnamefont{L.~G.}\ \bibnamefont{Sobotka}}, \bibinfo {author}
  {\bibfnamefont{J.}~\bibnamefont{Clifford}}, \bibinfo {author}
  {\bibfnamefont{D.}~\bibnamefont{Bazin}}, \bibinfo {author}
  {\bibfnamefont{A.}~\bibnamefont{Gade}}, \bibinfo {author}
  {\bibfnamefont{J.}~\bibnamefont{Lee}}, \bibinfo {author}
  {\bibfnamefont{S.~M.}\ \bibnamefont{Lukyanov}}, \bibinfo {author}
  {\bibfnamefont{W.~G.}\ \bibnamefont{Lynch}}, \bibinfo {author}
  {\bibfnamefont{M.}~\bibnamefont{Mocko}}, \bibinfo {author}
  {\bibfnamefont{S.~P.}\ \bibnamefont{Lobastov}}, \bibinfo {author}
  {\bibfnamefont{A.~M.}\ \bibnamefont{Rogers}}, \bibinfo {author}
  {\bibfnamefont{A.}~\bibnamefont{Sanetullaev}}, \bibinfo {author}
  {\bibfnamefont{M.~B.}\ \bibnamefont{Tsang}}, \bibinfo {author}
  {\bibfnamefont{M.~S.}\ \bibnamefont{Wallace}}, \bibinfo {author}
  {\bibfnamefont{S.}~\bibnamefont{Hudan}}, \bibinfo {author}
  {\bibfnamefont{C.}~\bibnamefont{Metelko}}, \bibinfo {author}
  {\bibfnamefont{M.~A.}\ \bibnamefont{Famiano}}, \bibinfo {author}
  {\bibfnamefont{A.~H.}\ \bibnamefont{Wuosmaa}},\ and\ \bibinfo {author}
  {\bibfnamefont{M.~J.}\ \bibnamefont{{van Goethem}}},\ }%
  \bibfield{journal}{%
  \bibinfo {journal} {Phys. Rev. C}\ }%
  \textbf{\bibinfo {volume} {76}},\ \bibinfo {pages} {064313} (\bibinfo {year}
  {2007})%
  \bibAnnoteFile{NoStop}{Charity07}%
\bibitem{Ziegler85}%
  \BibitemOpen
  \bibfield{author}{%
  \bibinfo {author} {\bibfnamefont{J.~F.}\ \bibnamefont{Ziegler}}, \bibinfo
  {author} {\bibfnamefont{J.~P.}\ \bibnamefont{Biersack}},\ and\ \bibinfo
  {author} {\bibfnamefont{U.}~\bibnamefont{Littmark}},\ }%
  \emph{\bibinfo {title} {The Stopping and Range of Ions in Solids}}\ (\bibinfo
  {publisher} {Pergamon Press},\ \bibinfo {address} {New York},\ \bibinfo
  {year} {1985})\ \bibinfo {note} {the code SRIM can be found at
  www.srim.org.}%
  \bibAnnoteFile{Stop}{Ziegler85}%
\bibitem{Anne88}%
  \BibitemOpen
  \bibfield{author}{%
  \bibinfo {author} {\bibfnamefont{R.}~\bibnamefont{Anne}}, \bibinfo {author}
  {\bibfnamefont{J.}~\bibnamefont{Herault}}, \bibinfo {author}
  {\bibfnamefont{R.}~\bibnamefont{Bimbot}}, \bibinfo {author}
  {\bibfnamefont{H.}~\bibnamefont{Gauvin}}, \bibinfo {author}
  {\bibfnamefont{C.}~\bibnamefont{Bastin}},\ and\ \bibinfo {author}
  {\bibfnamefont{F.}~\bibnamefont{Hubert}},\ }%
  \bibfield{journal}{%
  \bibinfo {journal} {Nucl. Instrum. Methods B}\ }%
  \textbf{\bibinfo {volume} {34}},\ \bibinfo {pages} {295} (\bibinfo {year}
  {1988})%
  \bibAnnoteFile{NoStop}{Anne88}%
\bibitem{Charity08}%
  \BibitemOpen
  \bibfield{author}{%
  \bibinfo {author} {\bibfnamefont{R.~J.}\ \bibnamefont{Charity}}, \bibinfo
  {author} {\bibfnamefont{S.~A.}\ \bibnamefont{Komarov}}, \bibinfo {author}
  {\bibfnamefont{L.~G.}\ \bibnamefont{Sobotka}}, \bibinfo {author}
  {\bibfnamefont{J.}~\bibnamefont{Clifford}}, \bibinfo {author}
  {\bibfnamefont{D.}~\bibnamefont{Bazin}}, \bibinfo {author}
  {\bibfnamefont{A.}~\bibnamefont{Gade}}, \bibinfo {author}
  {\bibfnamefont{J.}~\bibnamefont{Lee}}, \bibinfo {author}
  {\bibfnamefont{S.~M.}\ \bibnamefont{Lukyanov}}, \bibinfo {author}
  {\bibfnamefont{W.~G.}\ \bibnamefont{Lynch}}, \bibinfo {author}
  {\bibfnamefont{M.}~\bibnamefont{Mocko}}, \bibinfo {author}
  {\bibfnamefont{S.~P.}\ \bibnamefont{Lobastov}}, \bibinfo {author}
  {\bibfnamefont{A.~M.}\ \bibnamefont{Rogers}}, \bibinfo {author}
  {\bibfnamefont{A.}~\bibnamefont{Sanetullaev}}, \bibinfo {author}
  {\bibfnamefont{M.~B.}\ \bibnamefont{Tsang}}, \bibinfo {author}
  {\bibfnamefont{M.~S.}\ \bibnamefont{Wallace}}, \bibinfo {author}
  {\bibfnamefont{R.~G.~T.}\ \bibnamefont{Zegers}}, \bibinfo {author}
  {\bibfnamefont{S.}~\bibnamefont{Hudan}}, \bibinfo {author}
  {\bibfnamefont{C.}~\bibnamefont{Metelko}}, \bibinfo {author}
  {\bibfnamefont{M.~A.}\ \bibnamefont{Famiano}}, \bibinfo {author}
  {\bibfnamefont{A.~H.}\ \bibnamefont{Wuosmaa}},\ and\ \bibinfo {author}
  {\bibfnamefont{M.~J.}\ \bibnamefont{van Goethem}},\ }%
  \bibfield{journal}{%
  \Doi{10.1103/PhysRevC.78.054307}{\bibinfo {journal} {Phys. Rev. C}}\ }%
  \textbf{\bibinfo {volume} {78}},\ \bibinfo {pages} {054307} (\bibinfo {year}
  {2008})%
  \bibAnnoteFile{NoStop}{Charity08}%
\bibitem{Lane58}%
  \BibitemOpen
  \bibfield{author}{%
  \bibinfo {author} {\bibfnamefont{A.~M.}\ \bibnamefont{Lane}}\ and\ \bibinfo
  {author} {\bibfnamefont{R.~G.}\ \bibnamefont{Thomas}},\ }%
  \bibfield{journal}{%
  \Doi{10.1103/RevModPhys.30.257}{\bibinfo {journal} {Rev. Mod. Phys.}}\ }%
  \textbf{\bibinfo {volume} {30}},\ \bibinfo {pages} {257} (\bibinfo {year}
  {1958})%
  \bibAnnoteFile{NoStop}{Lane58}%
\bibitem{Barker99}%
  \BibitemOpen
  \bibfield{author}{%
  \bibinfo {author} {\bibfnamefont{F.~C.}\ \bibnamefont{Barker}},\ }%
  \bibfield{journal}{%
  \Doi{10.1103/PhysRevC.59.535}{\bibinfo {journal} {Phys. Rev. C}}\ }%
  \textbf{\bibinfo {volume} {59}},\ \bibinfo {pages} {535} (\bibinfo {year}
  {1999})%
  \bibAnnoteFile{NoStop}{Barker99}%
\bibitem{Biedenharn53}%
  \BibitemOpen
  \bibfield{author}{%
  \bibinfo {author} {\bibfnamefont{L.~C.}\ \bibnamefont{Biedenharn}}\ and\
  \bibinfo {author} {\bibfnamefont{M.~E.}\ \bibnamefont{Rose}},\ }%
  \bibfield{journal}{%
  \bibinfo {journal} {Rev. Mod. Phys.}\ }%
  \textbf{\bibinfo {volume} {25}},\ \bibinfo {pages} {729} (\bibinfo {year}
  {1953})%
  \bibAnnoteFile{NoStop}{Biedenharn53}%
\bibitem{Frauenfelder53}%
  \BibitemOpen
  \bibfield{author}{%
  \bibinfo {author} {\bibfnamefont{H.}~\bibnamefont{Frauenfelder}},\ }%
  \bibfield{journal}{%
  \bibinfo {journal} {Annu. Rev. Nucl. Sci.}\ }%
  \textbf{\bibinfo {volume} {2}},\ \bibinfo {pages} {129} (\bibinfo {year}
  {1953})%
  \bibAnnoteFile{NoStop}{Frauenfelder53}%
\bibitem{ENSDF}%
  \BibitemOpen
  \bibinfo {note} {Evaluated Nuclear Structure Data File (ENSDF),
  http://www.nndc.bnl.gov/ensdf/}%
  \bibAnnoteFile{NoStop}{ENSDF}%
\bibitem{Audi03}%
  \BibitemOpen
  \bibfield{author}{%
  \bibinfo {author} {\bibfnamefont{G.}~\bibnamefont{Audi}}, \bibinfo {author}
  {\bibfnamefont{A.~H.}\ \bibnamefont{Wapstra}},\ and\ \bibinfo {author}
  {\bibfnamefont{C.}~\bibnamefont{Thibault}},\ }%
  \bibfield{journal}{%
  \bibinfo {journal} {Nucl. Phys.}\ }%
  \textbf{\bibinfo {volume} {A729}},\ \bibinfo {pages} {337} (\bibinfo {year}
  {2003})%
  \bibAnnoteFile{NoStop}{Audi03}%
\bibitem{Robertson74}%
  \BibitemOpen
  \bibfield{author}{%
  \bibinfo {author} {\bibfnamefont{R.~G.~H.}\ \bibnamefont{Robertson}},
  \bibinfo {author} {\bibfnamefont{S.}~\bibnamefont{Martin}}, \bibinfo {author}
  {\bibfnamefont{W.~R.}\ \bibnamefont{Falk}}, \bibinfo {author}
  {\bibfnamefont{D.}~\bibnamefont{Ingham}},\ and\ \bibinfo {author}
  {\bibfnamefont{A.}~\bibnamefont{Djaloeis}},\ }%
  \bibfield{journal}{%
  \Doi{10.1103/PhysRevLett.32.1207}{\bibinfo {journal} {Phys. Rev. Lett.}}\ }%
  \textbf{\bibinfo {volume} {32}},\ \bibinfo {pages} {1207} (\bibinfo {year}
  {1974})%
  \bibAnnoteFile{NoStop}{Robertson74}%
\bibitem{Robertson76}%
  \BibitemOpen
  \bibfield{author}{%
  \bibinfo {author} {\bibfnamefont{R.~G.~H.}\ \bibnamefont{Robertson}},
  \bibinfo {author} {\bibfnamefont{W.}~\bibnamefont{Benenson}}, \bibinfo
  {author} {\bibfnamefont{E.}~\bibnamefont{Kashy}},\ and\ \bibinfo {author}
  {\bibfnamefont{D.}~\bibnamefont{Mueller}},\ }%
  \bibfield{journal}{%
  \Doi{10.1103/PhysRevC.13.1018}{\bibinfo {journal} {Phys. Rev. C}}\ }%
  \textbf{\bibinfo {volume} {13}},\ \bibinfo {pages} {1018} (\bibinfo {year}
  {1976})%
  \bibAnnoteFile{NoStop}{Robertson76}%
\bibitem{Tribble76}%
  \BibitemOpen
  \bibfield{author}{%
  \bibinfo {author} {\bibfnamefont{R.~E.}\ \bibnamefont{Tribble}}, \bibinfo
  {author} {\bibfnamefont{R.~A.}\ \bibnamefont{Kenefick}},\ and\ \bibinfo
  {author} {\bibfnamefont{R.~L.}\ \bibnamefont{Spross}},\ }%
  \bibfield{journal}{%
  \Doi{10.1103/PhysRevC.13.50}{\bibinfo {journal} {Phys. Rev. C}}\ }%
  \textbf{\bibinfo {volume} {13}},\ \bibinfo {pages} {50} (\bibinfo {year}
  {1976})%
  \bibAnnoteFile{NoStop}{Tribble76}%
\bibitem{Bochkarev89}%
  \BibitemOpen
  \bibfield{author}{%
  \bibinfo {author} {\bibfnamefont{O.~V.}\ \bibnamefont{Bochkarev}}, \bibinfo
  {author} {\bibfnamefont{L.~.~V.}\ \bibnamefont{Chulkov}}, \bibinfo {author}
  {\bibfnamefont{A.~A.}\ \bibnamefont{Korsheninnikov}}, \bibinfo {author}
  {\bibfnamefont{E.~A.}\ \bibnamefont{Kuz'min}}, \bibinfo {author}
  {\bibfnamefont{I.~G.}\ \bibnamefont{Mukha}},\ and\ \bibinfo {author}
  {\bibfnamefont{G.~B.}\ \bibnamefont{Yankov}},\ }%
  \bibfield{journal}{%
  \bibinfo {journal} {Nucl. Phys.}\ }%
  \textbf{\bibinfo {volume} {A505}},\ \bibinfo {pages} {215} (\bibinfo {year}
  {1989})%
  \bibAnnoteFile{NoStop}{Bochkarev89}%
\bibitem{Charity09}%
  \BibitemOpen
  \bibfield{author}{%
  \bibinfo {author} {\bibfnamefont{R.~J.}\ \bibnamefont{Charity}}, \bibinfo
  {author} {\bibfnamefont{T.~D.}\ \bibnamefont{Wiser}}, \bibinfo {author}
  {\bibfnamefont{K.}~\bibnamefont{Mercurio}}, \bibinfo {author}
  {\bibfnamefont{R.}~\bibnamefont{Shane}}, \bibinfo {author}
  {\bibfnamefont{L.~G.}\ \bibnamefont{Sobotka}}, \bibinfo {author}
  {\bibfnamefont{A.~H.}\ \bibnamefont{Wuosmaa}}, \bibinfo {author}
  {\bibfnamefont{A.}~\bibnamefont{Banu}}, \bibinfo {author}
  {\bibfnamefont{L.}~\bibnamefont{Trache}},\ and\ \bibinfo {author}
  {\bibfnamefont{R.~E.}\ \bibnamefont{Tribble}},\ }%
  \bibfield{journal}{%
  \Doi{10.1103/PhysRevC.80.024306}{\bibinfo {journal} {Phys. Rev. C}}\ }%
  \textbf{\bibinfo {volume} {80}},\ \bibinfo {pages} {024306} (\bibinfo {year}
  {2009})%
  \bibAnnoteFile{NoStop}{Charity09}%
\bibitem{Cohen65}%
  \BibitemOpen
  \bibfield{author}{%
  \bibinfo {author} {\bibfnamefont{S.}~\bibnamefont{Cohen}}\ and\ \bibinfo
  {author} {\bibfnamefont{D.}~\bibnamefont{Kurath}},\ }%
  \bibfield{journal}{%
  \bibinfo {journal} {Nuclear Physics}\ }%
  \textbf{\bibinfo {volume} {73}},\ \bibinfo {pages} {1} (\bibinfo {year}
  {1965})%
  \bibAnnoteFile{NoStop}{Cohen65}%
\bibitem{Cohen67}%
  \BibitemOpen
  \bibfield{author}{%
  \bibinfo {author} {\bibfnamefont{S.}~\bibnamefont{Cohen}}\ and\ \bibinfo
  {author} {\bibfnamefont{D.}~\bibnamefont{Kurath}},\ }%
  \bibfield{journal}{%
  \bibinfo {journal} {Nuclear Physics A}\ }%
  \textbf{\bibinfo {volume} {101}},\ \bibinfo {pages} {1 } (\bibinfo {year}
  {1967}),\ ISSN \bibinfo {issn} {0375-9474}%
  \bibAnnoteFile{NoStop}{Cohen67}%
\bibitem{Brown67}%
  \BibitemOpen
  \bibfield{author}{%
  \bibinfo {author} {\bibfnamefont{B.~A.}\ \bibnamefont{Brown}}\ and\ \bibinfo
  {author} {\bibfnamefont{F.~C.}\ \bibnamefont{Barker}},\ }%
  \bibfield{journal}{%
  \Doi{10.1103/PhysRevC.67.041304}{\bibinfo {journal} {Phys. Rev. C}}\ }%
  \textbf{\bibinfo {volume} {67}},\ \bibinfo {pages} {041304} (\bibinfo {year}
  {2003})%
  \bibAnnoteFile{NoStop}{Brown67}%
\bibitem{Barker02}%
  \BibitemOpen
  \bibfield{author}{%
  \bibinfo {author} {\bibfnamefont{F.~C.}\ \bibnamefont{Barker}},\ }%
  \bibfield{journal}{%
  \Doi{10.1103/PhysRevC.66.047603}{\bibinfo {journal} {Phys. Rev. C}}\ }%
  \textbf{\bibinfo {volume} {66}},\ \bibinfo {pages} {047603} (\bibinfo {month}
  {Oct}\ \bibinfo {year} {2002})%
  \bibAnnoteFile{NoStop}{Barker02}%
\bibitem{Barker03}%
  \BibitemOpen
  \bibfield{author}{%
  \bibinfo {author} {\bibfnamefont{F.~C.}\ \bibnamefont{Barker}},\ }%
  \bibfield{journal}{%
  \Doi{10.1103/PhysRevC.67.049902}{\bibinfo {journal} {Phys. Rev. C}}\ }%
  \textbf{\bibinfo {volume} {67}},\ \bibinfo {pages} {049902(E)} (\bibinfo
  {year} {2003})%
  \bibAnnoteFile{NoStop}{Barker03}%
\bibitem{McGrath67}%
  \BibitemOpen
  \bibfield{author}{%
  \bibinfo {author} {\bibfnamefont{R.~L.}\ \bibnamefont{McGrath}}, \bibinfo
  {author} {\bibfnamefont{J.}~\bibnamefont{Cerny}},\ and\ \bibinfo {author}
  {\bibfnamefont{E.}~\bibnamefont{Norbeck}},\ }%
  \bibfield{journal}{%
  \Doi{10.1103/PhysRevLett.19.1442}{\bibinfo {journal} {Phys. Rev. Lett.}}\ }%
  \textbf{\bibinfo {volume} {19}},\ \bibinfo {pages} {1442} (\bibinfo {year}
  {1967})%
  \bibAnnoteFile{NoStop}{McGrath67}%
\bibitem{Grigorenko02}%
  \BibitemOpen
  \bibfield{author}{%
  \bibinfo {author} {\bibfnamefont{L.}~\bibnamefont{Grigorenko}}, \bibinfo
  {author} {\bibfnamefont{R.}~\bibnamefont{Johnson}}, \bibinfo {author}
  {\bibfnamefont{I.}~\bibnamefont{Mukha}}, \bibinfo {author}
  {\bibfnamefont{I.}~\bibnamefont{Thompson}},\ and\ \bibinfo {author}
  {\bibfnamefont{M.}~\bibnamefont{Zhukov}},\ }%
  \bibfield{journal}{%
  \bibinfo {journal} {The European Physical Journal A - Hadrons and Nuclei}\ }%
  \textbf{\bibinfo {volume} {15}},\ \bibinfo {pages} {125} (\bibinfo {year}
  {2002})%
  \bibAnnoteFile{NoStop}{Grigorenko02}%
\bibitem{Robertson75}%
  \BibitemOpen
  \bibfield{author}{%
  \bibinfo {author} {\bibfnamefont{R.~G.~H.}\ \bibnamefont{Robertson}},
  \bibinfo {author} {\bibfnamefont{W.~S.}\ \bibnamefont{Chien}},\ and\ \bibinfo
  {author} {\bibfnamefont{D.~R.}\ \bibnamefont{Goosman}},\ }%
  \bibfield{journal}{%
  \Doi{10.1103/PhysRevLett.34.33}{\bibinfo {journal} {Phys. Rev. Lett.}}\ }%
  \textbf{\bibinfo {volume} {34}},\ \bibinfo {pages} {33} (\bibinfo {month}
  {Jan}\ \bibinfo {year} {1975})%
  \bibAnnoteFile{NoStop}{Robertson75}%
\bibitem{Woods88}%
  \BibitemOpen
  \bibfield{author}{%
  \bibinfo {author} {\bibfnamefont{C.~L.}\ \bibnamefont{Woods}}, \bibinfo
  {author} {\bibfnamefont{F.~C.}\ \bibnamefont{Barker}}, \bibinfo {author}
  {\bibfnamefont{W.~N.}\ \bibnamefont{Catford}}, \bibinfo {author}
  {\bibfnamefont{L.~K.}\ \bibnamefont{Fifield}},\ and\ \bibinfo {author}
  {\bibfnamefont{N.~A.}\ \bibnamefont{Orr}},\ }%
  \bibfield{journal}{%
  \bibinfo {journal} {Aust. J. Phys.}\ }%
  \textbf{\bibinfo {volume} {41}},\ \bibinfo {pages} {525} (\bibinfo {year}
  {1988})%
  \bibAnnoteFile{NoStop}{Woods88}%
\bibitem{Barker03a}%
  \BibitemOpen
  \bibfield{author}{%
  \bibinfo {author} {\bibfnamefont{F.~C.}\ \bibnamefont{Barker}},\ }%
  \bibfield{journal}{%
  \Doi{10.1103/PhysRevC.68.054602}{\bibinfo {journal} {Phys. Rev. C}}\ }%
  \textbf{\bibinfo {volume} {68}},\ \bibinfo {pages} {054602} (\bibinfo {year}
  {2003})%
  \bibAnnoteFile{NoStop}{Barker03a}%
\bibitem{Chakraborty10}%
  \BibitemOpen
  \bibfield{author}{%
  \bibinfo {author} {\bibfnamefont{N.}~\bibnamefont{Chakraborty}}, \bibinfo
  {author} {\bibfnamefont{B.~D.}\ \bibnamefont{Fields}},\ and\ \bibinfo
  {author} {\bibfnamefont{K.~A.}\ \bibnamefont{Olive}}}%
   (\bibinfo {year} {2010}),\
  \Eprint{http://arxiv.org/abs/nucl-th/1011.0722v1}{arXiv:nucl-th/1011.0722v1}%
  \bibAnnoteFile{NoStop}{Chakraborty10}%
\bibitem{Dixit91}%
  \BibitemOpen
  \bibfield{author}{%
  \bibinfo {author} {\bibfnamefont{S.}~\bibnamefont{Dixit}}, \bibinfo {author}
  {\bibfnamefont{W.}~\bibnamefont{Bertozzi}}, \bibinfo {author}
  {\bibfnamefont{T.~N.}\ \bibnamefont{Buti}}, \bibinfo {author}
  {\bibfnamefont{J.~M.}\ \bibnamefont{Finn}}, \bibinfo {author}
  {\bibfnamefont{F.~W.}\ \bibnamefont{Hersman}}, \bibinfo {author}
  {\bibfnamefont{C.~E.}\ \bibnamefont{Hyde-Wright}}, \bibinfo {author}
  {\bibfnamefont{M.~V.}\ \bibnamefont{Hynes}}, \bibinfo {author}
  {\bibfnamefont{M.~A.}\ \bibnamefont{Kovash}}, \bibinfo {author}
  {\bibfnamefont{B.~E.}\ \bibnamefont{Norum}}, \bibinfo {author}
  {\bibfnamefont{J.~J.}\ \bibnamefont{Kelly}}, \bibinfo {author}
  {\bibfnamefont{A.~D.}\ \bibnamefont{Bacher}}, \bibinfo {author}
  {\bibfnamefont{G.~T.}\ \bibnamefont{Emery}}, \bibinfo {author}
  {\bibfnamefont{C.~C.}\ \bibnamefont{Foster}}, \bibinfo {author}
  {\bibfnamefont{W.~P.}\ \bibnamefont{Jones}}, \bibinfo {author}
  {\bibfnamefont{D.~W.}\ \bibnamefont{Miller}}, \bibinfo {author}
  {\bibfnamefont{B.~L.}\ \bibnamefont{Berman}},\ and\ \bibinfo {author}
  {\bibfnamefont{D.~J.}\ \bibnamefont{Millener}},\ }%
  \bibfield{journal}{%
  \Doi{10.1103/PhysRevC.43.1758}{\bibinfo {journal} {Phys. Rev. C}}\ }%
  \textbf{\bibinfo {volume} {43}},\ \bibinfo {pages} {1758} (\bibinfo {year}
  {1991})%
  \bibAnnoteFile{NoStop}{Dixit91}%
\bibitem{Curtis08}%
  \BibitemOpen
  \bibfield{author}{%
  \bibinfo {author} {\bibfnamefont{N.}~\bibnamefont{Curtis}}, \bibinfo {author}
  {\bibfnamefont{N.~L.}\ \bibnamefont{Achouri}}, \bibinfo {author}
  {\bibfnamefont{N.~I.}\ \bibnamefont{Ashwood}}, \bibinfo {author}
  {\bibfnamefont{H.~G.}\ \bibnamefont{Bohlen}}, \bibinfo {author}
  {\bibfnamefont{W.~N.}\ \bibnamefont{Catford}}, \bibinfo {author}
  {\bibfnamefont{N.~M.}\ \bibnamefont{Clarke}}, \bibinfo {author}
  {\bibfnamefont{M.}~\bibnamefont{Freer}}, \bibinfo {author}
  {\bibfnamefont{P.~J.}\ \bibnamefont{Haigh}}, \bibinfo {author}
  {\bibfnamefont{B.}~\bibnamefont{Laurent}}, \bibinfo {author}
  {\bibfnamefont{N.~A.}\ \bibnamefont{Orr}}, \bibinfo {author}
  {\bibfnamefont{N.~P.}\ \bibnamefont{Patterson}}, \bibinfo {author}
  {\bibfnamefont{N.}~\bibnamefont{Soi\ifmmode~\acute{c}\else \'{c}\fi{}}},
  \bibinfo {author} {\bibfnamefont{J.~S.}\ \bibnamefont{Thomas}},\ and\
  \bibinfo {author} {\bibfnamefont{V.}~\bibnamefont{Ziman}},\ }%
  \bibfield{journal}{%
  \Doi{10.1103/PhysRevC.77.021301}{\bibinfo {journal} {Phys. Rev. C}}\ }%
  \textbf{\bibinfo {volume} {77}},\ \bibinfo {pages} {021301} (\bibinfo {year}
  {2008})%
  \bibAnnoteFile{NoStop}{Curtis08}%
\bibitem{Curtis10}%
  \BibitemOpen
  \bibfield{author}{%
  \bibinfo {author} {\bibfnamefont{N.}~\bibnamefont{Curtis}}, \bibinfo {author}
  {\bibfnamefont{N.~L.}\ \bibnamefont{Achouri}}, \bibinfo {author}
  {\bibfnamefont{N.~I.}\ \bibnamefont{Ashwood}}, \bibinfo {author}
  {\bibfnamefont{H.~G.}\ \bibnamefont{Bohlen}}, \bibinfo {author}
  {\bibfnamefont{W.~N.}\ \bibnamefont{Catford}}, \bibinfo {author}
  {\bibfnamefont{N.~M.}\ \bibnamefont{Clarke}}, \bibinfo {author}
  {\bibfnamefont{M.}~\bibnamefont{Freer}}, \bibinfo {author}
  {\bibfnamefont{P.~J.}\ \bibnamefont{Haigh}}, \bibinfo {author}
  {\bibfnamefont{B.}~\bibnamefont{Laurent}}, \bibinfo {author}
  {\bibfnamefont{N.~A.}\ \bibnamefont{Orr}}, \bibinfo {author}
  {\bibfnamefont{N.~P.}\ \bibnamefont{Patterson}}, \bibinfo {author}
  {\bibfnamefont{N.}~\bibnamefont{Soi\ifmmode~\acute{c}\else \'{c}\fi{}}},
  \bibinfo {author} {\bibfnamefont{J.~S.}\ \bibnamefont{Thomas}},\ and\
  \bibinfo {author} {\bibfnamefont{V.}~\bibnamefont{Ziman}},\ }%
  \bibfield{journal}{%
  \Doi{10.1103/PhysRevC.82.029907}{\bibinfo {journal} {Phys. Rev. C}}\ }%
  \textbf{\bibinfo {volume} {82}},\ \bibinfo {pages} {029907(E)} (\bibinfo
  {year} {2010})%
  \bibAnnoteFile{NoStop}{Curtis10}%
\end{thebibliography}
\end{document}